\documentclass[final,twocolumn,12pt]{elsarticle}

\usepackage{amssymb}
\usepackage{amsmath}

\usepackage{lineno}

\usepackage{adjustbox}
\usepackage{amsmath,amssymb,amsfonts}
\usepackage{graphicx}
\usepackage{textcomp}
\usepackage{multicol}
\usepackage{multirow}
\usepackage{subcaption}
\usepackage{enumitem}
\usepackage{booktabs}
\usepackage{array}
\newcolumntype{x}[1]{>{\centering\arraybackslash\hspace{0pt}}p{#1}}

\usepackage{pifont}

\usepackage{pst-all}
\usepackage{pstricks-add}
\usepackage{float}

\usepackage{xcolor}

\usepackage{lscape}

\usepackage{tikz}
\usepackage{pgfplots}

\usepackage{tikzscale}
\usepackage{transparent}

\usepackage{forest}
\usetikzlibrary{trees,positioning,shapes,shadows,arrows.meta}

\usepackage{algorithm}
\usepackage[noend]{algpseudocode}
\usepackage{hyperref}
\usepackage{comment}

\usepackage{fancyhdr}

\usepackage{flushend}

\usepackage{longtable}    
\usepackage{booktabs}     
\usepackage{colortbl}     

\definecolor{customblue}{HTML}{cfe2f3}
\definecolor{custombluedark}{HTML}{6fa8dc}
\definecolor{customred}{HTML}{f6e6e6}
\definecolor{customreddark}{HTML}{e06666}
\definecolor{customyellow}{HTML}{f9eed2}
\definecolor{customyellowdark}{HTML}{dbb12e}
\definecolor{customorange}{HTML}{fff1e3}
\definecolor{customorangedark}{HTML}{e69138}
\definecolor{customgray}{HTML}{efefef}
\definecolor{customgraydark}{HTML}{999999}
\definecolor{customgreen}{HTML}{e5f9de}
\definecolor{customgreendark}{HTML}{6aa84f}

\usepackage{longtable}
\usepackage{booktabs}

\journal{Computer Communications}

\begin{document}

\begin{frontmatter}



\title{AI-driven Intrusion Detection for UAV in Smart Urban Ecosystems: A Comprehensive Survey}

\author[label4]{Abdullah Khanfor} 

\affiliation[label4]{organization={College of Computer Science and Information Systems, Najran University},
            city={Najran},
            country={KSA}}

\author[label1]{Raby Hamadi}
\affiliation[label1]{organization={Saudi Technology and Security Comprehensive Control Company (Tahakom)},
             city={Riyadh},
             country={KSA}}

\author[label3]{Noureddine Lasla} 

\affiliation[label3]{organization={National School of Artificial Intelligence (ENSIA)},
            city={Algiers},
            country={Algeria}}

\author[label2]{Hakim Ghazzai} 

\affiliation[label2]{organization={Computer, Electrical, and Mathematical Sciences and Engineering (CEMSE) Division, King Abdullah University of Science and Technology (KAUST)},
            city={Thuwal},
            state={Makkah},
            country={KSA}}

\begin{abstract}
\textcolor{black}{UAVs have the potential to revolutionize urban management and provide valuable services to citizens. They can be deployed across diverse applications, including traffic monitoring, disaster response, environmental monitoring, and numerous other domains. However, this integration introduces novel security challenges that must be addressed to ensure safe and trustworthy urban operations. This paper provides a structured, evidence-based synthesis of UAV applications in smart cities and their associated security challenges as reported in the literature over the last decade, with particular emphasis on developments from 2019 to 2025. We categorize these challenges into two primary classes: 1) cyber-attacks targeting the communication infrastructure of UAVs and 2) unwanted or unauthorized physical intrusions by UAVs themselves. We examine the potential of Artificial Intelligence (AI) techniques in developing intrusion detection mechanisms to mitigate these security threats. We analyze how AI-based methods, such as machine/deep learning for anomaly detection and computer vision for object recognition, can play a pivotal role in enhancing UAV security through unified detection systems that address both cyber and physical threats. Furthermore, we consolidate publicly available UAV datasets across network traffic and vision modalities suitable for Intrusion Detection Systems (IDS) development and evaluation. The paper concludes by identifying ten key research directions, including scalability, robustness, explainability, data scarcity, automation, hybrid detection, large language models, multimodal approaches, federated learning, and privacy preservation. Finally, we discuss the practical challenges of implementing UAV IDS solutions in real-world smart city environments.}
\end{abstract}








\end{frontmatter}


\section{Introduction}

\thispagestyle{fancy}
\fancyhf{}
\fancyfoot[C]{\footnotesize This paper is accepted for publication in Computer Communications, Jan. 2026. \newline \textcopyright~2026 Elsevier B.V. . Personal use of this material is permitted. Permission from Elsevier must be obtained for all other uses, in any current or future media, including reprinting/republishing this material for advertising or promotional purposes, creating new collective works, for resale or redistribution to servers or lists, or reuse of any copyrighted component of this work in other works.}
\renewcommand{\headrulewidth}{0pt}
\renewcommand{\footrulewidth}{0pt}

Smart city initiatives represent a transformative approach to urban management, simultaneously enhancing infrastructure while generating new revenue sources, improving operational efficiency, and reducing costs for both governments and residents. Fundamentally, Smart cities utilize a network of interconnected IoT devices and other technologies to enhance the quality of life and promote economic growth. The utilization of smart cities can be achieved in the following four phases: 1) collecting real-time data through smart sensors, 2) analyzing this data for operational insights, 3) presenting findings to decision-makers, and 4) implementing service improvements in areas like energy distribution, waste management, and traffic control. Connected technologies, such as mobile devices, smart buildings, and vehicles, which result in reduced costs, promote sustainability, and improve air quality~\cite{bisio2022systematic, kumar2022drone, rodrigues2022drone, kaabi20222, naz2022investigation, cao2022mgdp}.

UAVs are becoming a critical component of IoT technologies, significantly influencing a wide range of smart city applications and services. That includes UAVs as communication infrastructure, pollution monitoring, security surveillance,  delivery services, disaster management, and Intelligent Transportation Systems (ITS)~\cite{MOHAMED2020119293, 6842265, 8746387, nguyen2021drone, alvares2021blockchain, payal2021drones, butilua2022urban}. One of the significant applications of UAVs is to contribute to the infrastructure needs of smart cities. In fact, UAVs offer numerous advantages over traditional ground infrastructure due to their Three-Dimensional (3D) mobility and flexibility~\cite{pourghasemian2022ai, mazaherifar2022uav, kang2022improving}. Moreover, using UAVs for security and safety enables monitoring from a different perspective, known as the aerial Field of View (FoV), while maintaining adequate transmission links with reduced path loss and a higher probability of Line-of-Sight (LoS) links~\cite{won2022survey, park2022survey}. Similarly, UAVs can serve as gliding IoT devices that can link networks to form Flying Ad Hoc Networks (FANETs), offering a wide range of connectivity options for data sharing, dissemination, and collection~\cite{lansky2022reinforcement, kakamoukas2022fanets, goumiri2022security, ahmad2022secure}.

In fact, UAVs play a critical role in various smart city applications, including traffic management, healthcare, and military operations, offering innovative solutions to diverse challenges. In traffic management, UAVs are used to mitigate road congestion caused by the increasing number of private automobiles. Huang et al. \cite{huang2021decentralized} proposed a distributed positioning strategy in which UAV networks identify obstructions and provide aerial surveillance to enhance ground vehicle monitoring, outperforming traditional ground-based methods. Similarly, Kumar et al.\cite{kumar2021novel} introduced a Software-Defined Drone Network (SDDN) to monitor roadway traffic, integrating collision avoidance and reducing computational overhead. In healthcare, UAVs enable low-power and secure interaction with body sensor hives to collect health data~\cite{zhao2021structural}, deliver self-collection kits for testing~\cite{munawar2021towards}, and execute UAV-based disinfectant spraying systems, proven 50 times more effective than manual disinfection~\cite{harfina2021disinfectant}. Furthermore, UAVs equipped with deep learning models and medical sensors have been utilized to detect COVID-19 cases in real time~\cite{9492907}. Beyond civilian applications and uses, UAVs are widely deployed in military operations for surveillance, vehicle detection, border monitoring~\cite{6761569, gupta2022edge}, and other strategic purposes~\cite{9663283, 9419470, gargalakos2021role}, showcasing their adaptability across sectors.

The inherent vulnerabilities of UAVs as cyber-physical systems stem from their reliance on wireless communication and their minimal cybersecurity integration during development, creating critical exposure points that adversaries increasingly exploit through sophisticated attack vectors~\cite{dahiya2019unmanned, 6568373, kim2012cyber, krishna2017review, 9599697}. Moreover, the amount and type of information stored by UAVs make them a desirable target for eavesdropping, tampering, and remote access using cyberattacks. According to~\cite{9599697}, there are six main categories of cyber-attacks on UAVs based on the entry point of the attack: channel jamming, message interception, message deletion, message injection, message spoofing, and on-board system attack. These attacks exploit security vulnerabilities within the UAV system, leading to potentially severe consequences. 

\textcolor{black}{Early publicized incidents, such as a reported military drone capture attributed to navigation deception \cite{shane2011drone, kerns2014unmanned}, illustrated how spoofed signals and link manipulation can result in operational losses.} As UAVs expand beyond military applications into widespread commercial and personal leisure use. In fact, UAVs are increasingly used for commercial and private purposes, and cyber-attacks can raise serious privacy concerns and pose various other threats ~\cite{parlin2018jamming}. Furthermore, the absence of authorities' policies and guidelines governing the safe use of UAVs raises safety concerns. The more UAVs are operated, the greater the risk of exposure to cyberattacks such as spoofing and jamming~\cite{silva2017gps}. \textcolor{black}{Recent advisories from aviation authorities, airport security, and municipal enforcement consistently note growth in unauthorized UAV sightings, indicating patterns highlighted in recent surveys on UAV cybersecurity and counter‑UAV operations \cite{9599697, park2021survey, yu2023cybersecurity} and motivating IDS designs that assume recurring intrusions.}

Along with the cyber-attack risks to UAVs, they can also pose a threat to the privacy and safety of people and property. Due to their dynamic mobility and flexibility, UAVs can easily invade people's privacy by capturing unauthorized photos or videos. Furthermore, UAVs can be particularly hazardous in causing crashes and damage, especially in densely populated areas, if proper precautionary measures are not in place. Unwanted UAVs can collide with the ground infrastructure, with other flying units operating in the area, or even with birds. Therefore, intelligent solutions are necessary to mitigate these intrusions and address the privacy, safety, and environmental concerns associated with unauthorized UAV operations.

\begin{table*}[ht]
\centering
\caption{Comparative Analysis of Existing Survey Literature on UAV Security and Machine Learning Applications}\label{existingsurveys}
{\scriptsize
\setlength{\tabcolsep}{2pt}
\begin{tabular}{p{4cm}cp{3.8cm}cccccc}
\hline
\textbf{Survey} & \textbf{Year} & \textbf{Venue} &
\textbf{\begin{tabular}[c]{@{}c@{}}\tiny ML/AI\\ \tiny Methods\end{tabular}} &
\textbf{\begin{tabular}[c]{@{}c@{}}\tiny UAV/Drone\\ \tiny Focus\end{tabular}} &
\textbf{\begin{tabular}[c]{@{}c@{}}\tiny Security\\ \tiny Coverage\end{tabular}} &
\textbf{\begin{tabular}[c]{@{}c@{}}\tiny RL\\ \tiny Integration\end{tabular}} &
\textbf{\begin{tabular}[c]{@{}c@{}}\tiny Dataset\\ \tiny Evaluation\end{tabular}} \\
\hline

{\tiny Recent Advances in ML-Driven Intrusion Detection in Transportation~\cite{bangui2021recent}} & 2021 & \tiny Procedia CS & \checkmark & \ding{55} & \checkmark & \ding{55} & \checkmark \\
{\tiny Survey on Anti-Drone Systems~\cite{park2021survey}} & 2021 & \tiny IEEE Access & \ding{55} & \checkmark & \checkmark & \ding{55} & \ding{55} \\
{\tiny Protect Your Sky: Counter UAV Systems~\cite{kang2020protect}} & 2020 & \tiny IEEE Access & \ding{55} & \checkmark & \checkmark & \ding{55} & \ding{55} \\
{\tiny Survey of intrusion detection for cyber-physical systems~\cite{mitchell2014survey}} & 2014 & \tiny  ACM Computing Surveys & \ding{55} & \ding{55} & \checkmark & \ding{55} & \checkmark \\
{\tiny IDS for Networked UAVs~\cite{choudhary2018intrusion}} & 2018 & \tiny IEEE IWCMC 18 & \ding{55} & \checkmark & \checkmark & \ding{55} & \checkmark \\
{\tiny Detection of Unauthorized Drones~\cite{9765451}} & 2022 & \tiny IEEE Sensors J. & \checkmark & \checkmark & \checkmark & \ding{55} & \checkmark \\
{\tiny IDS in IoT with ML Approach~\cite{9392075}} & 2020 & \tiny 2020 6th ICSITech & \checkmark & \ding{55} & \checkmark & \checkmark & \checkmark \\
{\tiny Cyber-security and RL survey~\cite{ADAWADKAR2022105116}} & 2022 & \tiny Eng. App. of AI & \checkmark & \ding{55} & \checkmark & \checkmark & \checkmark \\
{\tiny IDS models based on CIC-IDS2018~\cite{leevy2020survey}} & 2020 & \tiny J. Big Data & \checkmark & \ding{55} & \checkmark & \ding{55} & \checkmark \\
{\tiny ML approaches to IDS in UAVs~\cite{al2024machine}} & 2024 & \tiny Neural Comput. \& Apps & \checkmark & \checkmark & \checkmark & \ding{55} & \checkmark \\
{\tiny Security and privacy survey for UAVs~\cite{hadi2023comprehensive}} & 2023 & {\tiny Elsevier J. Net. \& Comp. App.} & \checkmark & \checkmark & \checkmark & \ding{55} & \ding{55} \\
{\tiny Cybersecurity of UAVs Survey~\cite{yu2023cybersecurity}} & 2023 & \tiny IEEE Aerospace \& Elec. Sys. Mag.& \checkmark & \checkmark & \checkmark & \ding{55} & \ding{55} \\
{\tiny Intelligent UAV security approaches~\cite{mohammed2023comprehensive}} & 2023 & \tiny Elsevier Comp. Net. & \checkmark & \checkmark & \checkmark & \ding{55} & \ding{55} \\
{\tiny Advancing UAV security with AI~\cite{tlili2024advancing}} & 2024 & \tiny Elsevier IoT Journal & \checkmark & \checkmark & \checkmark & \ding{55} & \ding{55} \\
{\tiny A survey of security in UAVS and FANETS~\cite{ceviz2024survey}} & 2024 & \tiny IEEE Comms. Surveys \& Tutorials & \ding{55} & \checkmark & \checkmark & \ding{55} & \ding{55} \\
\hline
\end{tabular}
}
\footnotesize \checkmark = Comprehensive coverage \quad \ding{55} = Limited coverage
\end{table*}

\textcolor{black}{In Table~\ref{existingsurveys}, we compare recent surveys along five evaluative columns central to UAV intrusion detection: ML and AI methods, UAV or drone focus, security coverage spanning cyber and physical intrusions, RL integration, and dataset evaluation across network and vision modalities; check marks indicate substantive treatment and x-marks denote limited or no coverage. The surveyed set was assembled through searches on IEEE Xplore, ACM Digital Library, and Google Scholar using combinations of “UAV” or “drone” with “intrusion detection,” “IDS,” “security,” “counter‑UAV,” or “anti‑UAV,” paired with “survey” or “review,” restricted to the last ten years; titles and abstracts were screened to retain archival English surveys with a core focus on UAV security, IDS for UAVs, or counter‑UAV sensing, while excluding application‑specific. Criteria are applied as follows: ML/AI Methods is marked comprehensive when machine or deep learning is synthesized beyond incidental mentions; UAV/Drone Focus when UAVs or counter‑UAV systems are the primary subject rather than a brief subsection; Security Coverage when both cyber threats on communication links and physical intrusions are discussed; RL Integration when reinforcement learning for detection, mitigation, or counter‑UAV control is treated; Dataset Evaluation when UAV‑relevant datasets are reviewed with guidance on usage or limitations across network traces and vision benchmarks. The comparison shows that most prior surveys emphasize one or two dimensions, commonly cyber IDS or anti‑UAV sensing, while offering limited unification. Our survey integrates the cyber and physical perspectives within a single IDS view, links AI choices to deployment tiers, and consolidates UAV‑specific datasets across network and vision, thereby addressing gaps in the existing literature.}

There is an urgent need to design innovative anti-intrusion systems for UAVs to address the dual challenges posed by cyber-attacks and unauthorized UAV operations. Alongside the risks of cyber-attacks targeting UAVs, these systems must also mitigate the threats UAVs themselves can pose to the privacy and safety of individuals and property. 
Significant research efforts over the last decade have focused on tackling these challenges within the complex UAV-IoT environment. Artificial Intelligence (AI) has further enhanced anti-intrusion systems, enabling greater intelligence and autonomy in combating cyber-attacks and addressing physical intrusions. 
This survey provides a structured, evidence-based overview of AI-enabled intrusion detection methods and resources documented in the literature and recent advancements for Unmanned Aerial Systems (UASs), which aim to mitigate cyber-attacks targeting the UAV fleet and mitigate various types of physical intrusions that could compromise the privacy and safety of people and property in smart cities. The contributions of this paper can be summarized as follows:
\begin{itemize}
\item We provide a focused review that integrates cyber-attacks over communication links with unwanted physical intrusions within a single scope, clarifying what prior surveys overlook.
\item We classify the intrusions into different categories according to their means: cyber-attacks over communication infrastructure and intrusions via unwanted/unauthorized UAVs
\item We investigate recent AI solutions and present a unified IDS taxonomy aligned to sensing, calibration, fusion, decision, and response across on‑board, cooperative, edge, and ground control tiers.
\item We provide an extensive list of existing open public datasets.
\item We propose open research directions that can further enhance anti-intrusion systems for UAVs.
\end{itemize}

\begin{figure*}[t]
    \centering  
    \includegraphics[width=\textwidth]{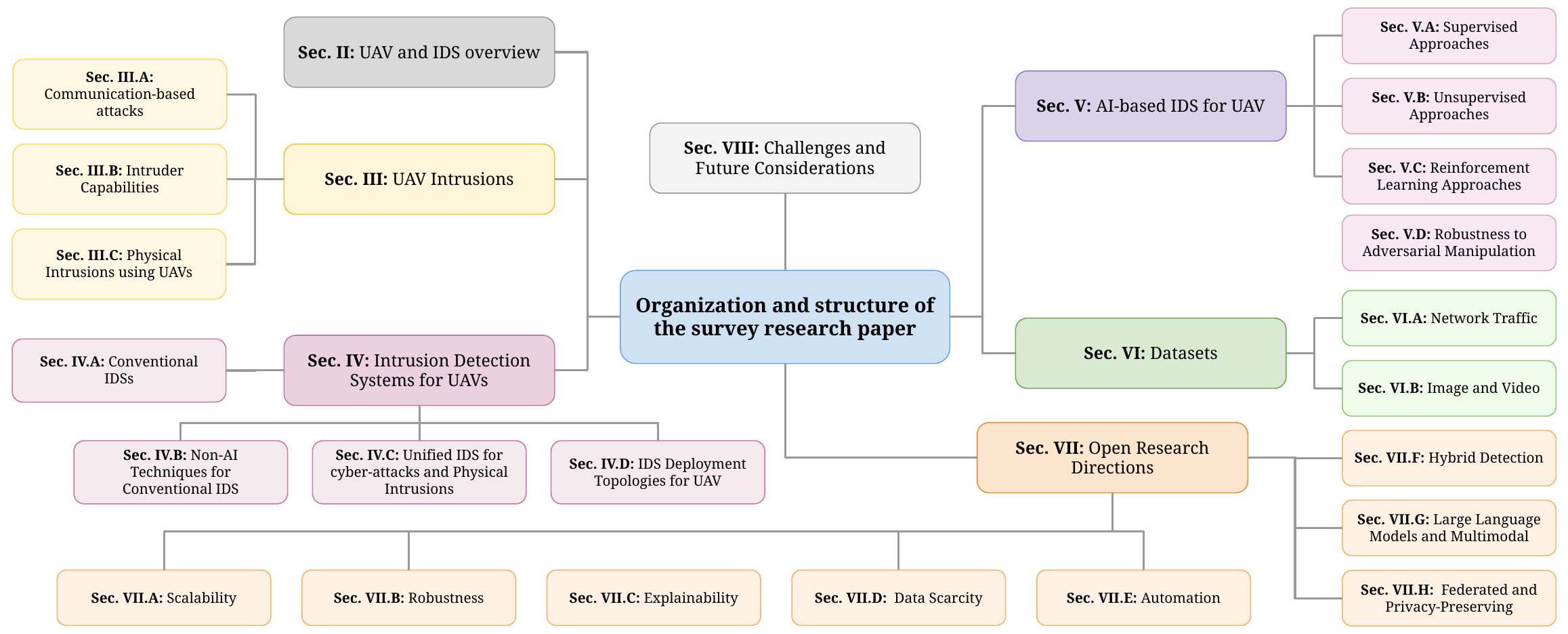}
    \caption{A detailed structure of the survey content.}
    \label{fig:structure}
\end{figure*}

The remainder of this paper is organized as follows. Section 2 provides a brief description of UAVs and their operational constraints for IDS deployment. Section 3 presents the intrusion attacks that challenge the usage of drones and their types. Section 4 discusses how these challenges can be solved using intrusion detection systems while presenting their types. Next, in Section 5, we demonstrate how AI is being utilized to enhance the performance of intrusion detection systems, thereby overcoming intrusion attacks. The datasets used to train those algorithms are presented in Section 6. Section 7 proposes some open research directions, followed by challenges and future considerations in Section 8, and concludes our survey with Section 9. An illustrative and more detailed structure of the survey is provided in Figure~\ref{fig:structure}.


\section{UAVs}
\label{background}

\begin{figure}[t]
    \centerline
    {\includegraphics[width=0.7\textwidth]{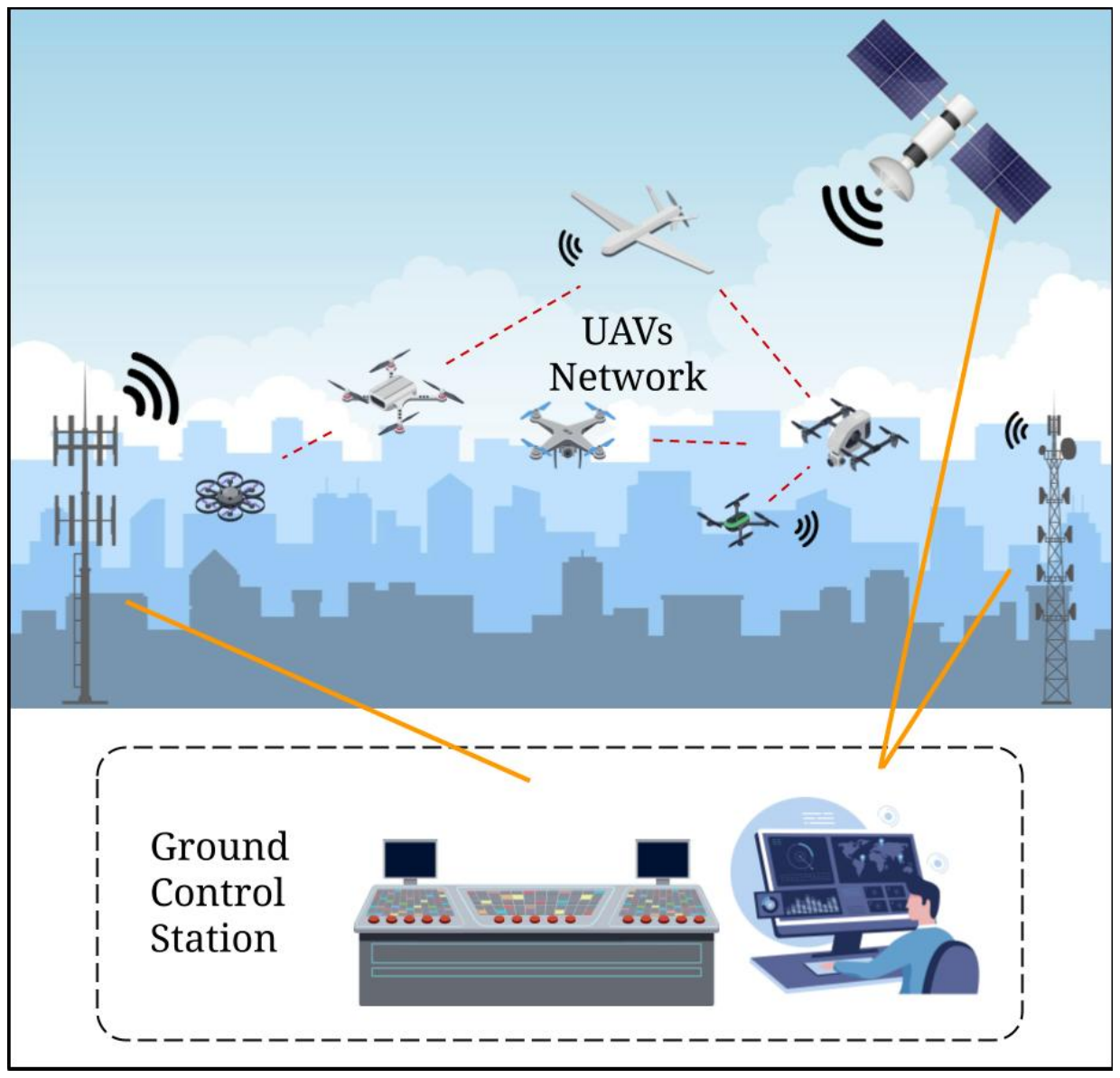}}
    \caption{The basic form of a UAV Architecture is composed of a fleet of UAVs, potentially equipped with a wide array of sensors and connected via wireless communication links. The UAVs can communicate with external entities such as ground cellular networks, satellites, and HAPs. A ground control station controls them to ensure the management and coordination of the fleet, as well as the processing of the data collected.}
\label{fig:UAS_arch}
\end{figure}
A UAV is an aerial vehicle that operates autonomously or through remote piloting, without onboard human operators. The main components of the UAV, depicted in Figure~\ref{fig:UAS_arch}, consist of the aircraft itself, various sensors, a wireless communications medium, and a Ground Control Station (GCS) that allows for remote control of the UAV. Besides that, when the UAV is controlled from the ground, it is also known as a remotely piloted vehicle (RPV) and must rely on stable wireless communication to function as intended~\cite{LAKSHMINARAYANAN2015163, valavanis2015handbook, newcome2004unmanned, 9696294}. UAVs can be equipped with a variety of sensors depending on their payload. They can be fitted with Inertial Measurement Units (IMUs), cameras, Light Detection and Ranging (LiDAR), and other sensors. They also include wireless communication interfaces that enable links with other UAVs, ground stations, and High-Altitude Platforms. Ultimately, UAVs can be equipped with sufficient intelligence to make on-site decisions. All of these features, in addition to their 3D mobility, make UAVs a key player in many applications and services, as they can simultaneously support the four functions of smart cities: collection, analysis, communication, and action.

\textcolor{black}{The design of placement-aware IDS for UAVs must address operational constraints that determine where and when detection functions can execute within the UAS architecture. UAVs operate with limited computational resources, onboard memory, and power constraints \cite{medhi2025lightweight}, necessitating lightweight IDS solutions that can function effectively in resource-constrained environments without compromising operational performance or flight time. Energy limitations necessitate efficient algorithms and selective data processing strategies. Communication link characteristics directly influence both threat exposure and detection capabilities. Command and telemetry channels operate with limited bandwidth and strict timing requirements, demanding protocol-compliant monitoring with minimal latency impact. Payload data links, particularly video streams, provide richer information sources suitable for advanced detection techniques at edge nodes. Flight patterns affect IDS performance requirements. Stationary operations enable comprehensive sensor fusion, while patrol missions require rapid assessment capabilities and conservative response thresholds to prevent false positives from movement-induced artifacts. Multi-UAV operations in congested airspace require cooperative validation mechanisms to reduce false alarm rates. These factors support a tiered IDS architecture that integrates onboard monitoring, edge processing, and ground station oversight with seamless handover capabilities.}

\begin{figure}[t]
    \centering  
    \includegraphics[width=0.5\textwidth]{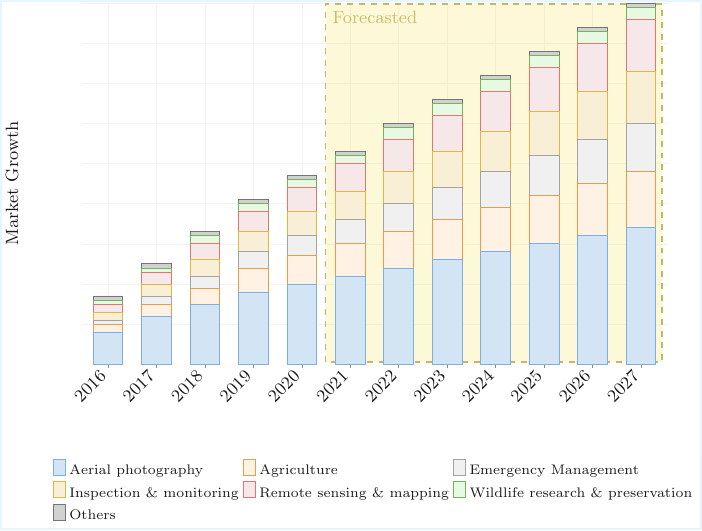}
    \caption{Growth of the European drone market size from 2016 up to 2027 in Million USD based on different applications - Graphical Research~\cite{EuropenDroneMarket}.}
    \label{fig:market_size}
\end{figure}

\begin{figure}[t]
    \centering  
    \includegraphics[width=0.5\textwidth]{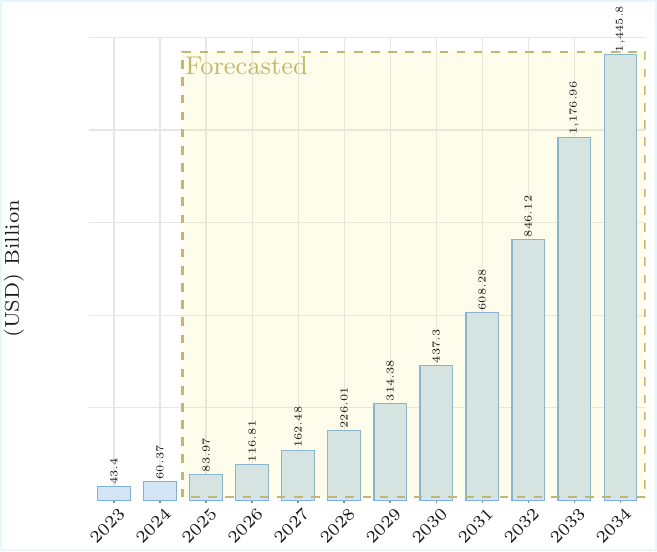}
    \caption{Growth of the global market size of drones per year from 2023 up to 2034 \cite{precedenceresearch}.}
    \label{fig:sold_drones}
\end{figure}

Initially, UAV applications were primarily limited to military use. Later, they became publicly available and were used for various civilian applications. The rapid growth in UAV usage is driven mainly by the low cost of acquiring them and their ease of deployment. On the one hand, the use of commercial drones has been rapidly increasing since 2016. It is projected to continue growing in the coming years as the Graphical Research study shows it illustrated in Figure ~\ref{fig:market_size}, where it has been demonstrated that the European commercial drone market has surpassed USD 3 billion in 2022 and is expected to reach USD 6 billion by 2027 with a significant portion of that growth coming from industries such as agriculture, transportation, and logistics~\cite{EuropenDroneMarket}. In sectors such as agriculture, logistics, and construction, drones are being utilized to enhance efficiency and productivity. For example, farmers are using drones to survey their crops, construction companies are using drones to inspect buildings and survey land, and retailers are experimenting with drones for package delivery. Additionally, the Federal Aviation Administration (FAA) has been working to create regulations allowing for the widespread use of commercial drones. As highlighted in Figure~\ref{fig:sold_drones}, the growth of market size in the drone industry has been rapidly increasing since 2023, and this trend is expected to continue in the upcoming years. According to market research, the global drone market is expected to grow at a compound annual growth rate (CAGR) of 18.4\% from 2020 to 2025, indicating a substantial increase in the number of drones sold annually. This trend is expected to grow significantly over the next few years. Factors such as technological advancements, increasing adoption in various industries, and the development of regulations that enable the usage of drones are driving this growth.

On the other hand, in many countries, regulatory authorities and laws governing the use of UAVs limit the development of practical deployment due to safety and security concerns~\cite{rs9050459}. \textcolor{black}{In the United States, FAA regulations for small unmanned aircraft are codified in 14 CFR Part 107, effective August 2016 with subsequent amendments, including Remote ID under Part 89~\cite{FAA-Part107,FAA-Part89}, while in the European Union, U‑space services are established by Commission Implementing Regulation (EU) 2021/664, applicable from January 2023 together with (EU) 2021/665 and (EU) 2021/666 for designated U‑space airspace~\cite{EU-2021-664,EU-2021-665,EU-2021-666}.} The anti-drone market is expanding due to the growing use of UAVs and their associated security risks. The increasing use of UAVs, both commercially and recreationally, has created a need for anti-drone solutions to ensure public safety and prevent malicious activities. The potential for UAVs to be used for terrorism, smuggling, and espionage has also heightened security concerns and increased demand for anti-drone solutions. Advancements in technology, such as AI and machine learning, have led to the development of more sophisticated anti-drone solutions, further driving the market's growth. With the growing number of UAV incidents, including close calls with commercial airlines and unauthorized intrusions into restricted airspace, awareness of the security risks posed by UAVs is increasing, driving demand for anti-drone solutions. The anti-drone market is expected to grow in the coming years, with analysts predicting a significant increase in demand across various industries.

\begin{figure*}[h]
    \centering  
    \includegraphics[width=\textwidth]{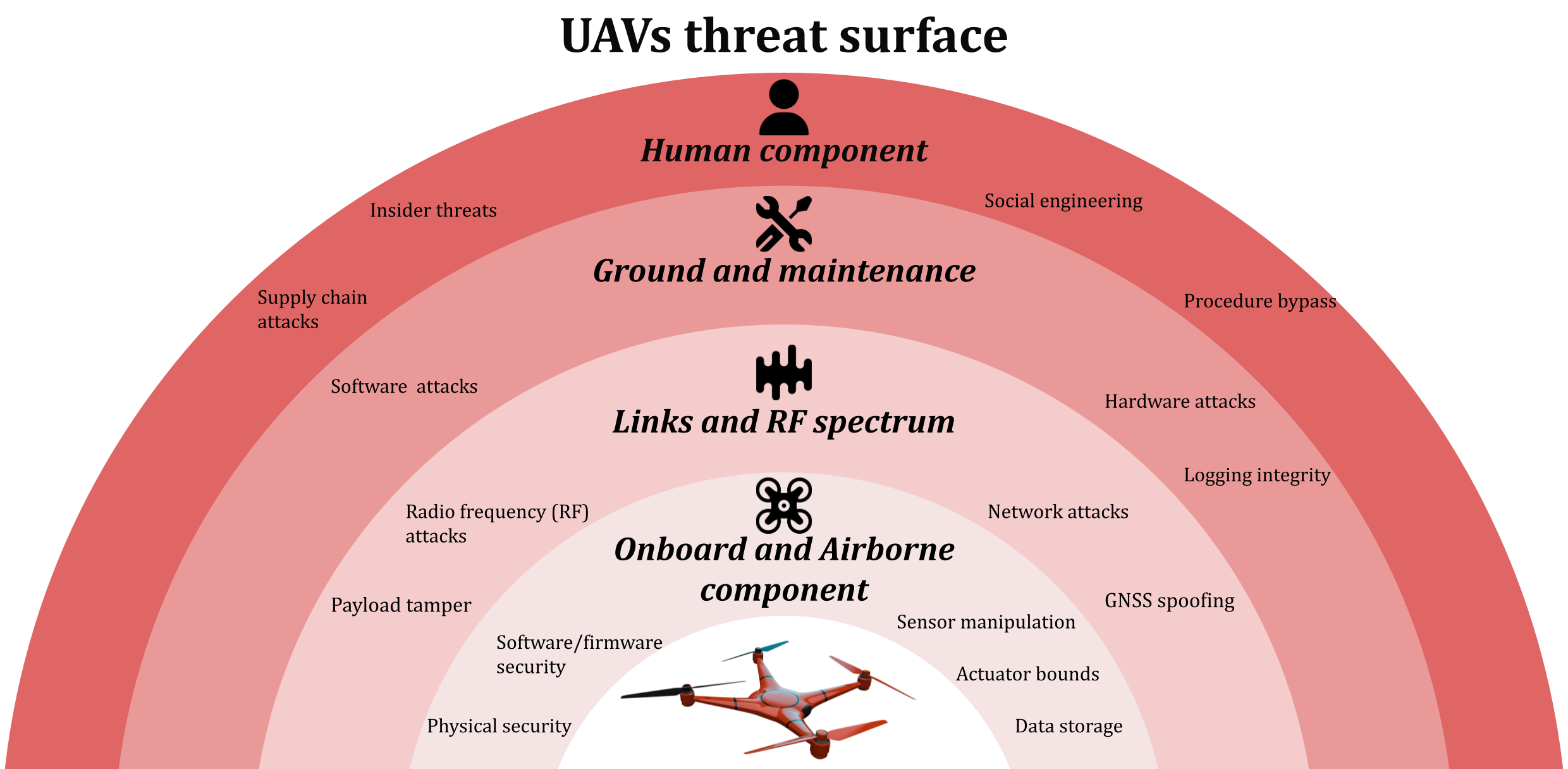}
    \caption{\textcolor{black}{Map of entry points and adversarial actions across airborne, links and RF, ground and maintenance, and human layers, indicating typical IDS tiers for onboard, edge fusion, and ground control correlation.}}
    \label{fig:uav_ids_mapping}
\end{figure*}

\textcolor{black}{These constraints define an IDS-oriented threat surface. Entry points include command, telemetry, and payload links; onboard sensors and firmware; ground control and maintenance; and the surrounding airspace where unauthorized drones intrude physically. Representative adversarial actions range from link jamming and spoofing to crafted traffic and replay, code or parameter tampering, and visual or RF evasion by small fast targets in clutter. Countermeasures are aligned by tier: onboard consistency checks over IMUs and navigation; edge fusion of RGB, IR, acoustic, and RF data to separate drones from birds; and ground-level correlation across fleet telemetry and policy for escalation and forensics. As summarized in Figure~\ref{fig:uav_ids_mapping}, the map links entry points to modalities and tiers and guides the placement and fusion choices elaborated in Section~\ref{unified_ids}.}

\section{UAV Intrusions}
\label{intrusion}
    \label{fig:anti-drine}



UAVs rely on wireless communication and GPS for navigation, making them inherently vulnerable to various types of attacks. This vulnerability has been increasingly exploited, as highlighted in~\cite{6459914}, which reports a significant rise in cyber-attacks since 2007 due to the growing popularity of UAVs. Intrusion attacks on UAVs can be broadly classified into two categories: those that leverage wireless communication and networks to target the UAV, and those that use the drone itself as an intruder to invade or spy on a region through various sensors. Hackers often exploit the radio links installed on UAVs, which contain critical navigation data. Once intercepted, this data can be manipulated, allowing attackers to take control of the drone. For instance, commercial Wi-Fi-based drones, as highlighted in~\cite{7795496}, are susceptible to simple security vulnerabilities. The authors demonstrated how a Parrot Bebop 2 drone could be instantly crashed by exploiting weaknesses in the Wi-Fi access point and ARDiscovery Connection procedure. \textcolor{black}{We focused on a concise attacker model for UAV IDS evaluation. The capabilities include what adversaries can realistically do, such as RF communication across command and data channels, traffic replay attacks, software tampering during maintenance, and deploying small UAVs with standard sensors. The goals examined include disrupting missions, gathering information, and maintaining hidden access while avoiding quick detection. The model incorporates key limitations, including short operational times near protected areas, communication range restrictions, partial knowledge of systems and procedures, and the risk of detection by monitoring systems. This framework enables each threat to be described by its entry method, requirements, observable signs, and resulting damage. These factors can guide the placement of IDS sensors and the combination of their data.}

\subsection{Communication-Based Attacks on UAVs}
\noindent$\bullet$ Jamming Attacks:
A common type of wireless-based attack is jamming, in which attackers broadcast radio signals to disrupt communication between the UAV and its controller, reducing the Signal-to-Interference-plus-Noise Ratio (SINR)~\cite{berg2008broadcasting, VADLAMANI201676}. This disruption can render the UAV uncontrollable, and in some cases, such as a 2012 incident, it was suspected that a GPS jamming attack caused a drone crash with casualties.
\\
$\bullet$ Spoofing Attacks:
Spoofing is another critical threat, tricking UAVs into following false navigation commands by broadcasting fake GPS signals. This can mislead the UAV into deviating from its intended path, jeopardizing its mission. 
\\
$\bullet$ Malware Attacks:
Malware attacks further expose UAV vulnerabilities by infecting the software or firmware, enabling attackers to control the drone or steal sensitive data.
\\
$\bullet$ Denial-of-Service (DoS) Attacks:
DoS attacks can overwhelm the UAV’s system with excessive traffic, disrupting its regular operations.\\
$\bullet$ Eavesdropping Attacks:
Eavesdropping allows attackers to intercept communications between the UAV and its controller to gain access to sensitive information, such as the drone’s location or payload. 
\\
$\bullet$ \textcolor{black}{Hardware vulnerabilities: UAV hardware presents potential security exposure points. For instance, a lack of firmware signature validation or unsecured debug interfaces can enable unauthorized firmware flashing or parameter tampering at maintenance depots and launch sites.}\\
$\bullet$ \textcolor{black}{Social engineering vulnerabilities: These exploits primarily focus on UAV operators through deceptive practices designed to extract sensitive information, including authentication credentials and operational data such as UAV positioning information. Additionally, such attacks may attempt to influence operators toward compromising operational procedures. Examples include impersonation of regulatory authorities or equipment vendors to manipulate traffic management protocols, as well as phishing campaigns that simulate legitimate Air Mission communications or distribute compromised mission-planning software.}\\
$\bullet$ \textcolor{black}{Mitigations: Procedural role-based access and least privilege, verified call-back for credential/config changes, two-person rule for geofencing, pre-approved channels for Notices to Air Missions and vendor communications, secure field Procedures (badge checks, visitor logs). Technical phishing-resistant measures encompass several well-established techniques, such as utilizing multi-factor authentication on the GCS, code signing and secure boot, Mobile Device Management with app allowlists on pilot tablets, tamper-evident logs for critical actions, and time-bound token access for mission changes. Moreover, periodic training is necessary for operator awareness, specifically designed to address UAV workflows and social drills.}

\subsection {Intruder Capabilities}
Traditionally, a cyber-attack can be described as any malicious and fraudulent behavior or policy violation that targets a single entity in a network or the network as a whole. Cyber-intrusions typically compromise the security of networks and/or their data, often resulting in the theft of valuable network resources and sensitive information. According to~\cite{fsdfsdf, dfsdfsdgsdhjghn}, the intruders, through their cyber-physical attacks, aim to gain one of the following capabilities: \\
$\bullet$ Revelation Capabilities: As the name suggests, the intruder is trying to reveal either encrypted data stored on some storage devices within the network or to intercept real-time information sent on data links.\\
$\bullet$ Knowledge Capabilities: This indicates the attacker's capacity to get prior information on the system's parameters. For example, this can manifest when the intruder gains access to the zones surveyed by a drone camera.\\
$\bullet$ Disruption Capabilities: refers to the intruder's capability to interfere with and even interrupt the system's normal functioning.

\subsection{Physical Intrusions Using UAVs}

\textcolor{black}{Physical intrusions use the airspace entry point with small unauthorized UAVs; preconditions are line-of-sight access and brief loiter windows, while impacts range from privacy violation and event disruption to safety hazards and sensing denial; observable cues include low-altitude flight profiles, target-centered loiter, and multimodal traces in RGB, IR, acoustic, and RF.}
Surveillance is another example, where the attacker uses the drone to gather information about a specific location by capturing video, audio surveillance, or even taking photos. Such physical attacks can lead to many damaging consequences and threats, including:\\
$\bullet$ Public Event Disruption: Unwanted UAVs can disrupt public events such as concerts, parades, or sports events. They can also violate privacy by capturing unauthorized footage and interfere with the event progress by causing noise or panic, distracting attendees and the crowd, and/or obstructing the view.\\
$\bullet$ Interference with Air Traffic: The airspace of smart cities is rapidly evolving with commercial flights, numerous UAVs, and futuristic electric Vertical Takeoff and Landing (eVTOLs). Unauthorized UAVs can increase the risk of collision, especially as they fly at low altitudes with eVTOLs. The risk of crashes can be significantly high if unauthorized UAVs are not prevented from occupying the airspace. Near-miss incidents can also pose a danger and cause panic for passengers, leading to economic losses due to delays and diversions.\\
$\bullet$ Hazardous Material Release: Unauthorized use of UAVs can lead to the release of hazardous materials, including chemical agents and harmful substances, posing a significant threat, especially in densely populated urban environments. These malicious activities can lead to dramatic consequences if no action is taken against the flying intruders.\\
$\bullet$ Environmental Disruption: In addition to the noise pollution that is caused by drones in urban zones, unauthorized flying units could cause disturbance to wildlife in cities, as well as damage the vegetation due to UAV landings or crashes. At night, when equipped with powerful lights, UAVs can cause light pollution for the residents. 

Given the potential cyber-attacks or physical invasions of urban space, it is essential for UAV stakeholders, including users, operators, manufacturers, and regulators, to prioritize these challenges and take the necessary regulatory measures and prevention techniques to mitigate the risks of intrusion. \textcolor{black}{In practice, IDS design and escalation must align with the applicable operational framework, for example, Part 107 operating limits and Remote ID in the United States and U‑space common services under Reg. (EU) 2021/664 in the European Union, which shape detection, identification, and response pathways. The structured model above is used in the remainder of the survey to motivate IDS placement, where onboard checks address integrity and consistency preconditions, edge fusion targets multimodal cues at the airspace entry point, and ground control correlates policy and operator actions.} Therefore, designing innovative Intrusion Detection Systems (IDS) is the key element to overcoming these attacks and protecting UAVs. With the development of IDS, drones can now not only identify intrusion attacks but also react to these threats efficiently and effectively, especially with the emergence of AI. AI is a crucial technology for mitigating the risks of physical intrusions or cyber-attacks on communication networks. Designing IDS is highly regarded in the literature, and a wide range of surveys exists to summarize the state-of-the-art approaches to developing them. As an example, Seongjoon et al.~\cite{9378538} presented a comprehensive study of non-military IDS, introducing its main components: detection, identification, and neutralization approaches. In our current survey, we specifically shed light on recent AI-based techniques for addressing the various problems caused by unauthorized UAV activities and potential cyber threats.

\section{Intrusion Detection Systems for UAVs}
\label{ids}
Conventional IDS have primarily focused on enhancing cybersecurity by monitoring and protecting the communication infrastructure against cyberattacks. In the UAV context, an evolved and advanced version of IDS, designed to comprehensively address security concerns in the realm of smart cities, extends its scope beyond cyber threats to encompass a broader spectrum of security challenges. This enhanced IDS not only remains vigilant against cyber-attacks targeting the communication interfaces of UAVs but also incorporates sophisticated measures to detect, deter, and mitigate unwanted or unauthorized physical intrusions by UAVs. By encompassing both cyber threats and physical security breaches, advanced IDSs aim to provide holistic and robust defense mechanisms tailored to modern challenges, ensuring the safety, integrity, and privacy of citizens while leveraging the benefits of UAV integration.

\subsection{Conventional IDSs}
Typically, an IDS is a device/software that plays the role of a scout or security guard by constantly monitoring a system to detect and report malicious, unauthorized behavior or policy breaches~\cite{9378538,agrawal2022federated}. To cope with cyber threats, we distinguish two main types of IDSs according to their data source and location: (i) HIDS (Host-based IDS), which can be installed on every device connected to the network to monitor and analyze internal events such as system calls, application logs, and file system activities, and (ii) NIDS (Network-based IDS) which exclusively monitors the network traffic using security hardware placed strategically in the network.

\begin{figure*}[t]
    \centering  
    \includegraphics[width=0.8\textwidth]{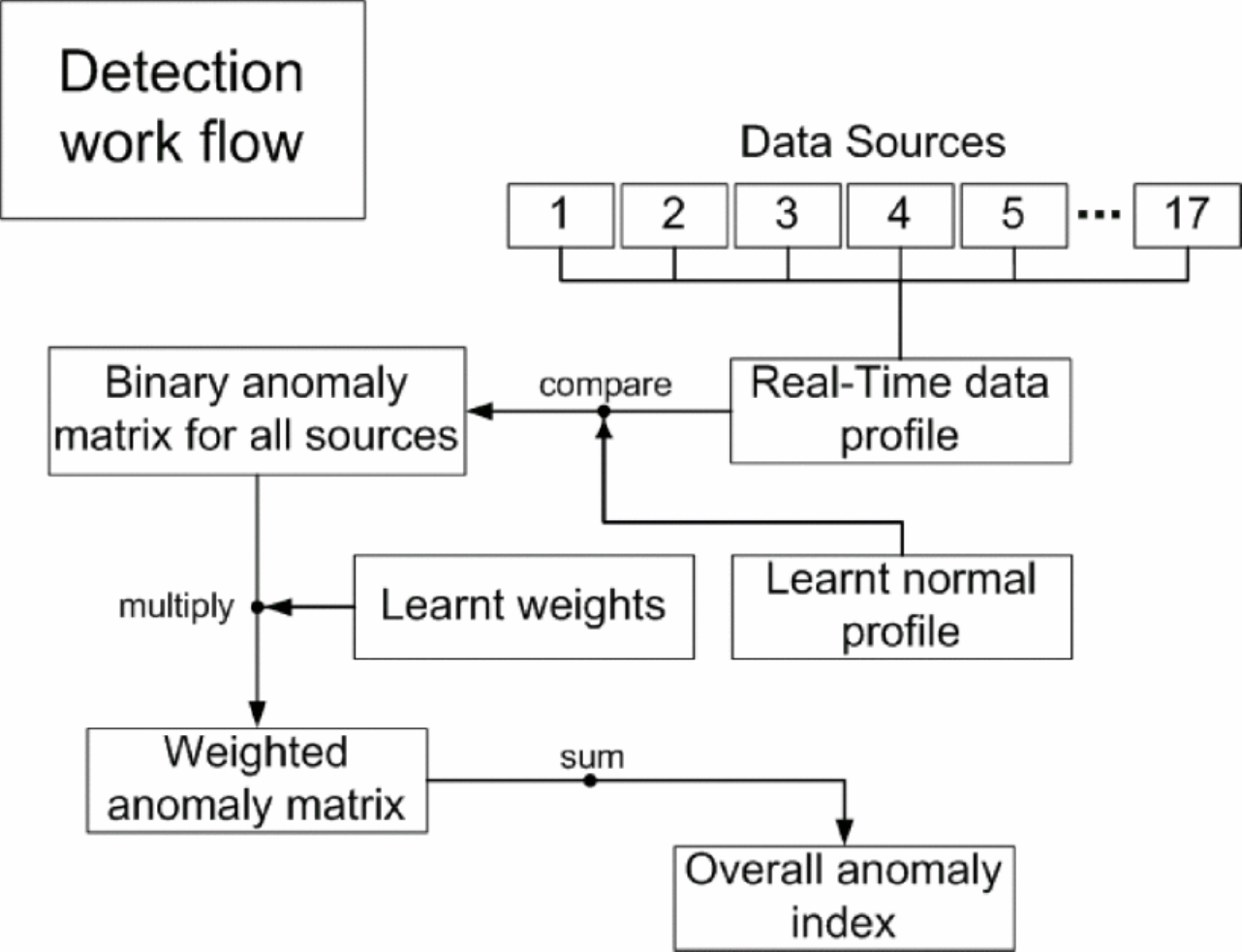}
    \caption{Anomaly-based IDS to detect Cyber-Physical Attacks on a Robotic Vehicle~\cite{7828584}}
    \label{fig:anomaly_based}
\end{figure*}

\begin{figure*}[t]
    \centerline
    {\includegraphics[width=\textwidth]{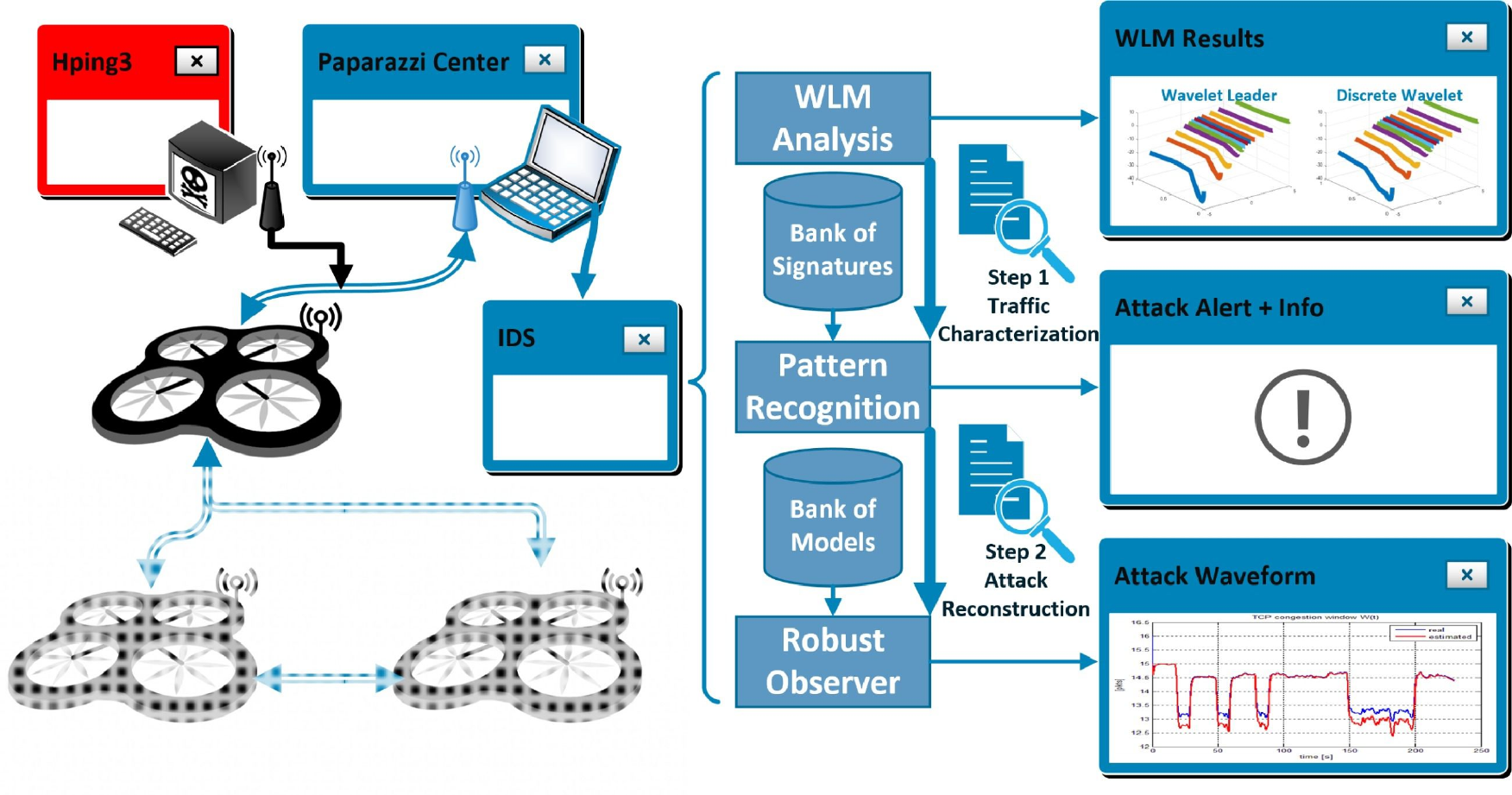}}
    \caption{A hybrid-based IDS for anomaly estimation inside UAV networks using spectral traffic analysis and robust controller observer~\cite{CONDOMINES2019101759}}
\label{fig:hybrid_based_IDS}
\end{figure*}

\textcolor{black}{Depending on the chosen type of IDS, the solution can rely on four key detection methods
~\cite{choudhary2018intrusion}}:
\begin{itemize}
\item Specification-based IDS: manually specified systems are designed to handle specific scenarios known in advance using a set of rules to follow when expected behaviors occur. These IDS have been widely used in communication networks~\cite{inproceedings123, inproceedings12, articlespecification} and cyber-physical systems security~\cite{6786500, 7684103, articledasjdhjad, NTALAMPIRAS2016164, 10.1145/2990499}. In~\cite{8715740}, the authors developed a misbehavior detection method called BRIoT for cyber-physical systems based on behavior rule specification. They tested this approach on a UAV and found that it was able to identify both known and unknown misbehaviors caused by cyber-attacks.
\item Signature-based IDS: These approaches aim to identify known threats based on well-known, predefined signatures and patterns, such as byte sequences or malware intrusion sequences. Upon detecting an anomaly, a classification operation is triggered to identify it based on its signature. These approaches minimize false positives, as the attacks are clearly defined in advance using their signatures, and they are generally easy to use. However, they are not very practical in the real world, as they are highly dependent on the attackers' knowledge and expertise. In~\cite{kumar2012signature}, the authors present a signature-based IDS that can detect intrusion in real-time and trigger alerts so that actions can be taken.
\item Anomaly-based IDS:  Based on a system breakdown or an illegitimate action seen in the system, anomalous behavior is recognized. This approach includes a learning mechanism where the classification of anomalies or normal behavior is based on heuristics or rules, rather than patterns or signatures, as opposed to the signature-based IDS, where detection occurs only when a signature has already been created. For example, the authors of~\cite{8937816} address the issue of the absence of security protocols in the controller area network bus of UAV by proposing a novel algorithm that detects data injection attacks using anomaly-based supervised learning. A sample from the anomaly-based UAV workflow is shown in Figure~\ref{fig:anomaly_based}.\\
$\bullet$ Hybrid-based IDS: To design a stronger and more efficient detection policy, hybrid-based IDS, as shown in Figure~\ref{fig:hybrid_based_IDS}, combines two or more detection methods from the ones defined above~\cite{CONDOMINES2019101759, AYDIN2009517}. \textcolor{black}{By leveraging the strengths of multiple approaches (e.g., the precision of signature-based detection and the adaptability of anomaly-based detection), these systems are particularly effective at identifying complex hybrid attacks that use multiple vectors or stages. Such approaches can reduce false positives while improving coverage of unknown threats. A detailed investigation of hybrid detection systems, including performance benchmarking against hybrid attacks, is identified as an important avenue for future research (Section\ref{hybridSetection}).} 
\end{itemize}

\subsection{Non-AI Techniques for Conventional IDS}
To improve the accuracy of detection methods in conventional IDSs, three main approaches are employed: statistics-based~\cite{lin2015cann}, knowledge-based~\cite{elhag2015combination, can2015survey}, and machine learning-based~\cite{buczak2015survey, meshram2017anomaly}. The statistics-based approach involves constructing a statistical model of typical user behavior by collecting and analyzing data to inform the model. The knowledge-based approach consists of identifying requested actions using existing system data, such as protocol specifications and network traffic instances. Machine learning methods utilize complex pattern-matching capabilities developed through training data. More details about these three classes, along with examples of their subclasses, are given by Ansam et al.~\cite{khraisat2019survey}.

Gaurav et al.~\cite{8450305}, briefly summarize state-of-the-art IDS mechanisms that deal with malicious attacks in UAV networks. In this survey, the authors classified existing IDSs primarily based on their detection techniques and the types of intrusions they handle. The authors also discussed the primary research challenges in developing IDS in highly resource-constrained and dynamic environments. In~\cite{BANGUI2021877}, the authors explore the recent advances in machine learning-based IDS in UAVs as an intelligent transportation system and a Vehicular Ad hoc network (VANET). In this survey, the authors discussed the different intrusion detection modes (signature-based, anomaly-based, and hybrid-based) and the possible strategies for their deployment (onboard, collaborative, and offloaded). Additionally, the authors investigated several ML algorithms used in designing IDS and the datasets and simulators used to train these algorithms. According to this survey, Artificial Neural Networks (ANNs) are the most commonly used in designing IDS, and collaborative deployment is the most widely used method of deployment, where UAVs participate in detecting threats. In the same paper, the authors found that jamming and spoofing are the most detected threats using these IDS. In~\cite{LOUKAS2019124}, the authors classify almost sixty-six IDS while accentuating their advantages and disadvantages to help identify techniques that can be adopted in the industry. To this end, the authors presented a taxonomy of IDS characteristics and architectures designed for vehicles, discussing deployment locations, degrees of autonomy, modes of operation, and levels of technological maturity for the IDS. The authors also examined the indicative security threats used to evaluate the designed IDS. They presented a comparative analysis of IDS for different types of vehicles, including UAVs, land vehicles, automobiles, VANETs, and watercraft.

To mitigate the risk of cyber-attacks against UAVs, several approaches have been investigated in the literature. In~\cite{condomines2019network}, the authors proposed a unified IDS that analyzes spectral traffic and estimates anomalies. The proposed model exchanges statistical signatures and uses them to estimate abnormal traffic. In~\cite{10.1145/2248326.2248334}, specification-based IDS is proposed to safeguard the UAVs' installed sensors and actuators. In~\cite{7890467}, the authors proposed an IDS that functions at the ground station levels and the UAV itself to identify attacks that might threaten the network. The proposed hierarchy monitors the UAV behavior and categorizes it into four different levels, from normal to malicious, according to the identified cyber-attack. In~\cite{9520850}, the authors presented an energy-harvesting IoT protocol that operates as an IDS. The protocol involves the cooperation of two UAVs and a ground-based station to detect intruders by exchanging confidential signals using harvested energy. First, a UAV relay cooperates with a UAV jammer to detect the intruder by harvesting energy from a source. In~\cite{7549080}, the authors focused not only on detecting intrusions of UAV behavior that degrade network performance but also on determining whether the intruder needs to be ejected. The phenomenon is modeled as a Bayesian game to accurately detect attacks. Bayesian game theory is also used in~\cite{sun2018intrusion} to detect intrusions in UAV networks. In \cite{khan2021blockchain}, the authors leveraged blockchain technology to achieve decentralized predictive analytics, thereby enhancing machine learning models in detecting UAV intruders.
\begin{figure*}[t]
    \centering  
    \includegraphics[width=\textwidth]{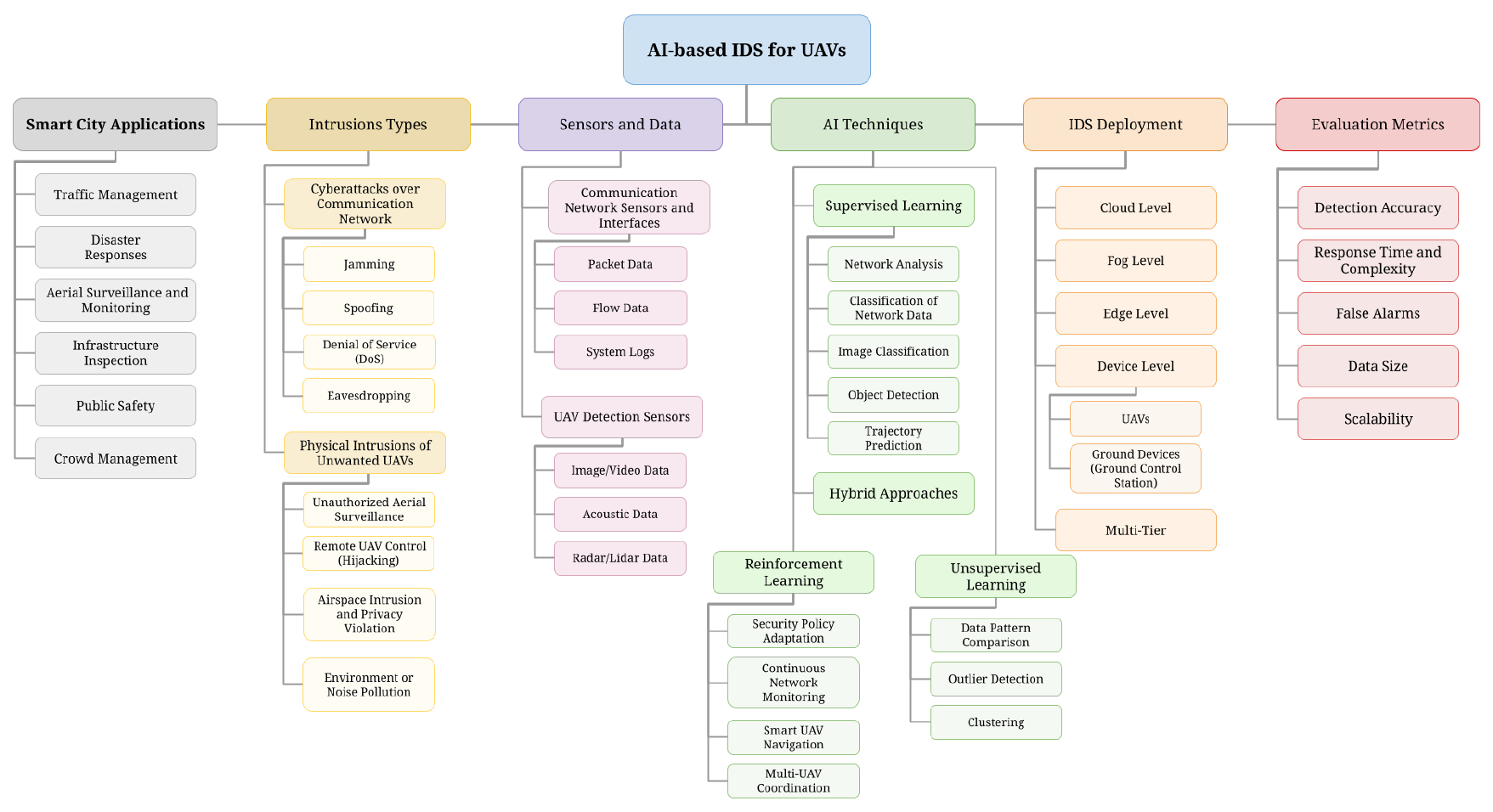}
    \caption{AI-based Intrusion Detection Systems for UAV.}
    \label{fig:taxonomy}
\end{figure*}

\subsection{Unified IDS for cyber-attacks and Physical Intrusions} \label{unified_ids}
The UAV-IDS also needs to integrate mechanisms to detect and track unwanted or unauthorized UAVs and alert to abnormal physical activities in the area of interest. A combination of sensor technologies, including radar, acoustic sensors, and video analytics, must be leveraged to establish comprehensive surveillance frameworks. In addition, counteraction strategies need to be devised to enable a single drone or a fleet of drones to mitigate intrusions and avoid any risk of collision. Machine learning algorithms, empowered by computer vision techniques, can enable IDSs to differentiate between authorized and unauthorized UAVs using flight patterns, behavioral analysis, and visual identification. Reinforcement learning and control techniques can be adapted to incorporate adaptive responses to attacks and mitigate the risk posed by unwanted UAVs.

\begin{figure*}[t]
    \centering  
    \includegraphics[width=\textwidth]{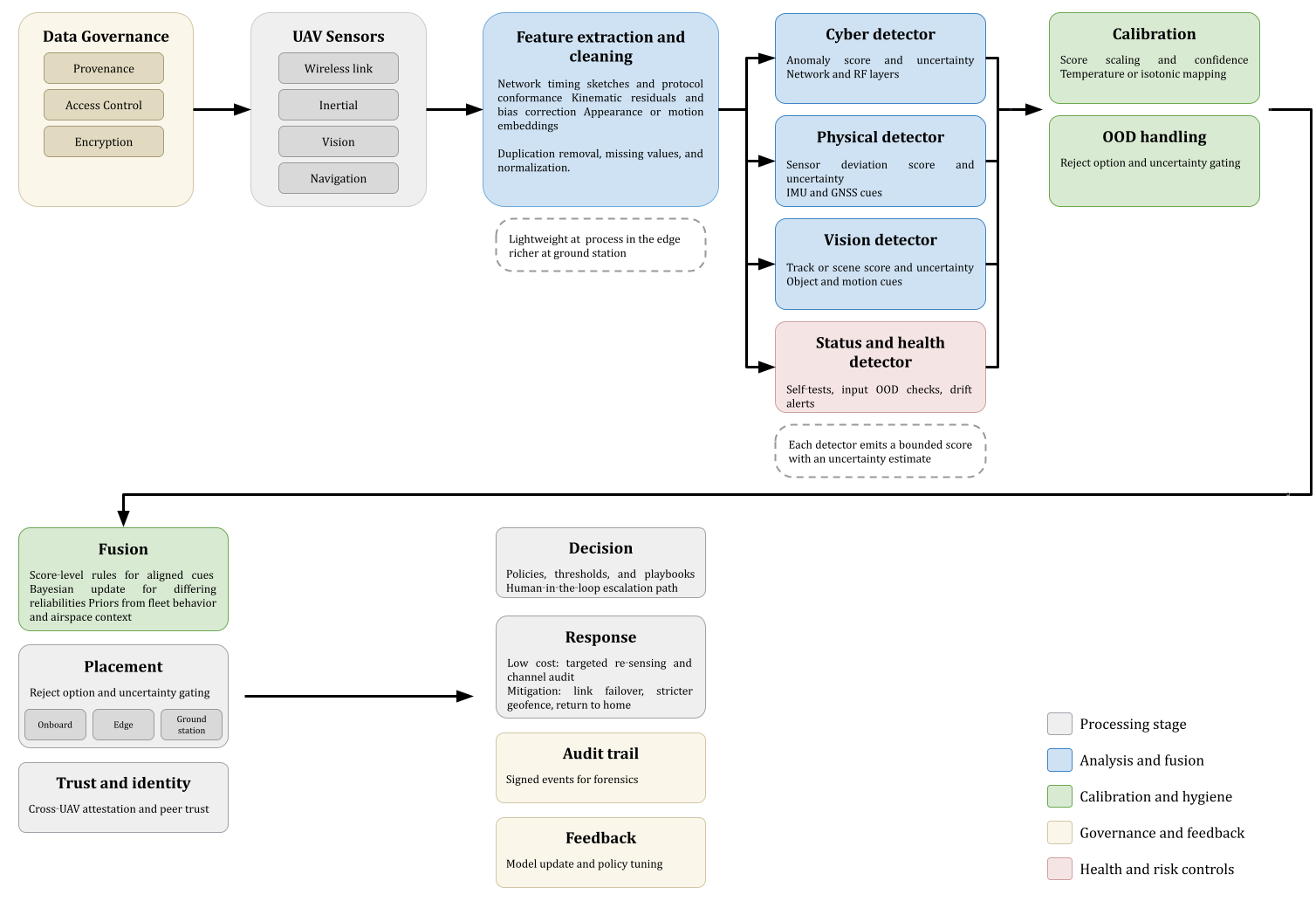}
    \caption{Unified IDS pipeline for UAVs, from governed sensor intake and feature cleaning to modular cyber/physical/vision detectors with health monitoring.}
    \label{fig:reference_pipeline}
\end{figure*}

\textcolor{black}{Figure~\ref{fig:reference_pipeline} presents a UAV security architecture that begins with data governance (provenance, access control, encryption) applied to multi-modal sensor inputs from wireless links, inertial navigation, and vision systems. Feature extraction and cleaning processes generate network timing sketches and protocol conformance for cyber channels, bias-corrected kinematic residuals for inertial/navigation data, and motion embeddings for vision, with duplication removal and normalization applied. Modular detectors, including cyber, physical, and visual components, emit anomaly scores and uncertainty estimates. A status and health detector monitors self-tests and drift alerts. The pipeline concludes with fusion using score-level rules and Bayesian updates, decision-making through policies and thresholds with human oversight. These response mechanisms include targeted re-sensing and geofencing, audit trail generation for forensic purposes, and feedback loops for model updates and policy tuning.}

\textcolor{black}{Following the detection phase, the pipeline implements calibration and fusion mechanisms to synthesize multi-modal outputs into actionable intelligence. Confidence calibration utilizes temperature scaling and isotonic mapping to normalize detector scores across modalities, leveraging historical logs and fleet telemetry. Additionally, OOD handling offers explicit reject options for uncertain cases. The fusion stage applies score-level rules for aligned cues and Bayesian updates when modality reliabilities diverge, incorporating priors from fleet behavior and airspace context at both vehicle edge and ground station levels. The decision layer executes policies and playbooks through human-in-the-loop escalation paths, progressing from low-cost targeted re-sensing and channel audits to more aggressive mitigation, including link failover, stricter geofencing, and return-to-home commands based on fused confidence levels. The pipeline incorporates placement flexibility through onboard, edge, and ground-station deployment options, with a reject option and uncertainty-gating capabilities. Trust and identity mechanisms enable cross-UAV attestation and the establishment of peer trust for multi-vehicle coordination. An audit trail component maintains a record of signed events for forensic analysis, while a feedback loop facilitates continuous model updates and policy tuning to enhance system performance over time.}

\textcolor{black}{A unified IDS emerges as a convergent architecture synthesizing specification-based, anomaly-based, supervised learning, and multi-modal detection methodologies from the reviewed literature; our presentation receives complementary support through the summary patterns documented in Table~\ref{PartialUnification}. Building on this architectural foundation, we align the unified taxonomy with discrete stages encompassing sensing, feature processing, detector outputs, calibration, fusion, decision-making, and response mechanisms, and indicate typical deployment configurations across the onboard, edge, and ground control tiers.}

Specifically tailored to address cyber threats and UAV-physical intrusions, the unified IDS integrates traditional network-based intrusion detection techniques with cutting-edge sensor-based and machine learning-driven mechanisms. \textcolor{black}{Unification manifests through four primary mechanisms: correlating network anomalies with physical signals by aligning RF link disruptions with IMU or GPS consistency checks and co-occurring vision sightings; enabling cross-modal alert handoff where cyber alarms trigger targeted video analysis or physical detections prompt command-and-control channel inspection; reconciling local indicators with shared telemetry, airspace context, and traffic management advisories at the fleet level; and normalizing heterogeneous events from networks, firmware, sensors, and vision through unified logging engines for joint reasoning. The framework emphasizes calibrated alerting and out-of-distribution handling to maintain score interpretability across modalities, employing on-board detectors with uncertainty estimates, calibrated thresholds, and reject options for inputs outside training support. Practical calibration employs lightweight techniques, such as temperature scaling on held-out splits. At the same time, out-of-distribution screening utilizes low-confidence flags, density margins, and cross-modal disagreement checks to gate high-impact actions. Robustness hooks incorporating calibrated reject options and cross-modal corroboration reduce adversarial errors, with RGB-IR coupling and acoustic-RF signatures providing complementary evidence that mitigates small-object, clutter, and bird confusions, as summarized in Table~\ref{PartialUnification}.} 


\begin{table*}[ht]
\centering
\caption{Partial unification of cyber and physical Intrusion Detection Systems in Unmanned Aerial Vehicle contexts.} \label{PartialUnification}
{\footnotesize
\begin{tabular}{p{0.25\textwidth}p{0.4\textwidth}p{0.25\textwidth}}
\hline
\textbf{Unification pattern} & \textbf{Primary modalities} & \textbf{Typical placement} \\
\hline

{\tiny Cross‑modal correlation~\cite{qin2014performance, yoon2017virtualdrone}} & \tiny Radio frequency or network with Inertial Measurement Unit, Global Navigation Satellite System, or vision & \tiny Onboard, Swarm cooperative \\

{\tiny Alert‑triggered sensing from cyber to physical~\cite{guo2020cyber}} & \tiny Network logs triggering vision analysis or radio frequency fingerprinting & \tiny Onboard, Edge, Ground Control Station \\

{\tiny Alert‑triggered auditing from physical to cyber~\cite{guo2020cyber}} & \tiny Vision or radio frequency detections prompting command‑and‑control and application inspection & \tiny Edge, Ground Control Station \\

{\tiny Fleet telemetry baselining~\cite{morais2025review}} & \tiny Cross--Unmanned Aerial Vehicle kinematics with link metrics and shared context & \tiny Swarm cooperative, Ground Control Station \\

{\tiny Unified event schema and policy~\cite{yang2025unified}} & \tiny Normalized network, firmware, sensor, and vision events in a standard schema & \tiny Ground Control Station, Cloud \\
\hline
\end{tabular}
}
\end{table*}

The existence of such systems has yet to be discussed in the literature. However, a wide range of AI-based techniques can be incorporated into the unified IDS, as discussed in the literature. \textcolor{black}{Building on this context, we present in Figure~\ref{fig:taxonomy} a taxonomy of an Artificial Intelligence–enabled unified Intrusion Detection System for Unmanned Aerial Vehicles that spans applications, entry points, cyber and physical modalities, and multi-tier execution, followed by evaluation guidance aligned with operational constraints.} We first provide several smart city applications where UAVs can be employed and where security threats and physical attacks may occur. The list includes traffic management, disaster response, aerial surveillance, and crowd management. Then, we categorize the intrusions into two types: i) cyber-attacks over communication networks, which encompass jamming, spoofing, and DoS, and ii) the physical intrusions of unwanted UAVs, which result in physical damage, hijacking, and spying, etc. The UAV entry points of these intrusion categories might differ depending on the types of attacks. They can be used through network interfaces or equipped with sensors, which can, in turn, detect intrusions. Several AI techniques can be employed with IDS, as discussed further in the next section. These AI modules can be executed at the device level or at various network levels, including fog and cloud, depending on factors such as the nature and complexity of the algorithms and the UAV's operational mode (e.g., remote or in the fleet). \textcolor{black}{IDS mechanisms require evaluation through a comprehensive metric suite that reflects both cyber-defense efficacy and UAV operational constraints. Beyond conventional measures such as accuracy, response time, and false alarm rates, assessments must explicitly incorporate latency-throughput-energy considerations and UAV-relevant multi-object tracking metrics for vision-driven counter-UAV functions. Latency should be reported end-to-end from event onset to mitigation actuation, stratified by deployment tier (onboard, swarm-cooperative, edge, and GCS) to capture placement effects. The energy footprint must be characterized on target platforms using indicative power draw and energy per inference, acknowledging that the IDS workload directly competes with flight endurance. For physical-intrusion detection pipelines, metrics including Multiple Object Tracking Accuracy (MOTA), Multiple Object Tracking Precision (MOTP), Identification F1 Score (IDF1), and track fragmentation quantify detection-association performance under occlusions, scale changes, and clutter. Complementary reports of model size, on-device inference time, and calibration quality enhance interpretability and reproducibility without relying on system-specific numerical claims.}

\subsection{\textcolor{black}{IDS Deployment Topologies for UAV}}
\textcolor{black}{Beyond generic Host-based Intrusion Detection System and Network-based Intrusion Detection System (HIDS/NIDS) taxonomies \cite{satilmics2024systematic}, we map IDS placement to UAV-specific platforms in the following. For Onboard IDS in the UAV, this includes IDS systems for the flight controller, avionics bus, and onboard radios. The strengths of embedding the IDS within the UAVs include low detection latency for flight-safety anomalies and link attacks, as well as resilience when disconnected. However, the limited energy and computational power within the UAVs limit the IDS systems implemented on them, as well as the risk of local overfitting and higher false alarms without context. The onboard IDS systems can perform integrity checks on the sensors and actuators in the UAVs, verifying firmware and code integrity, detecting link-layer jamming/spoofing cues, and utilizing lightweight anomaly models.}

\textcolor{black}{Cooperative IDS among UAVs via peer-to-peer or collaborative architectures. This approach enables the cross-validation of positioning data, facilitates distributed consensus on detected anomalies, and shares threat signatures to identify intruding UAVs. The cooperative model can reduce false alarms, detect location-specific spoofing attacks, and provide resilience against single-node compromises. However, implementing cooperative IDS faces several constraints, including limited UAV bandwidth, synchronization challenges, trust establishment requirements, and increased energy consumption for inter-UAV communication. To address these challenges, the system cross-validates Inertial Measurement Unit (IMU) and Global Positioning System (GPS) data consistency against planned flight paths. Additionally, it employs multi-UAV RF anomaly triangulation and collaborative detection of black-hole and gray-hole attacks within Flying Ad Hoc Networks (FANETs).}

\textcolor{black}{Edge and fog computing can enhance UAV intrusion detection systems by addressing the computational limitations and energy constraints associated with continuous monitoring in UAVs~\cite{sajid2023fog}. This can be implemented through ground relays, mobile edge units, or roadside units that possess greater processing capabilities to support efficient IDS operations. In edge-based IDS implementations, near-field computing reduces the round-trip times for data transmission. It enables more intensive analytics, including multi-sensor fusion and pre-filtering, before data is transmitted to the cloud. This approach allows the deployment of sophisticated AI models that can be processed at the edge. Additionally, edge computing enables more robust data fusion across UAV sensors, including radar, vision, and acoustic systems, while maintaining moderate latency and reducing onboard computational and energy requirements. The primary constraints of edge-based approaches include variable edge coverage and challenges related to data governance and management. This methodology proves particularly effective for anti-UAV applications that involve the fusion of vision and RF sensors, as well as establishing a baseline for fleet behavior.}

\textcolor{black}{GCS and cloud-deployed intrusion detection systems provide comprehensive fleet management capabilities through centralized portals, traffic management interfaces, and mission planning functionality. The cloud component of the IDS enables analytics without latency constraints, supporting threat intelligence and alert systems accessible through the GCS. The GCS and cloud-based IDS deployment offer a global operational view that reduces false alarms through historical context analysis and human intervention, while enabling scalable model training and signature distribution across the system. However, the primary constraints include high latency, dependence on reliable backhaul connectivity, and delayed response times for immediate in-flight threats from intruding objects. The GCS and cloud-deployed IDS components are particularly effective for applications that enable operators to detect social engineering attacks, ensure policy compliance, conduct post-flight forensic analysis, and deploy model updates across the fleet.}

\textcolor{black}{To summarize the trade-offs among different IDS deployment approaches, several key performance metrics reveal distinct characteristics. For RL-based policies, onboard action shields and edge or GCS oversight provide layered safety, enabling conservative on-device responses with supervised escalation and limited real-world exploration during staged rollout. In practice, offload is triggered when the on-board load, link quality, and remaining energy exceed conservative bounds, with a default to local, lower-fidelity inference and deferred synchronization when links are constrained. Regarding latency, on-board systems provide the fastest response times, while swarm-cooperative and edge/fog deployments offer comparable intermediate performance, and GCS/cloud implementations exhibit the highest latency due to communication overhead:}
\begin{align}
&\text{\textbf{Latency}: On-board} < \text{Swarm-cooperative} \nonumber \\
&\qquad \approx \text{Edge/Fog} < \text{GCS/Cloud}
\end{align}
\textcolor{black}{Energy consumption and computational load on UAVs follow an inverse pattern, where on-board and swarm-cooperative approaches demand the highest resources from individual UAVs, while edge/fog and GCS/cloud deployments provide significant computational offloading benefits:}
\begin{align}
\text{\textbf{Energy/Compute on UAV}:} &\nonumber \\
\text{On-board/Swarm-cooperative} &> \text{Edge/Fog} \nonumber \\
&\hspace*{-2.2cm} > \text{GCS/Cloud} \hspace*{\fill}
\end{align}
\textcolor{black}{False alarm rates are significantly reduced in context-aware deployments, including swarm-cooperative, edge, and GCS systems, compared to isolated on-board implementations. Hybrid architectures effectively address this limitation by combining multiple approaches to achieve a more comprehensive solution.}

\textcolor{black}{The recommended hybrid architecture integrates on-board fast guardrails for immediate threat response, cooperative validation among UAVs for distributed consensus, edge-based sensor fusion for enhanced analytics, and GCS oversight for system updates and comprehensive analysis. Fleet-level learning is naturally supported by the edge or GCS tier, where secure aggregation of on-board updates and scheduled synchronization windows reduces bandwidth and privacy exposure while refreshing models.}

\section{AI-enabled IDS for UAV}
\label{ai_ids_uav}


With the unprecedented data explosion that many smart cities are witnessing, AI is becoming one of the predominant technologies for novel solutions and services~\cite{ramu2022federated, pathik2022ai, bokhari2022use, alam2022application}. In IDS, AI, via machine learning, deep, and reinforcement learning techniques, has shown great potential in improving the performance of IDSs to link network features, identify patterns in large-scale datasets, detect anomalies, and autonomously react to many cyber threats and physical intrusions~\cite{10112557, 10.1145/2248326.2248334,9765451}. AI is providing a new level of protection to UAV systems by integrating with IDS. This integration enables more advanced and efficient methods for detecting and responding to potential security threats, including unauthorized access and malicious activity. With machine learning algorithms, AI-enabled IDS can analyze large amounts of data and detect patterns or anomalies that might indicate a security breach. This added layer of protection helps ensure the safety and security of UAV systems, preventing unauthorized access or attacks. AI-based IDS can be divided into three main categories: supervised, unsupervised, and reinforcement learning. All three approaches have their advantages and disadvantages, and the choice of which approach to use will depend on the specific requirements and constraints of the system. They are defined as follows:
\begin{itemize}
\item Supervised AI-based IDS uses labeled datasets to train machine learning algorithms to recognize specific patterns or anomalies that indicate a security threat, and respond to similar patterns or anomalies in new datasets. This approach is useful when there is a clear understanding of what constitutes a security threat and when labeled datasets are available to train the models.
\item Unsupervised AI-based IDS, conversely, uses unlabeled datasets to train the system to detect patterns or anomalies without prior knowledge of what constitutes a security threat. This approach is functional because it works for types of attacks that are not predefined and when labeled datasets are not available. The system can detect patterns, anomalies, or behaviors that may indicate a security breach or intrusion, but it requires further human analysis to confirm.
\item Reinforcement learning-based IDS uses a combination of trial-and-error learning and reward-based learning to train the system. The system is trained to learn from its environment and can adjust its behavior based on the outcomes of its actions. RL-based IDS is particularly useful in dynamic environments where the characteristics of security threats may change over time. RL can also be used to counteract or mitigate an attack.
\end{itemize}

\textcolor{black}{Placement-aware AI design requires careful consideration of computational constraints across deployment tiers. On-board models must remain lightweight, utilizing compressed CNNs, shallow autoencoders, and tiny Transformers with conservative thresholds to minimize energy consumption and limit false positives. Swarm-cooperative and edge/fog tiers can accommodate more resource-intensive deep learning applications, including vision and RF fusion and graph-based detectors, to validate on-board alerts and reduce false alarms. The GCS\/cloud tier performs slow-time-scale learning through federated or centralized approaches and distributes signed updates to lower tiers. We recognize that many cyber IDS studies for UAVs have been trained or validated on generic IT datasets rather than UAV-native traffic. This limits ecological validity, since control links, compact payloads, bursty telemetry, and RF impairments differ from enterprise flows. Consistent with our discussion on dataset scarcity in Section~\ref{datascarcity}, we therefore flag dataset provenance in our summaries and interpret such results as indicative patterns rather than deployable guarantees. When UAV-specific traces are unavailable, two practical mitigation paths reported in the literature are lightweight domain adaptation from generic corpora to UAV traces and transfer learning using small UAV captures for calibration. These caveats guide placement-aware use and realistic expectations. We therefore harmonize terminology across studies by consistently noting task, modality, dataset family, features, model family, compute tier, and typical metrics in prose rather than expanding into a benchmarking table.}

\subsection{Supervised Learning Approaches}

\textcolor{black}{Supervised IDSs for smart cities rely on labeled examples of malicious and benign activities to learn how to identify cyber and physical threats. These systems typically use machine learning techniques, such as decision trees, support vector machines (SVMs), and Artificial Neural Networks (ANNs), to analyze network traffic and classify it as either normal or malicious. These techniques enable the IDS to learn complex patterns in the data and enhance its ability to classify various types of activity. Another trend is the use of transfer learning, which involves training a model on a large dataset and then fine-tuning it for a specific task. This can be particularly useful in the context of smart cities, where data may be limited or specific to a particular location. In interpreting supervised results, we therefore frame findings as transferable hypotheses pending UAV-specific calibration or adaptation, in line with the scarcity caveats discussed in Section~\ref{datascarcity}.}

In~\cite{electronics10131549}, the authors proposed a classification-based IDS that enhances the UAV network security and protects it from various types of attacks by detecting malicious packets within the network. The proposed IDS identifies the malicious packets circulating in the network using ML classifiers, namely, Logistic Regression (LR), Linear Discriminant Analysis (LDA), K-Nearest-Neighbour (KNN), Decision Tree (DT), Gaussian Naive Bayes (GNB), Stochastic Gradient Descent (SGD), and K-means. These algorithms have been benchmarked for this purpose, and their performances were evaluated based on their accuracy, precision, recall, F1-score, and false negative rates. The simulation conducted in this paper demonstrated that DT outperformed all benchmarked algorithms in identifying malicious packets when tested against two types of attacks: denial-of-service (DoS) and Botnet attacks. The simulations done in this paper were based on a real-world dataset called ~\href{https://github.com/awslabs/open-data-registry/blob/main/datasets/cse-cic-ids2018.yaml}{CSE-CIC IDS-2018} publicly published by the Canadian Establishment for Cybersecurity (CIC)~\cite{Canadiandataset}.

Gaoyan et al.~\cite{8788766} proposed a GPS spoofing detection algorithm that leverages the air traffic control messages that are periodically broadcast by aerial vehicles. To this end, they designed an ML-based framework called GPS-Probe to determine the accurate location of the target aircraft and detect if its position data has been compromised by GPS spoofing attacks. The algorithm utilizes the adaptive KNN as a foundation to constantly analyze the Received Signal Strength Indicator (RSSI) and Timestamps at the Server (TSS) and determine position estimates based on these two characteristics. Then, an XGBoost binary classifier is trained based on the derived positions to detect spoofing. As reported in the paper results, at its best performance, the proposed framework achieves up to 89.7\% accuracy and 91.5\% precision in detecting GPS spoofing attacks in under 5 seconds over a 10 km range. 

In the same context of GPS spoofing attacks, Panice et al. proposed an SVM-based approach in~\cite{panice2017svm} to detect this malicious attack by analyzing navigation data. The authors demonstrated that the system can detect almost all spoofing attacks, unless the hijacker has absolute knowledge of the drone's trajectory in advance. The authors discussed some limitations of this approach and how its performance degrades significantly with long attacks, especially those that attempt to track the drone. To test their approach, the authors designed a simulation system where they generated noisy trajectory information and fed it to a simulated navigation system. 

Abdulhadi et al.~\cite{8323415}, proposed a framework that utilizes machine learning algorithms, including Random Forest (RF), KNN, SVM, Linear Discriminant (LD), and Quadratic Discriminant (QD), to analyze interactions with remotely controlled UAVs and detect hijacking attacks. The radio control signal sent from the pilot to the drone is captured and processed to identify the controlling pilot. The proposed approach focuses on how the pilot controls the remote device rather than what they actually do with it. The simplicity of the proposed approach enables its deployment onboard and makes it suitable for identifying hijacking onsets. Extensive simulation results were conducted, and the authors conclude that RF is the most suitable algorithm for the task, achieving 90\% accuracy in identifying controlling pilots. For simulation purposes, the authors gathered data from a Hummingbird quadrotor UAV, using a low-level protocol to acquire radio control data at a frequency of 10Hz and log them into a Comma-Separated Values (CSV) file. 

In~\cite{manesh2019performance}, the authors discussed jamming attacks and how they can interrupt plaintext messages exchanged between UAVs in a network over unencrypted data links. To mitigate this issue, the authors compare the performance of several ML algorithms, including SVM, KNN, ANN, and DT. For this comparison and to train these algorithms, the authors propose simulating a dataset of Automatic Dependent Surveillance-Broadcast (ADS-B) messages from a transmitter to a receiver and using it to feed the algorithms. The simulation results demonstrate that an ANN with two hidden layers and 15 neurons outperforms all the other benchmarked algorithms. 

\textcolor{black}{In~\cite{zhao2022vision}, the authors proposed a vision-based method for detecting and tracking UAVs in real time. The approach uses a combination of deep learning and computer vision techniques to detect UAVs in video frames and track their movements over time. The method is evaluated on the ICCV2021 Anti-UAV Challenge dataset and demonstrates good performance in detecting and tracking UAVs across various environments. This paper presents a promising solution for UAV detection and tracking, highlighting the potential of deep learning and computer vision techniques in this area.} 

\textcolor{black}{While recent vision baselines show promise, common failure modes persist in anti-UAV settings. Small and fast targets reduce apparent size and motion cues, leading to missed detections. Urban clutter, occlusions, and background motion can cause false positives and track fragmentation. Birds and airborne debris are frequent confounders due to their similar silhouettes and trajectories, especially at long ranges and low resolutions. Night, haze, and glare further degrade RGB performance. In practice, these limitations motivate complementary sensing and lightweight fusion, where RGB is paired with IR for thermal contrast in low-light conditions, and, where feasible, augmented with acoustic or RF cues to stabilize detection, reduce ambiguities, and support seamless handover between modalities.}

\begin{figure*}[t]
    \centering  
    \includegraphics[width=0.8\textwidth]{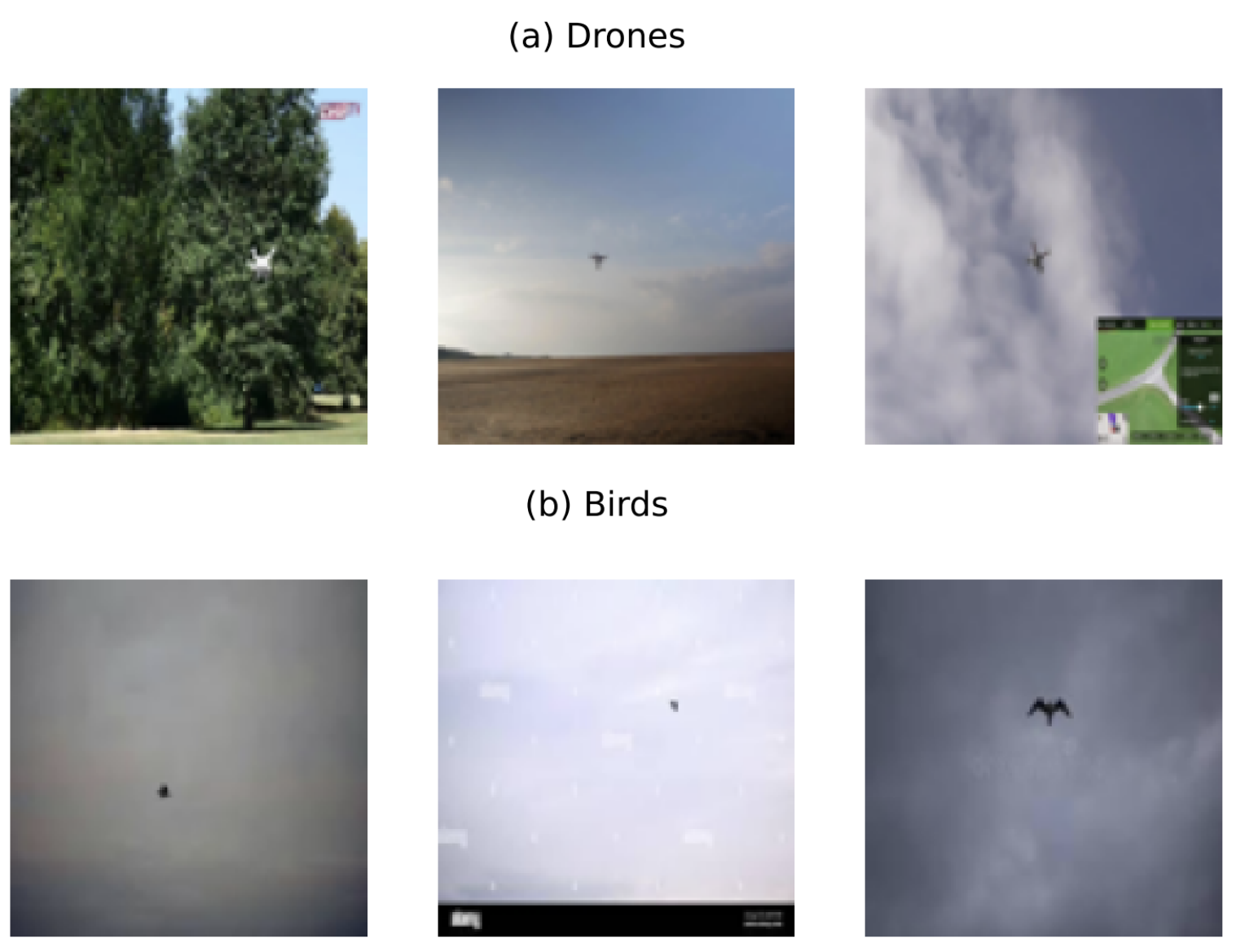}
    \caption{Samples from the Drone vs Bird datatset~\cite{9663844}}
    \label{fig:drone_vs_uav}
\end{figure*}

\textcolor{black}{In~\cite{9615243}, the authors presented a new dataset and benchmark for evaluating the performance of vision-based algorithms for detecting and tracking UAVs. The dataset comprises a variety of data types collected from multiple sources in diverse environments, encompassing both positive and negative examples of UAVs. The authors also present a new evaluation metric for this task, which considers the movement and orientation of the UAVs, as well as the location and orientation of the sensors. Using this dataset and metric, the authors demonstrate the performance of several state-of-the-art algorithms for UAV tracking and provide insights into the challenges and opportunities for further improvement in this field.} 

The authors of~\cite{8865525} presented an anti-UAV system that detects UAVs in the sky using an improved version of the YOLO v3 object detection algorithm. The improved algorithm is based on the YOLO v3 architecture, incorporating several modifications to enhance its performance, including adjustments to the anchor boxes and the use of multi-scale training. The system was tested on a dataset comprising a mix of UAV images collected from the internet and shot by their cameras. It achieved high accuracy and recall rates in detecting UAVs. The results demonstrated that the enhanced YOLO v3 algorithm is real-time using camera images in real-world scenarios, making it suitable for use in anti-UAV systems.

In~\cite{9368788}, the authors discussed the use of the YOLOv4 algorithm for object detection in UAV systems for anti-UAV applications. The YOLOv4 algorithm is a real-time object detection algorithm that accurately detects and classifies objects in images and videos. The paper describes how the YOLOv4 algorithm was trained and tested on a dataset of UAV images and videos, and how it accurately detected and classified UAVs in the test images and videos. The paper also discusses the potential applications of the YOLOv4 algorithm for anti-UAV systems, such as detecting and identifying UAVs that may be flying in restricted airspace, or detecting and tracking UAVs that may be used for malicious purposes.

In~\cite{shan2020image}, the authors described an image recognition method for an anti-UAV system that utilizes a Convolutional Neural Network (CNN). The system is designed to detect and identify UAVs in real-time using images captured by cameras. The CNN is trained on a dataset of images of UAVs and non-UAV objects, and it can classify new images as either UAVs or non-UAVs. The system is tested and evaluated using a dataset of real-world images, demonstrating high accuracy in detecting and identifying UAVs. The paper concludes that the CNN-based image recognition method is an effective solution for anti-UAV systems, and it has the potential to be integrated into real-world systems.

In~\cite{9824591}, the authors presented a new method, GASiam, for tracking and identifying small UAVs in infrared (IR) images using a graph-attention-based Siamese tracker. The proposed method employs a graph attention mechanism to model the interactions between UAVs and their surrounding environment, enabling more accurate tracking and identification of UAVs in IR images. The Siamese network architecture is utilized to learn the similarity between UAVs across consecutive frames, enabling robust tracking of UAVs even in challenging scenarios, such as occlusions or rapid motion. The proposed method was evaluated on a publicly available dataset, and the results show that it outperforms existing state-of-the-art methods for tracking and identifying UAVs in IR images.

In~\cite{s22103701}, the authors presented a method for long-term tracking of UAVs using a hybrid attention mechanism and hierarchical discriminator. The proposed method combines features from both the spatial and temporal dimensions to improve the accuracy of UAV tracking. The hybrid attention mechanism focuses on the important regions of the UAV by utilizing both spatial and temporal attention. At the same time, the hierarchical discriminator enhances the accuracy of UAV tracking by employing multiple levels of classification. Additionally, the method utilizes a combination of color and gradient information to enhance the robustness of the tracking algorithm across various lighting conditions. Experimental results demonstrated that the proposed method outperforms existing methods in terms of accuracy and robustness, enabling the tracking of UAVs over extended periods of time. The method can be applied in various settings, including surveillance, search and rescue operations, and security systems.

In~\cite{lei2022anti}, the authors presented a neural network-based early warning system for detecting and responding to unauthorized UAVs using high-performance computing and particle swarm optimization (PSO) algorithms. The system uses a combination of data from radar, cameras, and other sensors to detect UAVs and determine their flight paths. The PSO algorithm is utilized to optimize the parameters of the neural network, leading to enhanced accuracy and efficiency in detecting UAVs. The system also includes a decision-making module that can respond to detected UAVs in real time, such as by sending a signal to a jammer to disrupt the UAV's communication or control systems. The proposed system is shown to have a high detection rate and low false alarm rate in simulations and experimental tests.

\begin{table*}[ht]
\centering
\caption{Performance Comparison of Supervised Learning Approaches for UAV Intrusion Detection}\label{supervised_comparison}
{\scriptsize
\setlength{\tabcolsep}{2pt}
\begin{tabular}{p{1cm}p{2.8cm}p{2.4cm}p{1.5cm}p{1.5cm}p{1.8cm}p{1.2cm}}
\hline
\textbf{\begin{tabular}[c]{@{}c@{}}\tiny Ref.\end{tabular}} & 
\textbf{\begin{tabular}[c]{@{}c@{}}\tiny Method/Algorithm\end{tabular}} & 
\textbf{\begin{tabular}[c]{@{}c@{}}\tiny Dataset\end{tabular}} & 
\textbf{\begin{tabular}[c]{@{}c@{}}\tiny Accuracy\\ \tiny (\%)\end{tabular}} & 
\textbf{\begin{tabular}[c]{@{}c@{}}\tiny Precision\\ \tiny (\%)\end{tabular}} & 
\textbf{\begin{tabular}[c]{@{}c@{}}\tiny Attack Type\end{tabular}} & 
\textbf{\begin{tabular}[c]{@{}c@{}}\tiny Response\\ \tiny Time\end{tabular}} \\
\hline

{\tiny \cite{8788766}} & {\tiny GPS-Probe (Adaptive KNN + XGBoost)} & {\tiny OpenSky Network ATC data} & {\tiny 89.7} & {\tiny 91.5} & {\tiny GPS spoofing} & {\tiny 1.7s (mean), $<$ 5s} \\

{\tiny \cite{panice2017svm}} & {\tiny SVM (SVDD and v-SVM)} & {\tiny Simulated navigation data} & {\tiny Almost all spoofing detected} & {\tiny -} & {\tiny GPS spoofing} & {\tiny Within 30 seconds} \\

{\tiny \cite{8323415}} & {\tiny RF, KNN, SVM, LD, QD} & {\tiny Hummingbird quadrotor UAV} & {\tiny RF: 90} & {\tiny -} & {\tiny Hijacking} & {\tiny Real-time (10Hz)} \\

{\tiny \cite{manesh2019performance}} & {\tiny SVM, KNN, ANN, DT, LR} & {\tiny Simulated ADS-B messages} & {\tiny ANN (2 hidden, 15 neurons): 81} & {\tiny -} & {\tiny Jamming} & {\tiny -} \\

{\tiny \cite{zhao2022vision}} & {\tiny Deep learning + Computer vision} & {\tiny ICCV2021 Anti-UAV Challenge} & {\tiny Good performance} & {\tiny -} & {\tiny UAV detection} & {\tiny Real-time} \\

{\tiny \cite{9615243}} & {\tiny Deep learning + Computer vision} & {\tiny Anti-UAV benchmark (318 RGB-T pairs)} & {\tiny Good performance} & {\tiny -} & {\tiny UAV detection} & {\tiny Real-time} \\

{\tiny \cite{9368788}} & {\tiny Improved YOLO v4} & {\tiny UAV images (internet + own)} & {\tiny High accuracy} & {\tiny High recall} & {\tiny UAV detection} & {\tiny Real-time} \\

{\tiny \cite{shan2020image}} & {\tiny CNN + SVM} & {\tiny MNIST, UAV images, bird images} & {\tiny CNN: 95.9\% (UAV), SVM: 88.4\%} & {\tiny -} & {\tiny UAV detection} & {\tiny -} \\

{\tiny \cite{9824591}} & {\tiny GASiam} & {\tiny IR images} & {\tiny Outperforms SOTA} & {\tiny -} & {\tiny Small UAV tracking} & {\tiny -} \\

{\tiny \cite{s22103701}} & {\tiny Hybrid attention + Hierarchical} & {\tiny UAV tracking sequences} & {\tiny Outperforms existing} & {\tiny -} & {\tiny Long-term tracking} & {\tiny Extended} \\

{\tiny \cite{lei2022anti}} & {\tiny Neural Network + PSO} & {\tiny Radar, cameras, sensors} & {\tiny High detection rate} & {\tiny -} & {\tiny Unauthorized UAVs} & {\tiny Real-time} \\
\hline
\end{tabular}
}
\end{table*}

\textcolor{black}{Table~\ref{supervised_comparison} compares supervised learning methods for UAV intrusion detection across six dimensions: reference source, algorithm type, dataset used, accuracy/precision percentages, target attack type, and response time. The approaches range from traditional machine learning (SVM, KNN) to advanced deep learning methods (YOLO, CNN), addressing both cyber attacks (GPS spoofing, jamming, hijacking) and physical intrusions (UAV detection/tracking). Key findings show GPS-Probe achieving 89.7\% accuracy for spoofing detection in under 5 seconds, Random Forest reaching 90\% accuracy for hijacking detection at real-time 10Hz rates, and deep learning methods consistently outperforming traditional approaches for vision-based UAV detection. However, some studies lack specific numerical metrics, indicating a need for standardized evaluation frameworks in this field.}

\subsection{Unsupervised Learning Approaches}

Unsupervised IDSs for smart cities are designed to identify cyber threats without relying on prior knowledge or labeled examples of malicious activity. These systems typically use techniques from machine learning and data mining to analyze network traffic and identify patterns that may indicate an attack.
Unsupervised learning techniques enable the IDS to identify complex patterns in the data, thereby enhancing its ability to detect anomalies. Researchers also focus on improving the interpretability of unsupervised IDSs. This includes developing techniques for explaining the reasoning behind the IDS's decisions, as well as methods for identifying and mitigating any biases in the training data.
Altogether, the development of unsupervised IDSs for smart cities is an active area of research, aimed at creating more robust and effective systems for detecting and defending against cyber and physical threats.

In~\cite{8854240}, Tiep et al. proposed an unsupervised learning-based IDS that detects eavesdropping attacks in UAV wireless systems. The proposed IDS is based on One-Class Support Vector Machines (OC-SVM) and K-means clustering. In the same paper, the authors discuss how to prepare artificial training data to feed and train the ML algorithms based on statistical knowledge of channel state information. The performance of the algorithms was evaluated using accuracy, sensitivity, and specificity across various Signal-to-Noise Ratio (SNR) scenarios, ranging from 0 dB to 20 dB.

In~\cite{sedjelmaci2017hierarchical}, the authors designed a novel hierarchical IDS that can be deployed not only onboard the UAV itself but also on ground stations to detect several malicious anomalies such as false information dissemination, GPS spoofing, jamming, and black hole and gray hole attacks. The system was proposed for monitoring the actions of UAVs, categorizing them into four categories: normal, abnormal, suspicious, and malicious. To determine the status of the drone, a set of rules for detecting cyber-attacks is utilized, and an SVM-based anomaly detection algorithm is used to confirm any detected attacks. The simulations conducted by the authors demonstrate that anomalies are successfully detected, even in a large network of UAVs and attackers. 

In~\cite{park2020unsupervised}, Park et al. proposed an unsupervised learning-based IDS for UAVs that can identify an abnormal status of the UAV regardless of the type of attack. This behavior is obtained by fitting an autoencoder on only benign flight data that does not contain any anomalies or unusual behaviors. Hence, the model can identify any abnormalities with distributions that differ from the training data. At inference time, the reconstruction loss for benign data is low, whereas the reconstruction loss for data with anomalies is significantly higher; thus, the loss values are used for classification. 

In~\cite{10112531}, the study proposed a vision-based UAV detection and classification pipeline using ground cameras. UAVs are first detected using the YOLOv8 object detection model. For classification, an unsupervised approach is adopted in which Histogram of Oriented Gradients (HOG) features are extracted from the detected UAV regions. The extracted features are then embedded into a two-dimensional space using t-distributed Stochastic Neighbor Embedding (t-SNE) to enable separation of different UAV categories. Clustering is subsequently performed in the embedded space to group UAVs without relying on labeled training data. The framework is evaluated on an unlabeled anti-intrusion UAV dataset and compared against conventional machine-learning-based clustering methods.
\begin{table*}[ht]
\centering
\caption{Performance Comparison of Unsupervised Learning Approaches for UAV Intrusion Detection}\label{unsupervised_comparison}
{\scriptsize
\setlength{\tabcolsep}{2pt}
\begin{tabular}{p{0.6cm}p{2.3cm}p{2cm}p{2.4cm}p{2.2cm}p{2cm}p{1cm}}
\hline
\textbf{\begin{tabular}[c]{@{}c@{}}\tiny Ref.\end{tabular}} & 
\textbf{\begin{tabular}[c]{@{}c@{}}\tiny Method\end{tabular}} & 
\textbf{\begin{tabular}[c]{@{}c@{}}\tiny Dataset\end{tabular}} & 
\textbf{\begin{tabular}[c]{@{}c@{}}\tiny Evaluation\\ \tiny Metrics\end{tabular}} & 
\textbf{\begin{tabular}[c]{@{}c@{}}\tiny Accuracy\end{tabular}} &
\textbf{\begin{tabular}[c]{@{}c@{}}\tiny Attack Type\end{tabular}} & 
\textbf{\begin{tabular}[c]{@{}c@{}}\tiny SNR\\ \tiny Range\end{tabular}} \\
\hline

{\tiny \cite{8854240}} & {\tiny One-Class SVM, K-means++} & {\tiny Wireless signals from UAV relay} & {\tiny Accuracy, Sensitivity, Specificity} & {\tiny OC-SVM (20 dB - \%91.4) K-means (20 dB - \%100.4)} & {\tiny Eavesdropping} & {\tiny 0-20 dB} \\

{\tiny \cite{sedjelmaci2017hierarchical}} & {\tiny Rules-based + SVM} & {\tiny Simulated UAV network} & {\tiny Detection rate, False positive rate, Efficiency} & {\tiny Not specified} & {\tiny Multiple attacks} & {\tiny Not specified} \\

{\tiny \cite{park2020unsupervised}} & {\tiny Autoencoder} & {\tiny HITL UAV DoS \& GPS Spoofing dataset} & {\tiny Reconstruction loss} & {\tiny Not specified} & {\tiny DoS, GPS Spoofing} & {\tiny Not specified} \\
\hline
\end{tabular}
}
\end{table*}

\textcolor{black}{Table \ref{unsupervised_comparison} presents a comparative analysis of unsupervised learning approaches for UAV intrusion detection, examining three key studies across six evaluation dimensions: reference source, algorithmic methodology, dataset characteristics, evaluation metrics, targeted attack types, and signal-to-noise ratio (SNR) operating conditions. The approaches demonstrate diverse methodological strategies, ranging from clustering-based techniques (One-Class SVM with K-means) to rule-based hybrid systems and deep learning autoencoders. Notable observations include the One-Class SVM approach targeting eavesdropping attacks across a comprehensive SNR range of 0-20 dB, using standard classification metrics (accuracy, sensitivity, and specificity). In contrast, the autoencoder method employs reconstruction loss as a novel evaluation criterion for detecting DoS and GPS spoofing attacks. The limited number of studies and frequent absence of SNR specifications highlight the nascent state of unsupervised UAV intrusion detection research, indicating substantial opportunities for methodological advancement and standardized evaluation protocols in this emerging field.}

\subsection{Reinforcement Learning Approaches}

\begin{figure}[t]
    \centering  
    \includegraphics[width=0.5\textwidth]{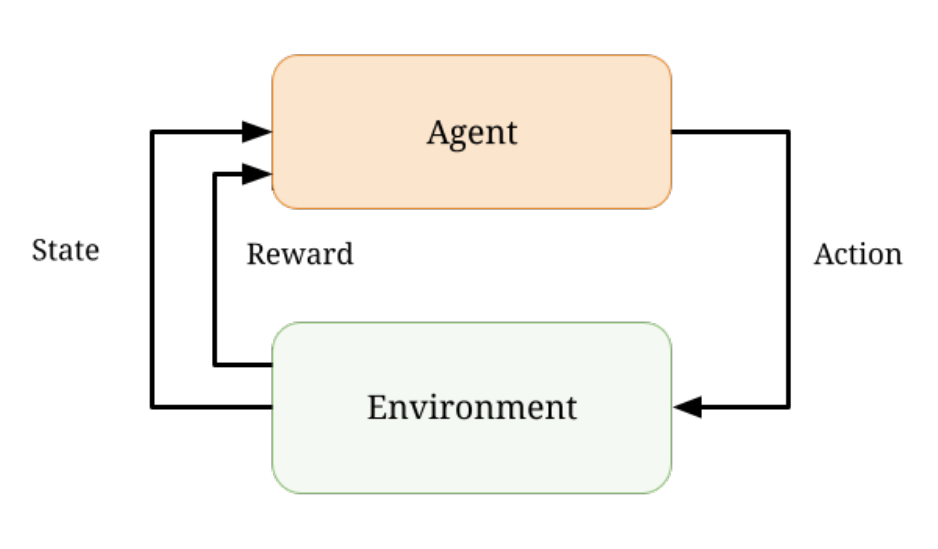}
    \caption{The reinforcement learning framework~\cite{make4010009}}
    \label{fig:rl}
\end{figure}

Reinforcement learning, a subfield of machine learning, involves teaching an agent to optimize its actions in a given environment to maximize a reward. It has recently been used to develop IDS for smart cities. As shown in Figure ~\ref{fig:rl} there are four main components of reinforcement learning: \\
$\bullet$ Agent: This is the entity or system that is learning and taking actions in the environment.\\
$\bullet$ Environment: This is the world in which the agent is interacting and taking action. \\
$\bullet$ Rewards: These are the positive or negative feedback signals that the agent receives as a result of its actions.\\
$\bullet$ Policy: This is the set of rules or strategies that the agent follows to maximize its reward. The policy determines the convenient action to take given a state.

\textcolor{black}{Beyond the fundamental agent-environment framework, deploying RL for UAV intrusion detection requires sophisticated environment modeling, domain transfer techniques, and rigorous safety protocols. Effective training environments must accurately represent communication channels, C2 links, and airspace dynamics while incorporating adaptive adversaries, diverse traffic patterns, and environmental factors like atmospheric interference or RF clutter that challenge policy robustness. The reward structure demands careful tuning to balance detection accuracy against latency constraints while minimizing false positives and power consumption, all critical considerations for resource-limited platforms. Addressing the sim-to-real gap requires comprehensive domain randomization across link characteristics, timing distributions, and sensor uncertainties, complemented by targeted on-platform adaptation. Operational safety relies on action masking to prevent hazardous commands, conservative fallback policies for degraded scenarios, and mandatory human oversight for critical responses. Validation protocols incorporate reward function audits, counterfactual evaluation, and adversarial testing to identify potential gaming behaviors and ensure alignment with mission requirements.}

Reinforcement learning is also widely used in IDS for UAVs, and that is not only for detection but also for prevention. For instance, Omar et al.~\cite{9463947} proposed a lightweight DL-based IDS for UAV networks as part of an intrusion detection and prevention system called IDPS. The system is based on deep reinforcement learning (DRL), particularly, Deep Q-learning (DQN). The proposed approach enables UAVs to survey and monitor their environment in real-time, identifying abnormalities and responding accordingly to protect the network. An offline learning approach is employed to ensure the system's autonomy and adaptability in production. 

In~\cite{cmc.2022.020066}, the authors proposed an Optimal Deep Reinforcement Learning (ODRL) system for IDS. According to the authors, optimality is achieved using an improved reinforcement learning-based Deep Belief Network (DBN) for intrusion detection, and its parameters are tuned using the Black Widow Optimization (BWO) algorithm. The primary objective of this study was to mitigate the false alarms generated by the trending IDS at the time, and it successfully achieved an F-1 score of 0.985. All the simulations done in this study were conducted using the NSL-KDD dataset~\cite{DIRO2018761}. 

In~\cite{CAMINERO201996}, the authors proposed a novel RL-based IDS called Adversarial Environment using Reinforcement Learning (AE-RL). The adversarial behavior originates from the use of two agents: an environment selector and a classifier, where the reward policies of the two agents are mutually antagonistic. The framework utilizes the DDQN algorithm, which combines elements of reinforcement learning and supervised learning. The simulations done in this paper were based on two datasets: NSL-KDD~\cite{5356528} and AWID~\cite{7041170}.  

In~\cite{9520324}, Tao et al. aimed to overcome the challenges that the openness and multi-connectivity of UAVs raise using Deep Reinforcement Learning (DRL) based IDS. The proposed approach aims to detect malicious attacks in aerial computing networks effectively by modeling the intrusion detection process as a Markov Decision Process (MDP). The designed IDS can be run and deployed on the UAV to observe the network environment and make decisions at each time step. 

\begin{table*}[ht]
\centering
\caption{Performance Comparison of Reinforcement Learning Approaches for UAV Intrusion Detection}\label{reinforcement_comparison}
{\scriptsize
\setlength{\tabcolsep}{2pt}
\begin{tabular}{p{1.5cm}p{2.8cm}p{2.2cm}p{1.5cm}p{2cm}p{4cm}}
\hline
\textbf{\begin{tabular}[c]{@{}c@{}}\tiny Ref.\end{tabular}} & 
\textbf{\begin{tabular}[c]{@{}c@{}}\tiny Method/Algorithm\end{tabular}} & 
\textbf{\begin{tabular}[c]{@{}c@{}}\tiny Dataset\end{tabular}} & 
\textbf{\begin{tabular}[c]{@{}c@{}}\tiny F1-Score\end{tabular}} & 
\textbf{\begin{tabular}[c]{@{}c@{}}\tiny Learning\\ \tiny Approach\end{tabular}} & 
\textbf{\begin{tabular}[c]{@{}c@{}}\tiny Key Features\end{tabular}} \\
\hline

{\tiny \cite{9463947}} & {\tiny DQN with periodic offline learning} & {\tiny CICIDS2017} & {\tiny 0.96} & {\tiny Deep Q-learning with periodic offline learning} & {\tiny Customized reward function, feature selection, energy-efficient} \\

{\tiny \cite{cmc.2022.020066}} & {\tiny DRL-BWO (Deep Reinforcement Learning with Black Widow Optimization)} & {\tiny NSL-KDD dataset} & {\tiny 0.988} & {\tiny Deep Reinforcement Learning} & {\tiny BWO hyperparameter optimization, DBN with softmax layer} \\

{\tiny \cite{CAMINERO201996}} & {\tiny AE-RL (Adversarial Environment Reinforcement Learning)} & {\tiny NSL-KDD, AWID} & {\tiny $>$ 0.79} & {\tiny Adversarial Deep Reinforcement Learning} & {\tiny Dual-agent adversarial training, intelligent environment sampling, unbalanced dataset handling} \\

{\tiny \cite{9520324}} & {\tiny DDPG (Deep Deterministic Policy Gradient)} & {\tiny Simulation data} & {\tiny Not specified} & {\tiny Deep Reinforcement Learning} & {\tiny Actor-critic networks, replay memory, UAV aerial computing networks} \\
\hline
\end{tabular}
}
\end{table*}

\textcolor{black}{Table~\ref{reinforcement_comparison} presents four reinforcement learning approaches for UAV intrusion detection, evaluated across algorithm type, dataset, F1-score, learning paradigm, and key features. The methods range from Deep Q-Networks with offline learning to adversarial dual-agent frameworks, achieving F1-scores between 0.79-0.988 on standard cybersecurity datasets (CICIDS2017, NSL-KDD, AWID). The Deep Reinforcement Learning with Black Widow Optimization (DRL-BWO) achieves a 0.988 F1-score through hyperparameter optimization, while Adversarial Environment Reinforcement Learning (AE-RL) demonstrates robustness across multiple datasets via dual-agent training. Key innovations include energy-efficient reward functions, intelligent environment sampling, and actor-critic architectures. However, the reliance on generic cybersecurity datasets rather than UAV-specific data indicates a gap in domain-tailored evaluation frameworks.}


\subsection{Data Fusion for UAV IDS}
\textcolor{black}{Fusing diverse sensors and systems can help mitigate environmental challenges such as weather and topography, as the environment for detecting unwanted UAVs can vary from plains or fields with minimal obstacles to complex environments with many obstacles, including buildings, trees, and crowded areas. Weather conditions can affect the reliability of IDS for UAVs, necessitating the use of multiple sensors to accurately detect unwanted UAVs. Relying on a single or a few sensors for intrusion detection can be overwhelmed by environmental noise, limiting their capabilities. Combined vision and acoustic detection systems should be utilized to determine whether an intrusion has occurred without relying solely on a single detection method, enabling IDS systems to use different wavelengths of light, including a full spectrum of colors, as well as radar and LiDAR capabilities. In contrast, acoustic detection provides an extra layer of assurance for the correct identification of unwanted drones in the surveillance area. Typical vision failure cases in urban skies include small targets, occlusions, and bird look-alikes, where a pragmatic approach triggers IR or acoustic corroboration when RGB confidence drops and uses RF presence cues to gate false alarms during cluttered scenes.}

\textcolor{black}{Recently, more research has focused on data fusion within IDS to diversify intrusion-detection sources by combining data from multiple sensors, thereby enhancing the accuracy and completeness of telemetry, camera, and acoustic-sensor data, as well as external sources such as weather data or other UAVs. By combining this data, hybrid IDS can construct a more comprehensive situational picture and better identify multi-stage or coordinated attacks. Data fusion in UAV IDS can operate across three architectural levels: early fusion combining raw sensor data, preserving maximum information but requiring precise temporal alignment, feature-level fusion operating on extracted features, offering computational efficiency while reducing dimensionality, and decision-level fusion combining outputs from independent classifiers, providing robustness against single-point failures.}

\textcolor{black}{While data fusion is well-established in computer vision and sensor networks, its systematic application to UAV IDS remains unexplored, with very few works addressing this direction. Guo et al.\cite{guo2020cyber} analyzed cyber-physical attack threats against UAVs from a CPS perspective, highlighting the need for integrated detection approaches, while Hassler et al.\cite{hassler2023cyber} demonstrated a practical cyber-physical data fusion approach by combining 37 cyber features with 16 physical features via temporal synchronization and min-max normalization, achieving up to 96.13\% F1-score and consistently outperforming single-modality approaches across multiple machine learning models. The authors in~\cite{10368002} proposed an end-to-end cyber-physical IDS that analyzes cyber telemetry alongside physical-state information through a hybrid architecture that performs lightweight, time-synchronized processing onboard to generate low-latency alerts, while fusing modality-specific detector outputs via meta-classification to capture coordinated attack patterns such as GNSS spoofing combined with anomalous control behavior.}

\textcolor{black}{The study in~\cite{10368002} provided implementations of the baseline detectors and the multimodal fusion pipeline, along with preprocessing scripts, example model checkpoints, and experiment configurations, enabling the reproduction of representative results and serving as a valuable foundation for benchmarking multimodal UAV IDS approaches. In a complementary study, Mane et al.~\cite{Mane2025} investigated a system-level perspective on cyber-physical IDS design for UAVs, synthesizing common fusion strategies and mapping them to different deployment tiers, including onboard, edge, and ground-station processing, while distinguishing between feature-level, decision-level, and hybrid fusion paradigms and proposing escalation policies for graded defensive responses. These multi-sensor fusion approaches represent a critical advancement in UAV IDS design, enabling robust detection capabilities that maintain effectiveness across diverse operational environments and attack scenarios while addressing the inherent limitations and environmental vulnerabilities of single-modality detection systems.}

\textcolor{black}{\subsection{Robustness to Adversarial Manipulation in UAV IDS}\label{robustness_adversarial_RL}}
\textcolor{black}{UAV intrusion detection systems face multifaceted adversarial threats targeting both sensor perception and communication pathways. Visual sensors are vulnerable to small adversarial perturbations, physical patch attacks, and decoy objects that can trigger misclassifications or cause tracking failures, particularly when detecting small aerial targets against complex backgrounds. Network and RF-based detection faces crafted traffic injection, timing manipulation, and feature obfuscation designed to evade anomaly detection algorithms, while training data poisoning during system updates can compromise detection thresholds and degrade overall performance. Evaluation methodologies should emphasize operationally realistic threat scenarios rather than theoretical extremes. Visual system testing must encompass small target detection under occlusion, dynamic backgrounds, and simple physical overlays or reflective patterns that simulate natural airborne clutter and avian interference across varying illumination conditions. Network-based assessments should incorporate timing variations, controlled payload modifications, replay attacks with minor alterations, and protocol-compliant evasion techniques that test model robustness without assuming adversarial access to system internals.}

\textcolor{black}{Defensive strategies prove most effective when implemented in layered, computationally efficient architectures. Visual processing benefits from input preprocessing with adaptive filtering, temporal consistency checks, and multi-spectral sensor fusion to mitigate single-frame attacks. Network feature analysis employs invariant feature extraction, protocol validation, and traffic rate limiting to counter straightforward evasion attempts. Cross-modal approaches utilize confidence-based rejection mechanisms, anomaly detection for out-of-distribution inputs, and conservative decision thresholds at the platform level to prevent erroneous autonomous actions. Data integrity measures, including cryptographic signatures, chain-of-custody tracking, and validation datasets, protect against supply chain compromises during fleet-wide updates. Sensor fusion frameworks and human operator oversight for critical decisions provide operational robustness while maintaining acceptable computational overhead.}

\section{Investigated Datasets}
\label{data}

\begin{figure}[t]
    \centering  
    \includegraphics[width=0.5\textwidth]{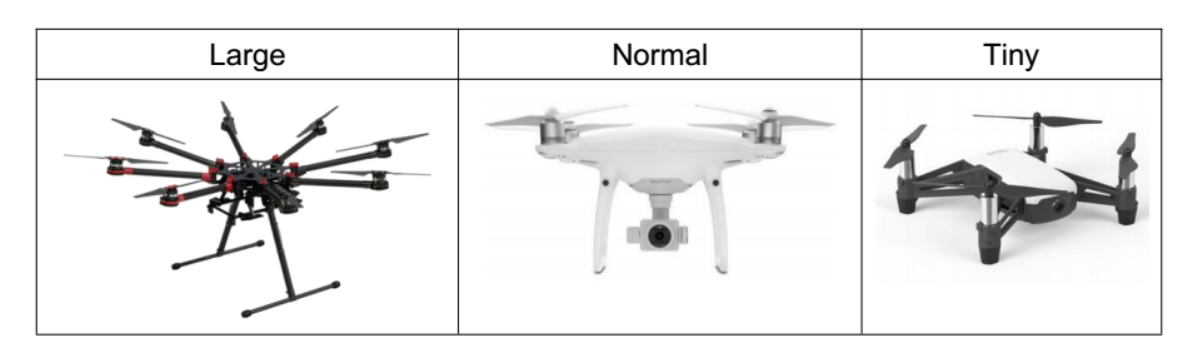}
    \caption{UAVs scales spanned by the Anti-UAV dataset~\cite{9615243}}
    \label{fig:anti_uav_sample}
\end{figure}

To effectively detect and prevent intrusions, an IDS relies on various datasets that provide the necessary information and context for the system to analyze and make informed decisions. \textcolor{black}{To make the connection between datasets and study objectives explicit, we provide a compact dataset-to-task mapping and reference, indicating which studies in Section~\ref{ai_ids_uav} used which corpus or subset for cyber intrusion versus physical drone detection or tracking.} \textcolor{black}{We examine multiple datasets and summarize those most relevant to UAV IDS tasks as an overview, with a compact cross‑reference provided in Table~\ref{datasetable}. To ease selection and interpretation, we present the dataset summary by modality and task}. Understanding the datasets employed in IDS research enables better comprehension of system capabilities and limitations while identifying opportunities for advancement. \textcolor{black}{A focused benchmark with multi-sensor urban scenarios, standard splits, and unified metrics would align future studies; we outline this in Section~\ref{robustness}}. \textcolor{black}{While traditional IT network datasets serve useful purposes for initial algorithmic validation, UAV-specific IDS requires data sources that capture command-and-control protocols, telemetry patterns, and air-to-ground communication characteristics. In our summaries, we emphasize datasets that approximate UAV operational environments where available and explicitly note when studies depend on general-purpose network corpora. When UAV-specific data remains unavailable, a practical approach involves model adaptation through focused UAV calibration phases, utilizing limited proprietary traces or publicly available counter-UAV datasets for feature adaptation and detection threshold optimization. This approach aligns with our discussion in Section~\ref{datascarcity} regarding the scarcity of UAV IDS datasets and the necessity for careful domain adaptation.}

\begin{table*}[!htbp]
\centering
\caption{Datasets organized by modality and task, with key attributes for UAV IDS studies}
\label{datasetable}
\footnotesize
\begin{tabular}{p{2.5cm}p{0.8cm}p{1.2cm}p{2.2cm}p{2.5cm}p{3.5cm}}
\toprule
\textbf{Dataset} & 
\textbf{Lic.$^{\dagger}$} & 
\textbf{Splits} & 
\textbf{Sensor Geometry} & 
\textbf{Environmental Conditions} & 
\textbf{Description} \\ 
\midrule

\multicolumn{6}{>{\columncolor{gray!5}}c}{\textbf{RF/Network -- Cyber Intrusion Detection}} \\
\midrule

KDD Cup 1999~\cite{hettich1999kdd} & 
PA & 
Train/test & 
Enterprise LAN & 
Generic IT traffic & 
Legacy flow records for IDS screening \\

NSL-KDD~\cite{5356528} & 
PA & 
Subsets & 
Enterprise LAN & 
Generic IT traffic & 
Curated KDD; balanced classes \\

CIC-IDS2017~\cite{sharafaldin2018detailed} & 
PA & 
Day-based & 
Lab network & 
Mixed benign/attack & 
Labeled flows for activities \\

CIC-IDS2018~\cite{Sharafaldin2018TowardGA} & 
PA & 
Scenario days & 
Lab network & 
Mixed benign/attack & 
Expanded attacks for IDS \\

UNSW-NB15~\cite{moustafa2015unsw} & 
PA & 
Subsets & 
Lab network & 
Modern protocols & 
Modern flows with attacks \\

ISOT Botnet~\cite{5971980} & 
PA & 
Research & 
Backbone traces & 
Botnet traffic & 
Real botnet traces for C\&C \\

DReLAB~\cite{venturi2021drelab} & 
PA & 
Research & 
Simulated & 
Adversarial & 
Adversarial botnet benchmark \\

\midrule
\multicolumn{6}{>{\columncolor{gray!5}}c}{\textbf{RGB/IR -- Drone Detection \& Tracking}} \\
\midrule

Anti-UAV~\cite{9615243} & 
RL & 
Train/val/test & 
Static ground camera & 
Day/night; weather & 
RGB/IR sequences with annotations \\

DUT Anti-UAV~\cite{zhao2022vision} & 
RL & 
Train/val/test & 
Ground cameras & 
Urban/rural; clutter & 
Sequences for outdoor scenes \\

UAV Dataset~\cite{makrigiorgis2022uav} & 
PA & 
Research & 
Mixed views & 
Varied illumination & 
Annotated images for detection \\

MAV-VID~\cite{isaac2021unmanned} & 
RL & 
Research & 
UAV/ground views & 
Outdoor backgrounds & 
Videos for small-drone detection \\

Drone vs Bird~\cite{9663844} & 
RL & 
Challenge & 
Sky-facing setup & 
Daylight; sky & 
Images for drone/bird distinction \\

\midrule
\multicolumn{6}{>{\columncolor{gray!5}}c}{\textbf{Audio -- Drone Detection}} \\
\midrule

Drone Audio~\cite{DBLP:journals/corr/abs-2007-07396} & 
PA & 
Research & 
Ground microphones & 
Indoor/outdoor; noise & 
Acoustic signatures for detection \\

\midrule
\multicolumn{6}{>{\columncolor{gray!5}}c}{\textbf{Multimodal -- Drone Detection}} \\
\midrule

Multimodal Drone~\cite{DBLP:journals/corr/abs-2007-07396} & 
PA & 
Research & 
Co-located sensors & 
Day/night; weather & 
Synchronized RGB/thermal/audio \\

Drone Monitoring~\cite{8282120} & 
RL & 
Research & 
Ground cameras & 
Synthetic/real; varied & 
Real/synthetic for multimodal \\

\bottomrule
\end{tabular}
\vspace{0.2cm}
\raggedright
$^{\dagger}$ PA=Public Academic, RL=Research License
\end{table*}

\subsection{Network Traffic Datasets}
The KDD Cup 1999~\cite{hettich1999kdd} dataset is a collection of data used in the annual KDD Cup competition, a data mining and machine learning competition organized by the Knowledge Discovery and Data Mining (KDD) conference. The dataset consists of data from various sources, including the Defense Advanced Research Projects Agency (DARPA), the Lincoln Laboratory at MIT, and the Defense Research and Engineering Network (DREN). The dataset is designed to simulate a real-world scenario and contains data on network traffic, including information on network connections, packet sizes, and protocol types. It also includes data on cyber-attacks, such as denial-of-service (DoS) attacks and other types of network intrusions. The dataset is divided into two sets: a training set and a test set. The training set is used to build and train models, while the test set is used to evaluate the performance of these models. The KDD Cup 1999 dataset is widely used in research and has been used in numerous papers and publications.

The NSL-KDD~\cite{5356528} dataset is a refined version of the KDD Cup 1999 dataset, which was previously used in the Third International Knowledge Discovery and Data Mining Tools Competition. The original KDD Cup 1999 dataset was created to evaluate machine learning algorithms for the task of network intrusion detection. The NSL-KDD dataset was designed to address some of the known issues with the KDD Cup 1999 dataset, such as the presence of redundant records and the need for a more balanced class distribution. It consists of approximately 25,000 labeled network connection records, each representing a network connection with features such as protocol type, service, flag, and various other connection-specific details. The goal of using the NSL-KDD dataset is to train a machine learning model to classify each connection record as either normal or an attack, based on the features of the record. The NSL-KDD dataset comprises a total of 23 types of attacks, including denial-of-service, unauthorized access, and others.

The Aegean Wi-Fi Intrusion Dataset (AWID~\cite{7041170}) is a collection of wireless network traffic data collected over several weeks in a real-world wireless network environment. It was designed for research on detecting wireless network intrusions and anomalies, and has been utilized in numerous studies and papers on the subject. The dataset comprises several million records of wireless network traffic, each representing a wireless packet captured within the network. The records include various features, such as the source and destination MAC addresses, the type of wireless frame, and other packet characteristics. The dataset also contains labels indicating whether each packet was normal or an attack. In the case of an attack, the type of attack is also specified. The attacks in the AWID include a variety of different types, such as unauthorized access, denial-of-service attacks, and spoofing. The dataset is valuable for research on detecting and mitigating these types of attacks in wireless networks.

The UNSW-NB15~\cite{moustafa2015unsw} dataset is a large dataset of network traffic data that was created by the University of New South Wales (UNSW) in 2015. It is designed for research on network intrusion detection and has been utilized in numerous studies and papers on the subject. The dataset comprises approximately 2.5 million records of network traffic, each representing a distinct network connection. The dataset contains connection features and labels indicating whether each connection was normal or an attack. In the case of an attack, the type of attack is also specified. The UNSW-NB15 dataset comprises a total of nine types of attacks, including denial-of-service, unauthorized access, and others. It is a valuable resource for researchers working on developing and improving techniques for detecting and mitigating network attacks.

The ISOT Botnet~\cite{5971980} dataset is a collection of network traffic data that was collected from a real-world botnet (a network of compromised computers controlled by a hacker). The dataset hosted by the University of Victoria is intended for research on detecting and mitigating botnets, and it has been utilized in several studies and papers on the subject. The dataset consists of several million records of network traffic, each representing a single network connection. The records include various features such as the source and destination IP addresses, the protocol used, and different statistical measures of the connection. The dataset also contains labels indicating whether each connection was normal or part of the botnet. The ISOT Botnet dataset is a valuable resource for detecting and mitigating botnets.

The CIC-IDS2017~\cite{sharafaldin2018detailed} dataset is a collection of network traffic data labeled as either normal or malicious. It was created by the Canadian Institute for Cybersecurity (CIC) at the University of New Brunswick. It was used in the "Intrusion Detection System (IDS) Based on Machine Learning Techniques" competition on Kaggle in 2017. The dataset comprises a total of 78,964 network flow records, collected using the CICFlowMeter tool. Each record contains 115 features that represent various characteristics of the network flow, such as the protocol type, packet counts, and packet lengths. The dataset consists of two classes: benign and malicious. The benign class contains normal, non-threatening network traffic, while the malicious class contains network traffic that is potentially harmful or malicious in nature. The CIC-IDS2017 dataset is often used to evaluate the performance of machine learning algorithms for intrusion detection tasks.

The CIC-IDS2018~\cite{Sharafaldin2018TowardGA} dataset is the result of a collaborative project between the Communications Security Establishment (CSE) and CIC, aiming to design a framework that generates diverse benchmark datasets for IDS. The dataset includes information about seven different types of cyber attacks, including Brute-force, Heartbleed, and Web attacks. During the attack, the perpetrator utilized 50 machines within their infrastructure, whereas the targeted organization had five departments and a total of 420 machines and 30 servers. The data gathered for this incident includes network traffic and system logs from every machine, as well as 80 features extracted from the traffic using CICFlowMeter-V3 software.

The Deep Reinforcement Learning Adversarial Botnet (DReLAB~\cite{venturi2021drelab}) dataset is a collection of data related to botnets, which are networks of compromised computers that can be controlled remotely by an attacker. The dataset includes information on the actions taken by botnets, as well as the rewards and punishments received by the bots as a result of their actions. The data in this dataset is generated using deep reinforcement learning algorithms, which allow the bots to learn and adapt their behavior based on the rewards and punishments they receive. This makes the bots more intelligent and capable of evading detection, making them a significant cybersecurity threat. The dataset includes information on the various actions taken by the botnets, such as command and control communications, data exfiltration, and network scanning. It also contains information about the rewards and punishments received by the bots, such as the success or failure of their actions, and the impact on the overall botnet network. This dataset is useful for researchers studying botnets and the methods for detecting and mitigating them. It can also be used by cybersecurity professionals to develop and test countermeasures against botnets.

\subsection{Image/Video Datasets}
Utilizing computer vision and object recognition techniques can aid in identifying malicious and unwanted drones. \textcolor{black}{For each vision or multimodal dataset, we summarize access terms, typical splits, indicative resolution or range, basic sensor geometry, and common conditions to facilitate fair benchmarking across settings.} Makrigiorgis et al.~ \cite{makrigiorgis2022uav} curated a dataset containing 1,535 images in addition to 1,620 total drone annotation images. Besides that, the Anti-UAV~\cite{9615243} dataset is a part of the anti-UAV project of the Institute of North Electronic Equipment, Beijing, China. The dataset contains full HD RGB and IR video sequences captured by a static ground camera of UAVs of different scales. This large-scale multi-modal benchmark dataset was used in the CVPR 2020 Workshop on the 1st Anti-UAV Challenge. The data are densely annotated with each frame's drone bounding boxes, speed change, and other attributes. The sizes of the UAVs spanned by this dataset are shown in Figure~\ref{fig:anti_uav_sample}.

\begin{figure*}[t]
    \centering  
    \includegraphics[width=0.8\textwidth]{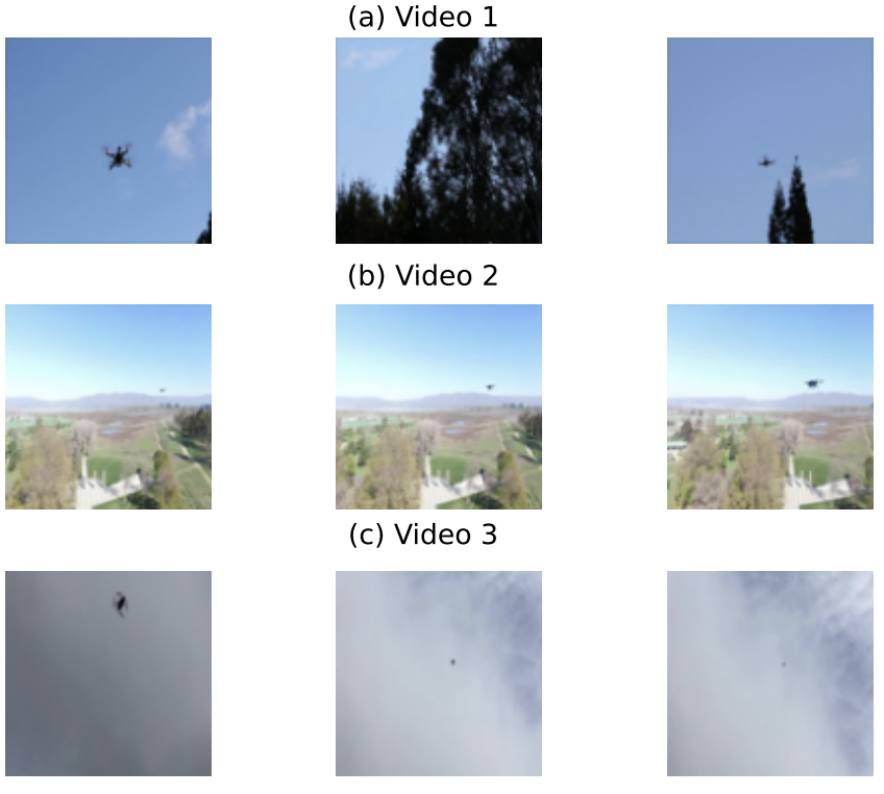}
    \caption{Samples from the DUT Anti-UAV~\cite{zhao2022vision}}
    \label{fig:tracking_uav}
\end{figure*}

The DUT Anti-UAV~\cite{zhao2022vision} is a dataset comprising data related to the detection and tracking of UAVs, encompassing a variety of data types, including video, audio, and sensor data from multiple sources, such as cameras, microphones, and radar. The data was collected from a variety of different environments, including urban, rural, and indoor settings, to provide a diverse range of challenges for participants in the challenge. The dataset comprises both positive and negative examples of UAVs, accompanied by annotated bounding boxes and class labels that indicate the presence or absence of a UAV in each frame. It also includes metadata such as the location, orientation, and speed of the UAVs, as well as the location and orientation of the sensors that collected the data. Samples from this dataset are shown in Figure~\ref {fig:tracking_uav}.

\section{Open Research Directions}
\label{ord}
With significant progress in AI-driven intrusion detection for UAVs in the context of smart cities, as discussed in the relevant literature, there remain challenges and opportunities that need to be addressed in future research. This section highlights the research directions that could further advance the body of knowledge in UAVs and intrusion detection systems.

\subsection{Scalability}
One promising research direction in the field of AI is to enhance the scalability of models for deployment in drones. The scalability of AI models is a pivotal factor in the effective utilization of drones and UAV networks. It holds significant potential to substantially improve the capabilities of these systems, thereby unlocking new and diverse applications. As drones are increasingly used for a wide range of applications, such as package delivery, surveillance, and search and rescue, the amount of data they collect can be massive. Therefore, the AI models used to process and analyze this data must be capable of handling large amounts of information in real-time. This requires developing new techniques for training and deploying AI models that can efficiently handle the high volume and velocity of data generated by drones. Hence, it is essential to develop models that can operate with limited resources, as drones often operate in remote and dynamic environments where charging stations are sometimes inaccessible. Research in this direction can focus on designing algorithms that handle the distributed and parallel processing of data, operate on limited computational resources, and consume minimal energy when deployed on a UAV. \textcolor{black}{While practitioners often seek concrete budgets for power, latency, and memory, we treat these as open research targets that vary by airframe, mission, and regulatory. On-board tiers should aim for a low single-digit watt draw under sustained inference, sub-frame-scale decision latencies on typical video rates, and compact memory footprints compatible with embedded compute. When these bounds are exceeded or confidence is low, the pipeline should favor edge or GCS offload using link-aware triggers such as available bandwidth, round‑trip delay, and current energy reserve, with graceful degradation if connectivity drops. We view the precise thresholds as deployment- and platform-specific, and encourage reporting resource envelopes alongside accuracy to advance reproducible scalability studies.}

\subsection{Robustness} \label{robustness}
Robustness is another crucial aspect of AI models for deployment in drones and UAV networks, enabling these systems to operate effectively in challenging environments. As drones are often deployed in dynamic and uncertain environments, it is crucial that the models can perform reliably under a wide range of conditions. \textcolor{black}{To support calibrated alerting and OOD detection in future studies, we outline lightweight evaluation practices suited to this review context: keep a small calibration split distinct from training and testing; report reliability diagrams or a summary of calibration error on the test split; include coverage versus accuracy when a reject option is used; and probe OOD with simple, realistic shifts such as background or lighting changes for vision, benign bird-like distractors, and timing jitter or protocol-conformant perturbations for network traces. These practices indicate whether alerts remain calibrated as conditions drift while keeping evaluation practical.} This requires developing models that can handle a variety of inputs and adapt to changing conditions. Additionally, drones frequently operate remotely and in hazardous locations. Hence, it is essential to develop models that can handle failures and/or partial data. In this direction, researchers can focus on integrating existing methods for robustness training, such as adversarial training or robust optimization, or design and develop new approaches to address this concern. Models designed for UAVs should also consider the uncertain environments in which a UAV might operate, which require the ability to adapt to changing conditions and handle uncertainty. Additionally, it could involve developing new techniques for testing and validating models in both simulated and real-world environments to ensure they perform robustly across diverse operating scenarios. \textcolor{black}{In addition to environmental uncertainty, robustness encompasses resistance to adversarial manipulation and data poisoning, which can affect perception and links. We outline practical tests and lightweight defenses in Section~\ref{robustness_adversarial_RL} to guide evaluation and deployment. We recommend a community benchmark tailored to UAV IDS. Scenarios should reflect smart-city sensors, including RGB and IR, with optional acoustic and RF capabilities. Urban flight profiles should include hovering, patrolling, and transiting over streets, parks, and building corridors. Conditions should cover day and night, weather, occlusions, and small-object ranges. A unified metric set should pair cyber and physical detection quality with calibrated alerting summaries, reject coverage, and placement-aware latency and energy descriptors. Lightweight splits for training, calibration, validation, and testing, along with a shared event schema, can enable fair comparison without requiring model fixes.}

\subsection{Explainability}
Exploring the explainability of AI models stands as a potential research direction for IDS in UAV networks. It is imperative that AI models, particularly those employed in security measures like IDS, possess a high degree of explainability to ensure their reliability and effectiveness.
The complexity and opacity of AI models make it challenging to understand why the system makes certain decisions, which can lead to mistrust and a lack of confidence in the model's ability to accurately identify and respond to security threats. Therefore, explaining AI models in security, and specifically in IDS, is necessary. Such an explanation is essential, as it enables users to understand how the model processes and analyzes data, which can help build trust and confidence in the system. It also allows users to identify any biases or inaccuracies in the model's decision-making process, which can help improve the model's overall performance. Additionally, explanations also help ensure that the decisions made by the model align with the user's goals and values, thereby preventing unwanted or harmful outcomes. In IDS, explainability is even more crucial as it helps detect and prevent malicious activities, and it also allows the UAV to understand the reason behind the alert and respond accordingly.

\subsection{Data Scarcity}\label{datascarcity}
The scarcity of IDS datasets in UAVs poses a significant challenge to the development of effective and reliable security systems. Without adequate datasets, it is difficult to train and evaluate IDS models, hindering the identification and mitigation of potential threats and attack patterns. The scarcity of these datasets stems from the fact that UAVs come in many forms and sizes and operate across a wide range of environments. This makes it difficult to standardize data collection and labeling. Additionally, UAVs often collect sensitive information, including images and videos, which raises concerns about privacy and security. This makes it challenging to release datasets publicly, thereby limiting their availability. This highlights the importance of developing automated tools for generating and labeling datasets that can be used to build AI-based IDS for UAVs. These tools are essential for collecting, processing, and organizing large amounts of data from UAVs. The importance of such automated tools lies in the fact that they can significantly reduce the time and effort required to collect and label data manually. They can also improve the accuracy and consistency of the data, which is essential for training and evaluating IDS models. Additionally, these tools can also help to overcome the limitations of manual data collection and labeling, such as human error and bias. Furthermore, automated tools for generating and labeling datasets can also help to improve the scalability and adaptability of IDS models. As UAV technology continues to evolve, new types of threats and attack patterns are likely to emerge. With automated tools, it's easier to update and adapt the datasets accordingly, which is crucial for maintaining the effectiveness of IDS models over time.

\subsection{Automation}
A promising area for future research is advancing IDS automation by developing Automated Intrusion Detection Systems (AIDS) that integrate sophisticated drone behavior management mechanisms. AIDS is designed to detect potential attacks on the drone, such as unauthorized access attempts, and respond to them in a predetermined way, such as avoiding the attacker, neutralizing the threat, or maintaining a safe distance from the attacker. This type of system can be helpful in various contexts, such as military or surveillance operations, where drones are used to gather information or conduct surveillance in potentially hostile environments. However, these systems are still in the early stages of development, and more research and testing are needed to fully understand their capabilities and limitations.


\subsection{Hybrid Detection}\label{hybridSetection}
Another challenge to UAV system safety is detecting hybrid attacks, those that combine multiple attack types, such as physical and cyber attacks, to target UAV systems. \textcolor{black}{These attacks can be particularly challenging to detect, as they often involve correlated but subtle indicators across different data sources. Future research should explore the use of \textit{hybrid IDS}, which combines multiple detection approaches (e.g., signature-based, anomaly-based, and specification-based) to improve detection coverage and reduce false positives. On the other hand, in-situ decision-making, which refers to making decisions based on the current situation rather than relying on pre-programmed responses, can leverage AI algorithms to analyze data in real time and inform decisions. For example, an AI algorithm could analyze sensor data to detect unusual activity and then make a decision to avoid that area or take other measures to protect the UAV. Hybrid detection for UAVs will likely require a multi-layered IDS that integrates diverse data sources, combines complementary detection methods, and supports real-time response capabilities. A detailed benchmarking of such systems against realistic hybrid attack scenarios is identified as an important step for future work.}

\subsection{Complexity Reduction}
\textcolor{black}{The computational demands of AI-driven intrusion detection present a fundamental barrier to practical UAV deployment, as detection models must operate within severe power, memory, and latency constraints while maintaining sufficient accuracy to identify genuine threats. Model compression techniques offer immediate opportunities to reduce this overhead through knowledge distillation, enabling compact networks to approximate larger teacher models suitable for offline training and transferring detection capability to lightweight architectures appropriate for on-board execution. Recent work by Wisanwanichthan and Thammawichai~\cite{wisanwanichthan2025lightweight} demonstrates that knowledge distillation achieves 92-95\% parameter reduction and 7-11\% faster inference while maintaining or improving detection performance across multiple IDS benchmarks, including UAV-specific datasets targeting GPS spoofing and jamming attacks. Structured pruning removes entire channels or layers based on learned importance scores, yielding models that execute efficiently on embedded hardware, while quantization reduces numerical precision from 32-bit floating-point to 8-bit integers or lower, thereby decreasing memory bandwidth and enabling integer-only arithmetic on resource-constrained processors.}

\textcolor{black}{Adaptive inference mechanisms allow detection systems to modulate computational effort based on input complexity and operational context through early-exit architectures that attach auxiliary classifiers at intermediate network layers, enabling confident predictions to terminate computation before reaching deeper stages. When sensor inputs are unambiguous, early exits conserve energy while maintaining responsiveness, whereas ambiguous inputs proceed through the full network to leverage the maximum representational capacity. Temporal redundancy in video streams offers additional opportunities through keyframe architectures, which apply full detection at reduced temporal resolution while utilizing lightweight propagation for intermediate frames. These optimization strategies collectively enable efficient UAV IDS deployment by balancing computational requirements with detection accuracy, as validated in IoT and UAV contexts where distilled student models reduce inference latency by over 7\% while achieving F1 scores exceeding 98\% on UAV intrusion detection tasks~\cite{wisanwanichthan2025lightweight}.}

\subsection{Large Language Models for UAV IDS}
Large Language Models (LLMs) and Generative Artificial Intelligence (GenAI) are emerging as transformative technologies across diverse applications, offering significant potential for UAV intrusion detection systems. Recent research demonstrates LLMs' ability to enable natural language interaction between operators and UAV IDS platforms. Alhammadi et al.\cite{alhammadi2025llm} illustrate how LLMs can facilitate UAV execution of human intentions while identifying key challenges and opportunities for UAV operations in smart city environments. Similarly, Samma et al.\cite{samma2025uav} investigated LLM applications for UAV path planning in GPS-denied environments, proposing a camera-based method that analyzes visual data and object positions to generate navigation feedback. Their approach demonstrated superior performance compared to deep reinforcement learning models, achieving reduced collision rates and optimized travel distances.

However, integrating LLMs into UAV systems introduces critical security vulnerabilities that must be addressed within IDS frameworks. Piggott et al.~\cite{piggott2023net} demonstrated these risks through their Net-GPT tool, which successfully generates realistic network packets for UAV transmission, exposing potential attack vectors in UAV-GCS communication channels. Such capabilities highlight how unauthorized users executing man-in-the-middle attacks could compromise LLM feeds, necessitating robust defensive measures within UAV IDS architectures.

Practical implementation of LLM-assisted UAV IDS requires comprehensive safety protocols and human oversight mechanisms. Natural language interfaces enable operators to interact with IDS systems through intuitive textual commands, while alert systems can present technical information in accessible formats through textual or audible descriptions. Critical system requirements include a clear distinction between AI-generated recommendations and operator directives, ensuring decision traceability and reliability. Emergency override capabilities must be integrated to allow operators to disable LLM-generated automation when necessary, while fail-safe mechanisms should prevent unexpected system behaviors. These considerations emphasize the need for careful design and validation when incorporating LLM capabilities into safety-critical UAV IDS deployments, balancing operational efficiency with security and reliability requirements.

\subsection{Federated and Privacy-Preserving Learning for Fleet IDS}
\textcolor{black}{Federated learning and distributed edge-learning architectures enable UAV fleets to enhance IDS models while maintaining data locality and operational security. In UAV deployments, gradient computations occur onboard individual platforms, with aggregation performed at edge nodes or ground control stations, thereby minimizing the exposure of sensitive telemetry data, reconnaissance imagery, and communication traces. Communication constraints imposed by intermittent connectivity and power limitations necessitate efficient update mechanisms. Practical implementations employ sparse gradient transmission, quantized parameter updates, adaptive client participation based on link conditions and remaining battery capacity, and schedule model synchronization cycles rather than operate continuously.}

\textcolor{black}{Maintaining system integrity and operational security requires robust aggregation protocols and privacy-preserving mechanisms to counter gradient interception, model poisoning attacks, and parameter inference vulnerabilities. Defensive measures include cryptographic client authentication, gradient anomaly detection, signed update verification, and validation against held-out reference datasets at aggregation points before deploying the fleet-wide model. Heterogeneous sensor configurations and diverse mission profiles benefit from personalized model adaptations or clustered learning approaches that customize local decision boundaries while maintaining shared feature representations. Fleet learning integrates naturally within the layered detection architecture, where onboard rapid-response detectors handle time-critical threat assessment. At the same time, federated model updates provide long-term adaptation capabilities across the fleet without compromising sensitive operational data.}

\section{Practical Implementation Challenges}

\subsection{Resource Constraints}
Practical IDS implementation for UAVs must contend with severe constraints in energy, processing power, and communication bandwidth, while addressing the heterogeneity of integrated systems and sensors in smart city environments. Onboard computing platforms offer limited capacity for continuous inference and model updates, with energy consumption scaling nonlinearly with both payload and computation demands. As discussed in earlier sections, lightweight architectures such as pruned CNNs, quantized models, and transformer distillation are being explored to balance accuracy and efficiency, while offloading to edge or fog nodes can further reduce on-device load through adaptive scheduling and secure, low-latency communication channels. The heterogeneity of integrated systems and sensors in smart cities necessitates the establishment of standardization and methods to ensure uniformity, thereby achieving reliable identification of unwanted UAVs across diverse operational environments. Critical system performance criteria must be addressed when deploying IDS in real-world applications: real-time threat response capabilities, reliable detection across diverse operational requirements and conditions, and effective false alarm management in complex urban environments. Periodic training of detection systems to accurately identify new drone designs and shapes ensures continued effectiveness despite evolving threats. At the same time, lightweight frameworks become essential for IDS deployment on energy-constrained platforms, particularly in applications where the IDS itself operates aboard UAVs with limited power resources. Future research should emphasize resource-aware AI design, where detection performance dynamically adjusts to operational energy budgets and mission-critical priorities, enabling sustained protection without compromising flight endurance or operational capabilities.

\subsection{Scalability and Integration Challenges}
\textcolor{black}{As UAV-based IDS moves from isolated prototypes to city-scale deployments, scalability and interoperability become central challenges. Large fleets may involve heterogeneous hardware, diverse communication protocols, and mixed network topologies, complicating consistent monitoring. Effective integration demands standardized data schemas and communication interfaces, allowing IDS modules on different UAVs or infrastructures to exchange alerts and updates seamlessly. Federated or cooperative learning paradigms can support distributed adaptation while preserving local autonomy. Moreover, interoperability between UAV IDS and terrestrial systems, such as traffic management, IoT monitoring, or emergency response networks, is critical for realizing a unified urban defense layer. Achieving this requires both technical standardization and institutional coordination across agencies and vendors.}

\subsection{Robustness and Resilience Considerations}
\textcolor{black}{IDS deployed in urban UAV operations must remain robust against diverse and evolving threats. Adversarial attacks on machine learning models, signal interference, or environmental variability can degrade detection accuracy and reliability. Robustness, therefore, hinges on adaptive calibration, redundancy, and continual learning under realistic constraints. Resilient IDS architectures should combine multiple sensing modalities—RF, vision, telemetry, and network flow—to mitigate single-point failures and enhance detection under uncertainty. Furthermore, resilience planning should account for degraded communication or compromised nodes by introducing redundancy in communication links and cooperative fallback mechanisms. Future efforts should advance methods for adversarial robustness evaluation and self-healing adaptation, ensuring dependable performance in dynamic and contested urban environments.}

\subsection{Regulatory and Privacy Challenges}
UAV-based IDS must navigate complex regulatory frameworks governing both airspace usage and data protection in smart cities while operating under severe constraints in energy, processing power, and communication bandwidth. Onboard computing platforms offer limited capacity for continuous inference and model updates, with energy consumption scaling nonlinearly with both payload and computation demands. Privacy concerns arise from continuous sensing, recording, and communication of potentially identifiable data, particularly when fusing cyber, RF, and vision modalities, necessitating compliance with standards such as the EU General Data Protection Regulation (GDPR) and national civil aviation authorities through adoption of data minimization, purpose limitation, and accountability principles. Compliance spans operational rules, including FAA 14 CFR Part 107 and Remote ID provisions in the United States, as well as the EU's U-space framework under Regulation (EU) 2021/664, applicable from January 2023, which defines operational envelopes, identification, and traffic-coordination constraints relevant to IDS deployment and evaluation. As discussed in earlier sections, lightweight architectures such as pruned CNNs, quantized models, and transformer distillation are being explored to balance accuracy with efficiency, while offloading strategies to edge or fog nodes can reduce onboard load through adaptive scheduling and secure, low-latency communication channels. Legal ambiguity persists regarding responsibility when autonomous IDS actions affect public spaces, underscoring the need for coordination among technical design, policy frameworks, and municipal governance to ensure trust and transparency in IDS-enabled urban air operations. Future research should emphasize resource-aware AI design that dynamically balances detection performance with operational energy budgets, regulatory compliance requirements, and mission-critical priorities, enabling sustained protection while maintaining legal adherence and public trust in smart city deployments.

\subsection{Human Oversight and Ethical Operation}
UAV IDS increasingly incorporates autonomous reasoning and self-adaptive learning capabilities, yet human oversight remains essential for operational trust and accountability in smart city deployments. Overreliance on automated alerts or mitigation actions may lead to misinterpretation or disproportionate responses, necessitating human-in-the-loop supervision, transparent decision interfaces, and graded response mechanisms aligned with policy frameworks. Ethical operation demands that UAV-based surveillance and anomaly detection respect societal norms and privacy expectations, while social and public acceptance prove vital to ensuring successful IDS implementation. Clear audit trails, operator training, and explainable AI mechanisms can ensure responsible use and maintain public confidence in smart city security systems. Although IDS for UAVs involves limited direct human interaction when detecting unwanted UAVs, these systems require adequate public education and awareness programs when implemented in public areas to ensure community acceptance and protect civil rights. Effective deployment, therefore, necessitates a balanced approach that integrates technical capabilities with human judgment, transparent operations, and comprehensive stakeholder engagement to maintain both security effectiveness and public trust in urban environments.







\section{Conclusion}
\label{cln}
\textcolor{black}{In this paper, we presented a comprehensive survey of AI-driven Intrusion Detection Systems (IDS) for Unmanned Aerial Vehicles (UAVs) operating in smart city environments. We systematically reviewed UAV applications and identified the dual nature of intrusion threats, encompassing both cyberattacks targeting communication and control infrastructures, as well as physical intrusions carried out by unauthorized UAVs. We examined existing IDS paradigms, proposed a unified taxonomy that integrates cyber and physical detection perspectives, reviewed AI-based detection techniques across supervised, unsupervised, and reinforcement learning, as well as data fusion techniques, and analyzed available public datasets relevant to UAV intrusion detection. In addition, we discussed deployment architectures spanning onboard, cooperative, edge, and ground-control tiers and outlined open research directions to guide future developments.}

\textcolor{black}{Several important takeaways emerge. First, while a large body of work exists on cyber intrusion detection and counter-UAV sensing separately, only limited efforts attempt to unify these dimensions within a single IDS framework. Second, AI techniques—particularly learning-based anomaly detection and computer vision—have demonstrated strong potential, but their effectiveness is highly dependent on deployment constraints such as computational resources, latency, and energy availability. Third, the lack of standardized datasets and evaluation protocols remains a key obstacle to fair comparison and reproducibility. Finally, no single IDS design fits all applications; instead, effective solutions must be tailored to specific scenarios, sensing modalities, and operational requirements.}

\textcolor{black}{Despite this progress, significant challenges remain for real-world deployment. Practical UAV-based IDS must operate under stringent energy, computation, and communication constraints while maintaining low false-alarm rates and timely responses. The heterogeneity of UAV platforms, sensing configurations, and regulatory frameworks further complicates deployment. Additionally, physical intrusion detection in dense urban environments remains particularly challenging due to clutter, occlusions, small target sizes, and the potential for adversarial evasion. Addressing these challenges requires robust calibration, uncertainty handling, multimodal fusion, and careful system integration across multiple operational tiers.}

\textcolor{black}{Looking ahead, future UAV-based IDS deployments should be designed as integral components of broader smart city ecosystems. As UAVs become increasingly embedded in daily urban operations, IDS solutions must interoperate seamlessly with IoT sensor networks, edge and cloud computing infrastructures, traffic management systems, and public safety platforms. Emerging technologies such as multimodal data fusion, federated and privacy-preserving learning, large language models for decision support, and cooperative multi-UAV intelligence offer promising directions to enhance scalability, robustness, and autonomy. By advancing these capabilities, UAV-based IDS can evolve from isolated security mechanisms into key enablers of resilient, trustworthy, and intelligent smart cities.}

\bibliographystyle{elsarticle-num}
\bibliography{references}

@INPROCEEDINGS{10112531,
  author={Hamadi, Raby and Ghazzai, Hakim and Massoud, Yehia},
  booktitle={2023 IEEE International Conference on Smart Mobility (SM)}, 
  title={Image-based Automated Framework for Detecting and Classifying Unmanned Aerial Vehicles}, 
  year={2023},
  volume={},
  number={},
  pages={149-153},
  doi={10.1109/SM57895.2023.10112531}}

@INPROCEEDINGS{10112557,
  author={Hamadi, Raby and Ghazzai, Hakim and Massoud, Yehia},
  booktitle={2023 IEEE International Conference on Smart Mobility (SM)}, 
  title={Reinforcement Learning Based Intrusion Detection Systems for Drones: A Brief Survey}, 
  year={2023},
  volume={},
  number={},
  pages={104-109},
  doi={10.1109/SM57895.2023.10112557}}

@techreport{EuropenDroneMarket,
  title={{Europe Commercial Drone Market Forecast 2027}},
  author={{Graphical Research}},
  institution={Graphical Research},
  year={2022},
  month={January},
  number={GR1016},
  type={Market Research Report},
  address={Delaware, USA},
  url={https://www.graphicalresearch.com/industry-insights/1016/europe-commercial-drone-market},
  pages={150},
  language={English}
}

@INPROCEEDINGS{8746387,
  author={Lasla, Noureddine and Ghazzai, Hakim and Menouar, Hamid and Massoud, Yehia},
  booktitle={2019 IEEE 89th Vehicular Technology Conference (VTC2019-Spring)},
  title={Exploiting Land Transport to Improve the UAV's Performances for Longer Mission Coverage in Smart Cities},
  year={2019},
  month={April},
  pages={1--7},
  address={Kuala Lumpur, Malaysia},
  publisher={IEEE},
  doi={10.1109/VTCSpring.2019.8746387},
  isbn={978-1-7281-0539-8},
  issn={2577-2465},
  organization={IEEE Vehicular Technology Society}
}

@article{MOHAMED2020119293,
  title={Unmanned Aerial Vehicles Applications in Future Smart Cities},
  author={Mohamed, Nader and Al-Jaroodi, Jameela and Jawhar, Imad and Idries, Ahmed and Mohammed, Farhan},
  journal={Technological Forecasting and Social Change},
  volume={153},
  pages={119293},
  year={2020},
  month={April},
  publisher={Elsevier},
  issn={0040-1625},
  doi={10.1016/j.techfore.2018.05.004},
  language={English},
  note={Article 119293}
}

@INPROCEEDINGS{6842265,
  author={Mohammed, Farhan and Idries, Ahmed and Mohamed, Nader and Al-Jaroodi, Jameela and Jawhar, Imad},
  booktitle={2014 International Conference on Unmanned Aircraft Systems (ICUAS)},
  title={UAVs for smart cities: Opportunities and challenges},
  year={2014},
  month={May},
  pages={267--273},
  address={Orlando, FL, USA},
  publisher={IEEE},
  doi={10.1109/ICUAS.2014.6842265},
  isbn={978-1-4799-2376-2},
  issn={2575-7296},
  organization={Association for Unmanned Vehicle Systems International (AUVSI)},
  note={Conference dates: 27-30 May 2014}
}

@incollection{nguyen2021drone,
  title={Drone application in smart cities: The general overview of security vulnerabilities and countermeasures for data communication},
  author={Nguyen, Huu Phuoc Dai and Nguyen, Dinh Dung},
  booktitle={Development and Future of Internet of Drones (IoD): Insights, Trends and Road Ahead},
  pages={185--210},
  year={2021},
  publisher={Springer},
  address={Cham, Switzerland},
  isbn={978-3-030-63338-9},
  editor={Krishnamurthi, Raghvendra and Nayyar, Anand and Hassanien, Aboul Ella},
  series={Studies in Systems, Decision and Control},
  volume={342},
  chapter={8},
  language={English}
}

@article{alvares2021blockchain,
  title={Blockchain-Based Solutions for UAV-Assisted Connected Vehicle Networks in Smart Cities: A Review, Open Issues, and Future Perspectives},
  author={Álvares, Paulo and Silva, Lion and Magaia, Naercio},
  journal={Telecom},
  volume={2},
  number={1},
  pages={108--140},
  year={2021},
  month={March},
  publisher={MDPI},
  address={Basel, Switzerland},
  issn={2673-4001},
  doi={10.3390/telecom2010008},
  language={English},
  note={Article number 8},
  license={CC BY 4.0}
}

@incollection{payal2021drones,
  title={Drones in Smart Cities},
  author={Payal, Manju and Dixit, Pooja and Dutt, Vishal},
  booktitle={AI and IoT-Based Intelligent Automation in Robotics},
  pages={205--220},
  year={2021},
  publisher={John Wiley \& Sons},
  address={Hoboken, NJ, USA},
  isbn={978-1-119-71139-0},
  editor={Sharma, Ashutosh and Sharma, Abhishek and Choudhary, Ajay and Agarwal, Bhupesh and Chhabra, Jasdeep Kaur},
  chapter={11},
  language={English},
  note={Part of Intelligent Data-Centric Systems series}
}

@article{butilua2022urban,
  title={Urban Traffic Monitoring and Analysis Using Unmanned Aerial Vehicles (UAVs): A Systematic Literature Review},
  author={Butil{\u{a}}, Eugen Valentin and Boboc, Răzvan Gabriel},
  journal={Remote Sensing},
  volume={14},
  number={3},
  pages={620},
  year={2022},
  month={February},
  publisher={MDPI},
  address={Basel, Switzerland},
  issn={2072-4292},
  doi={10.3390/rs14030620},
  language={English},
  note={Article number 620},
  license={CC BY 4.0}
}

@article{huang2021decentralized,
  title={Decentralized autonomous navigation of a UAV network for road traffic monitoring},
  author={Huang, Hailong and Savkin, Andrey V. and Huang, Chao},
  journal={IEEE Transactions on Aerospace and Electronic Systems},
  volume={57},
  number={4},
  pages={2558--2564},
  year={2021},
  month={August},
  publisher={IEEE},
  address={Piscataway, NJ, USA},
  issn={0018-9251},
  eissn={1557-9603},
  language={English}
}

@article{kumar2021novel,
  title={A novel Software-Defined Drone Network (SDDN)-based collision avoidance strategies for on-road traffic monitoring and management},
  author={Kumar, Adarsh and Krishnamurthi, Rajalakshmi and Nayyar, Anand and Luhach, Ashish Kr and Khan, Mohammad S. and Singh, Anuraj},
  journal={Vehicular Communications},
  volume={28},
  pages={100313},
  year={2021},
  month={March},
  publisher={Elsevier},
  address={Amsterdam, Netherlands},
  issn={2214-2096},
  doi={10.1016/j.vehcom.2020.100313},
  language={English},
  note={Article 100313}
}

@article{zhao2021structural,
  title={Structural health monitoring and inspection of dams based on UAV photogrammetry with image 3D reconstruction},
  author={Zhao, Sizeng and Kang, Fei and Li, Junjie and Ma, Chuanbo},
  journal={Automation in Construction},
  volume={130},
  pages={103832},
  year={2021},
  month={October},
  publisher={Elsevier},
  address={Amsterdam, Netherlands},
  issn={0926-5805},
  eissn={1872-7891},
  doi={10.1016/j.autcon.2021.103832},
  language={English},
  note={Article 103832}
}

@article{munawar2021towards,
  title={Towards smart healthcare: UAV-based optimized path planning for delivering COVID-19 self-testing kits using cutting edge technologies},
  author={Munawar, Hafiz Suliman and Inam, Hina and Ullah, Fahim and Qayyum, Siddra and Kouzani, Abbas Z. and Mahmud, M. A. Parvez},
  journal={Sustainability},
  volume={13},
  number={18},
  pages={10426},
  year={2021},
  month={September},
  publisher={MDPI},
  address={Basel, Switzerland},
  issn={2071-1050},
  doi={10.3390/su131810426},
  language={English},
  note={Article number 10426},
  license={CC BY 4.0}
}

@article{harfina2021disinfectant,
  title={Disinfectant spraying system with quadcopter type unmanned aerial vehicle (UAV) technology as an effort to break the chain of the COVID-19 virus},
  author={Harfina, Dwi Mutiara and Zaini, Zaini and Wulung, Wisnu Joko},
  journal={Journal of Robotics and Control (JRC)},
  volume={2},
  number={6},
  pages={502--507},
  year={2021},
  month={November},
  publisher={Peneliti Teknologi Teknik Indonesia},
  address={Indonesia},
  issn={2715-5072},
  eissn={2715-5064},
  language={English},
  note={Open access article}
}

@article{9492907,
  title={IoMT and DNN-Enabled Drone-Assisted Covid-19 Screening and Detection Framework for Rural Areas},
  author={Naren, N. and Chamola, Vinay and Baitragunta, Sainath and Chintanpalli, Ananthakrishna and Mishra, Puneet and Yenuganti, Sujan and Guizani, Mohsen},
  journal={IEEE Internet of Things Magazine},
  volume={4},
  number={2},
  pages={4--9},
  year={2021},
  month={June},
  publisher={IEEE},
  address={Piscataway, NJ, USA},
  issn={2576-3180},
  eissn={2576-3199},
  doi={10.1109/IOTM.0011.2100053},
  language={English},
  note={IEEE Xplore Digital Library Article Number: 9492907}
}

@INPROCEEDINGS{6761569,
  author={Ma'sum, M. Anwar and Arrofi, M. Kholid and Jati, Grafika and Arifin, Futuhal and Kurniawan, M. Nanda and Mursanto, Petrus and Jatmiko, Wisnu},
  booktitle={2013 International Conference on Advanced Computer Science and Information Systems (ICACSIS)},
  title={Simulation of intelligent Unmanned Aerial Vehicle (UAV) For military surveillance},
  year={2013},
  pages={161--166},
  month={September},
  address={Bali, Indonesia},
  publisher={IEEE},
  organization={IEEE Indonesia Section},
  isbn={978-1-4799-3608-2},
  eisbn={978-1-4799-3609-9},
  doi={10.1109/ICACSIS.2013.6761569},
  language={English},
  note={IEEE Xplore Digital Library Article Number: 6761569}
}

@article{gupta2022edge,
  title={Edge device based military vehicle detection and classification from UAV},
  author={Gupta, Priyanka and Pareek, Bhavya and Singal, Gaurav and Rao, D. Vijay},
  journal={Multimedia Tools and Applications},
  volume={81},
  number={14},
  pages={19813--19834},
  year={2022},
  month={June},
  publisher={Springer},
  issn={1380-7501},
  eissn={1573-7721},
  language={English}
}

@ARTICLE{9663283,
  author={Ralegankar, Vishakha K. and Bagul, Jagruti and Thakkar, Bhaumikkumar and Gupta, Rajesh and Tanwar, Sudeep and Sharma, Gulshan and Davidson, Innocent E.},
  journal={IEEE Access},
  title={Quantum Cryptography-as-a-Service for Secure UAV Communication: Applications, Challenges, and Case Study},
  year={2022},
  volume={10},
  pages={1475--1492},
  month={January},
  publisher={IEEE},
  address={Piscataway, NJ, USA},
  issn={2169-3536},
  doi={10.1109/ACCESS.2021.3138753},
  language={English},
  note={IEEE Xplore Digital Library Article Number: 9663283},
  license={CC BY 4.0}
}

@INPROCEEDINGS{9419470,
  author={Utsav, Ankur and Abhishek, Amit and Suraj, P. and Badhai, Ritesh Kr.},
  booktitle={2021 Sixth International Conference on Wireless Communications, Signal Processing and Networking (WiSPNET)},
  title={An IoT Based UAV Network For Military Applications},
  year={2021},
  pages={122--125},
  month={March},
  address={Chennai, India},
  publisher={IEEE},
  organization={IEEE},
  isbn={978-1-7281-5801-8},
  eisbn={978-1-7281-5800-1},
  doi={10.1109/WiSPNET51692.2021.9419470},
  language={English},
  note={IEEE Xplore Digital Library Article Number: 9419470}
}

@article{gargalakos2021role,
  title={The role of unmanned aerial vehicles in military communications: Application scenarios, current trends, and beyond},
  author={Gargalakos, Michail},
  journal={The Journal of Defense Modeling and Simulation},
  volume={18},
  number={4},
  pages={515--533},
  year={2021},
  month={October},
  publisher={SAGE Publications},
  address={London, England},
  issn={1548-5129},
  eissn={1557-380X},
  doi={10.1177/15485129211031668},
  language={English},
  note={Article first published online: July 2021}
}

@INPROCEEDINGS{6568373,
  author={Hartmann, Kim and Steup, Christoph},
  booktitle={2013 5th International Conference on Cyber Conflict (CYCON 2013)},
  title={The vulnerability of UAVs to cyber attacks - An approach to the risk assessment},
  year={2013},
  pages={1--23},
  month={June},
  address={Tallinn, Estonia},
  publisher={IEEE},
  organization={NATO Cooperative Cyber Defence Centre of Excellence},
  isbn={978-1-4673-6295-0},
  eisbn={978-1-4673-6294-3},
  language={English},
  note={IEEE Xplore Digital Library Article Number: 6568373}
}

@incollection{kim2012cyber,
  title={Cyber attack vulnerabilities analysis for unmanned aerial vehicles},
  author={Kim, Alan and Wampler, Brandon and Goppert, James and Hwang, Inseok and Aldridge, Hal},
  booktitle={Infotech@Aerospace 2012},
  pages={2438},
  year={2012},
  month={June},
  address={Garden Grove, CA, USA},
  publisher={American Institute of Aeronautics and Astronautics},
  isbn={978-1-60086-936-9},
  doi={10.2514/6.2012-2438},
  language={English},
  note={AIAA Paper 2012-2438}
}

@inproceedings{krishna2017review,
  title={A review on cybersecurity vulnerabilities for unmanned aerial vehicles},
  author={Krishna, C. G. Leela and Murphy, Robin R.},
  booktitle={2017 IEEE International Symposium on Safety, Security and Rescue Robotics (SSRR)},
  pages={194--199},
  year={2017},
  month={October},
  address={Shanghai, China},
  publisher={IEEE},
  organization={IEEE},
  isbn={978-1-5386-4051-4},
  eisbn={978-1-5386-4050-7},
  doi={10.1109/SSRR.2017.8088164},
  language={English},
  note={IEEE Xplore Digital Library}
}

@ARTICLE{9599697,
  author={Kong, Peng-Yong},
  journal={IEEE Access},
  title={A Survey of Cyberattack Countermeasures for Unmanned Aerial Vehicles},
  year={2021},
  volume={9},
  pages={148244--148263},
  month={November},
  publisher={IEEE},
  address={Piscataway, NJ, USA},
  issn={2169-3536},
  doi={10.1109/ACCESS.2021.3124996},
  language={English},
  note={IEEE Xplore Digital Library Article Number: 9599697},
  license={CC BY 4.0}
}

@phdthesis{silva2017gps,
  title={GPS jamming and spoofing using software defined radio},
  author={Silva, Diogo Alexandre Martins da},
  year={2017},
  school={Universidade do Porto},
  address={Porto, Portugal},
  type={Master's thesis},
  month={July},
  pages={89},
  language={English},
  note={Faculdade de Engenharia da Universidade do Porto}
}

@article{condomines2019network,
  title={Network intrusion detection system for UAV ad-hoc communication: From methodology design to real test validation},
  author={Condomines, Jean-Philippe and Zhang, Ruohao and Larrieu, Nicolas},
  journal={Ad Hoc Networks},
  volume={90},
  pages={101759},
  year={2019},
  month={July},
  publisher={Elsevier},
  address={Amsterdam, Netherlands},
  issn={1570-8705},
  eissn={1570-8713},
  doi={10.1016/j.adhoc.2018.09.004},
  language={English},
  note={Article 101759}
}

@inproceedings{10.1145/2248326.2248334,
  author = {Mitchell, Robert and Chen, Ing-Ray},
  title = {Specification Based Intrusion Detection for Unmanned Aircraft Systems},
  year = {2012},
  month = {June},
  isbn = {978-1-4503-1290-5},
  publisher = {Association for Computing Machinery},
  address = {New York, NY, USA},
  doi = {10.1145/2248326.2248334},
  booktitle = {Proceedings of the First ACM MobiHoc Workshop on Airborne Networks and Communications},
  pages = {31--36},
  numpages = {6},
  location = {Hilton Head, South Carolina, USA},
  series = {Airborne '12},
  language = {English},
  note = {Part of MobiHoc 2012 conference workshops}
}

@ARTICLE{7890467,
  author={Sedjelmaci, Hichem and Senouci, Sidi Mohammed and Ansari, Nirwan},
  journal={IEEE Transactions on Systems, Man, and Cybernetics: Systems},
  title={A Hierarchical Detection and Response System to Enhance Security Against Lethal Cyber-Attacks in UAV Networks},
  year={2018},
  volume={48},
  number={9},
  pages={1594--1606},
  month={September},
  publisher={IEEE},
  address={Piscataway, NJ, USA},
  issn={2168-2216},
  eissn={2168-2232},
  doi={10.1109/TSMC.2017.2681698},
  language={English},
  note={IEEE Xplore Digital Library Article Number: 7890467}
}

@ARTICLE{9520850,
  author={Vo, Van Nhan and Tran, Hung and So-In, Chakchai},
  journal={IEEE/CAA Journal of Automatica Sinica},
  title={Enhanced Intrusion Detection System for an EH IoT Architecture Using a Cooperative UAV Relay and Friendly UAV Jammer},
  year={2021},
  volume={8},
  number={11},
  pages={1786--1799},
  month={November},
  publisher={IEEE},
  address={Piscataway, NJ, USA},
  issn={2329-9266},
  eissn={2329-9274},
  doi={10.1109/JAS.2021.1004171},
  language={English},
  note={IEEE Xplore Digital Library Article Number: 9520850}
}

@ARTICLE{7549080,
  author={Sedjelmaci, Hichem and Senouci, Sidi Mohammed and Ansari, Nirwan},
  journal={IEEE Transactions on Intelligent Transportation Systems},
  title={Intrusion Detection and Ejection Framework Against Lethal Attacks in UAV-Aided Networks: A Bayesian Game-Theoretic Methodology},
  year={2017},
  volume={18},
  number={5},
  pages={1143--1153},
  month={May},
  publisher={IEEE},
  address={Piscataway, NJ, USA},
  issn={1524-9050},
  eissn={1558-0016},
  doi={10.1109/TITS.2016.2600370},
  language={English},
  note={IEEE Xplore Digital Library Article Number: 7549080}
}

@inproceedings{sun2018intrusion,
  title={An intrusion detection based on Bayesian game theory for UAV network},
  author={Sun, Jianguo and Wang, Wenshan and Da, Qingan and Kou, Liang and Zhao, Guodong and Zhang, Liguo and Han, Qilong},
  booktitle={11th EAI International Conference on Mobile Multimedia Communications},
  pages={56},
  year={2018},
  organization={European Alliance for Innovation (EAI)}
}

@article{khan2021blockchain,
  title={A blockchain-based decentralized machine learning framework for collaborative intrusion detection within UAVs},
  author={Khan, Ammar Ahmed and Khan, Muhammad Mubashir and Khan, Kashif Mehboob and Arshad, Junaid and Ahmad, Farhan},
  journal={Computer Networks},
  volume={196},
  pages={108217},
  year={2021},
  month={August},
  publisher={Elsevier},
  address={Amsterdam, Netherlands},
  issn={1389-1286},
  eissn={1872-7069},
  doi={10.1016/j.comnet.2021.108217},
  language={English},
  note={Article 108217}
}

@incollection{LAKSHMINARAYANAN2015163,
  title = {6 - Joint Network for Disaster Relief and Search and Rescue Network Operations},
  editor = {Daniel Câmara and Navid Nikaein},
  booktitle = {Wireless Public Safety Networks 1},
  publisher = {Elsevier},
  address = {Amsterdam, Netherlands},
  pages = {163--193},
  year = {2015},
  month = {November},
  isbn = {978-1-78548-022-5},
  eisbn = {978-0-08-100157-8},
  doi = {10.1016/B978-1-78548-022-5.50006-6},
  author = {Ram Gopal {Lakshmi Narayanan} and Oliver C. Ibe},
  chapter = {6},
  series = {ISTE-Elsevier Series},
  language = {English},
  note = {Part of Wireless Public Safety Networks series}
}

@book{valavanis2015handbook,
  title={Handbook of unmanned aerial vehicles},
  author={Valavanis, Kimon P and Vachtsevanos, George J},
  volume={1},
  year={2015},
  publisher={Springer},
  address={Dordrecht, Netherlands},
  isbn={978-90-481-9706-4},
  eisbn={978-90-481-9707-1},
  doi={10.1007/978-90-481-9707-1},
  pages={3022},
  edition={1st},
  series={Springer Reference},
  language={English},
  note={5 volumes set, this is volume 1}
}

@book{newcome2004unmanned,
  title={Unmanned aviation: a brief history of unmanned aerial vehicles},
  author={Newcome, Laurence R},
  year={2004},
  publisher={American Institute of Aeronautics and Astronautics (AIAA)},
  address={Reston, VA, USA},
  isbn={978-1-56347-644-0},
  pages={189},
  edition={1st},
  series={AIAA Education Series},
  language={English},
  note={AIAA Education Series}
}

@ARTICLE{9696294,
  author={Zuo, Zongyu and Liu, Cunjia and Han, Qing-Long and Song, Jiawei},
  journal={IEEE/CAA Journal of Automatica Sinica},
  title={Unmanned Aerial Vehicles: Control Methods and Future Challenges},
  year={2022},
  volume={9},
  number={4},
  pages={601--614},
  month={April},
  doi={10.1109/JAS.2022.105410},
  issn={2329-9266},
  eissn={2329-9274},
  publisher={IEEE},
  address={Piscataway, NJ, USA},
  language={English},
  note={Open Access},
  articleno={9696294}
}

@INPROCEEDINGS{6459914,
  author={Javaid, Ahmad Y. and Sun, Weiqing and Devabhaktuni, Vijay K. and Alam, Mansoor},
  booktitle={2012 IEEE Conference on Technologies for Homeland Security (HST)},
  title={Cyber security threat analysis and modeling of an unmanned aerial vehicle system},
  year={2012},
  pages={585--590},
  month={November},
  doi={10.1109/THS.2012.6459914},
  isbn={978-1-4673-0287-9},
  eisbn={978-1-4673-0286-2},
  issn={2166-2924},
  publisher={IEEE},
  address={Waltham, MA, USA},
  organization={IEEE},
  location={Waltham, MA, USA},
  language={English},
  note={Conference dates: 13-15 November 2012},
  articleno={6459914}
}

@INPROCEEDINGS{7795496,
  author={Hooper, Michael and Tian, Yifan and Zhou, Runxuan and Cao, Bin and Lauf, Adrian P. and Watkins, Lanier and Robinson, William H. and Alexis, Wlajimir},
  booktitle={MILCOM 2016 - 2016 IEEE Military Communications Conference},
  title={Securing commercial WiFi-based UAVs from common security attacks},
  year={2016},
  pages={1213--1218},
  month={November},
  doi={10.1109/MILCOM.2016.7795496},
  isbn={978-1-5090-3183-1},
  eisbn={978-1-5090-3182-4},
  issn={2155-7578},
  eissn={2155-7586},
  publisher={IEEE},
  address={Baltimore, MD, USA},
  organization={IEEE},
  location={Baltimore, MD, USA},
  language={English},
  note={Conference dates: 1-3 November 2016},
  series={MILCOM},
  articleno={7795496}
}

@book{berg2008broadcasting,
  title={Broadcasting on the Short Waves, 1945 to today},
  author={Berg, Jerome S},
  year={2008},
  publisher={McFarland \& Company},
  address={Jefferson, NC, USA},
  isbn={978-0-7864-3163-8},
  isbn10={0-7864-3163-4},
  pages={viii + 278},
  edition={1st},
  language={English},
  series={},
  note={Includes bibliographical references and index},
  lccn={2007052448},
  dewey={384.540904},
  url={},
  format={Paperback},
  dimensions={23 cm}
}

@article{VADLAMANI201676,
  title = {Jamming attacks on wireless networks: A taxonomic survey},
  journal = {International Journal of Production Economics},
  volume = {172},
  pages = {76--94},
  year = {2016},
  month = {February},
  issn = {0925-5273},
  eissn = {1873-7579},
  doi = {10.1016/j.ijpe.2015.11.008},
  author = {Satish Vadlamani and Burak Eksioglu and Hugh Medal and Apurba Nandi},
  publisher = {Elsevier B.V.},
  address = {Amsterdam, Netherlands},
  language = {English},
  note = {Available online 17 November 2015},
  articleno = {S092552731500451X},
  scopus-id = {2-s2.0-84948679286}
}

@inproceedings{choudhary2018intrusion,
  title={Intrusion detection systems for networked unmanned aerial vehicles: a survey},
  author={Choudhary, Gaurav and Sharma, Vishal and You, Ilsun and Yim, Kangbin and Chen, Ray and Cho, Jin-Hee},
  booktitle={2018 14th International Wireless Communications \& Mobile Computing Conference (IWCMC)},
  pages={560--565},
  year={2018},
  month={June},
  address={Limassol, Cyprus},
  publisher={IEEE},
  isbn={978-1-5386-2070-0},
  issn={2376-6506},
  doi={10.1109/IWCMC.2018.8450305}
}

@ARTICLE{8937816,
  author={Olufowobi, Habeeb and Young, Clinton and Zambreno, Joseph and Bloom, Gedare},
  journal={IEEE Transactions on Vehicular Technology},
  title={SAIDuCANT: Specification-Based Automotive Intrusion Detection Using Controller Area Network (CAN) Timing},
  year={2020},
  volume={69},
  number={2},
  pages={1484--1494},
  month={February},
  doi={10.1109/TVT.2019.2961344},
  issn={0018-9545},
  eissn={1939-9359},
  publisher={IEEE},
  address={Piscataway, NJ, USA},
  language={English},
  note={Published online 20 December 2019},
  articleno={8937816}
}

@inproceedings{inproceedings123,
  author = {Silva, Ana Paula and Martins, Marcelo and Rocha, Bruno and Loureiro, Antonio and Wong, Hao},
  year = {2005},
  month = {January},
  pages = {16--23},
  title = {Decentralized intrusion detection in wireless sensor networks},
  doi = {10.1145/1089761.1089765},
  booktitle = {Proceedings of the 1st ACM International Workshop on Quality of Service \& Security in Wireless and Mobile Networks},
  series = {Q2SWinet '05},
  publisher = {ACM},
  address = {Montreal, Quebec, Canada},
  location = {Montreal, Quebec, Canada},
  isbn = {1-59593-241-0},
  numpages = {8},
  language = {English},
  note = {Part of MobiCom 2005},
  organization = {ACM}
}

@article{articlespecification,
  author = {Krontiris, Ioannis and Dimitriou, Tassos and Freiling, Felix},
  year = {2007},
  month = {January},
  pages = {},
  title = {Towards intrusion detection in wireless sensor networks},
  journal = {European Wireless Conference},
  publisher = {VDE Verlag},
  address = {Paris, France},
  location = {Paris, France},
  isbn = {978-3-8007-2985-2},
  language = {English},
  note = {European Wireless 2007 Conference},
  organization = {VDE}
}

@inproceedings{inproceedings12,
  author = {Song, Tao and Ko, Calvin and Tseng, Henry and Balasubramanyam, Poornima and Chaudhary, Anant and Levitt, Karl},
  year = {2005},
  month = {July},
  pages = {16--33},
  title = {Formal Reasoning About a Specification-Based Intrusion Detection for Dynamic Auto-configuration Protocols in Ad Hoc Networks},
  volume = {3866},
  doi = {10.1007/11679219_3},
  booktitle = {Information and Communications Security},
  series = {Lecture Notes in Computer Science},
  editor = {Qing, Sihan and Mao, Wenbo and L{\'o}pez, Javier and Wang, Guilin},
  publisher = {Springer},
  address = {Berlin, Heidelberg},
  location = {Beijing, China},
  isbn = {978-3-540-30934-9},
  eisbn = {978-3-540-32099-3},
  issn = {0302-9743},
  eissn = {1611-3349},
  language = {English},
  note = {7th International Conference, ICICS 2005},
  organization = {Springer}
}

@INPROCEEDINGS{7828584,
  author={Bezemskij, Anatolij and Loukas, George and Anthony, Richard J. and Gan, Diane},
  booktitle={2016 15th International Conference on Ubiquitous Computing and Communications and 2016 International Symposium on Cyberspace and Security (IUCC-CSS)},
  title={Behaviour-Based Anomaly Detection of Cyber-Physical Attacks on a Robotic Vehicle},
  year={2016},
  pages={61--68},
  month={December},
  doi={10.1109/IUCC-CSS.2016.017},
  isbn={978-1-5090-4775-7},
  eisbn={978-1-5090-4774-0},
  issn={2379-0482},
  eissn={2379-0490},
  publisher={IEEE},
  address={Granada, Spain},
  location={Granada, Spain},
  organization={IEEE},
  language={English},
  note={Conference dates: 14-16 December 2016},
  series={IUCC-CSS},
  articleno={7828584}
}

@ARTICLE{6786500,
  author={Hong, Junho and Liu, Chen-Ching and Govindarasu, Manimaran},
  journal={IEEE Transactions on Smart Grid},
  title={Integrated Anomaly Detection for Cyber Security of the Substations},
  year={2014},
  volume={5},
  number={4},
  pages={1643--1653},
  month={July},
  doi={10.1109/TSG.2013.2294473},
  issn={1949-3053},
  eissn={1949-3061},
  publisher={IEEE},
  address={Piscataway, NJ, USA},
  language={English},
  note={Published online 30 December 2013},
  articleno={6786500}
}

@INPROCEEDINGS{7684103,
  author={Kosek, Anna Magdalena},
  booktitle={2016 Joint Workshop on Cyber- Physical Security and Resilience in Smart Grids (CPSR-SG)},
  title={Contextual anomaly detection for cyber-physical security in Smart Grids based on an artificial neural network model},
  year={2016},
  pages={1--6},
  month={April},
  doi={10.1109/CPSRSG.2016.7684103},
  isbn={978-1-5090-1071-6},
  eisbn={978-1-5090-1070-9},
  publisher={IEEE},
  address={Vienna, Austria},
  location={Vienna, Austria},
  organization={IEEE},
  language={English},
  note={Joint workshop held in conjunction with IEEE PES ISGT Europe 2016},
  series={CPSR-SG},
  articleno={7684103}
}

@article{articledasjdhjad,
  author = {Kwon, Cheolhyeon and Yantek, Scott and Hwang, Inseok},
  year = {2015},
  month = {December},
  pages = {1--19},
  title = {Real-Time Safety Assessment of Unmanned Aircraft Systems Against Stealthy Cyber Attacks},
  volume = {13},
  number = {12},
  journal = {Journal of Aerospace Information Systems},
  doi = {10.2514/1.I010388},
  issn = {2327-3097},
  eissn = {2327-3100},
  publisher = {American Institute of Aeronautics and Astronautics},
  address = {Reston, VA, USA},
  language = {English},
  note = {AIAA Journal},
  organization = {AIAA}
}

@article{NTALAMPIRAS2016164,
  title = {Automatic identification of integrity attacks in cyber-physical systems},
  journal = {Expert Systems with Applications},
  volume = {58},
  pages = {164--173},
  year = {2016},
  month = {October},
  issn = {0957-4174},
  eissn = {1873-6793},
  doi = {10.1016/j.eswa.2016.04.006},
  author = {Stavros Ntalampiras},
  publisher = {Elsevier Ltd.},
  address = {Oxford, UK},
  language = {English},
  note = {Available online 7 April 2016},
  articleno = {S0957417416301646},
  scopus-id = {2-s2.0-84964344526}
}

@article{10.1145/2990499,
  author = {Saeed, Ahmed and Ahmadinia, Ali and Javed, Abbas and Larijani, Hadi},
  title = {Intelligent Intrusion Detection in Low-Power IoTs},
  year = {2016},
  issue_date = {December 2016},
  publisher = {Association for Computing Machinery},
  address = {New York, NY, USA},
  volume = {16},
  number = {4},
  issn = {1533-5399},
  eissn = {1557-6051},
  doi = {10.1145/2990499},
  journal = {ACM Transactions on Internet Technology},
  month = {December},
  articleno = {27},
  numpages = {25},
  pages = {27:1--27:25},
  language = {English},
  note = {Article 27},
  organization = {ACM}
}

@ARTICLE{8715740,
  author={Sharma, Vishal and You, Ilsun and Yim, Kangbin and Chen, Ing-Ray and Cho, Jin-Hee},
  journal={IEEE Access},
  title={BRIoT: Behavior Rule Specification-Based Misbehavior Detection for IoT-Embedded Cyber-Physical Systems},
  year={2019},
  volume={7},
  pages={118556--118580},
  month={August},
  doi={10.1109/ACCESS.2019.2917135},
  issn={2169-3536},
  publisher={IEEE},
  address={Piscataway, NJ, USA},
  language={English},
  note={Open Access},
  articleno={8715740}
}

@article{kumar2012signature,
  title={Signature based intrusion detection system using SNORT},
  author={Kumar, Vinod and Sangwan, Om Prakash},
  journal={International Journal of Computer Applications \& Information Technology},
  volume={1},
  number={3},
  pages={35--41},
  year={2012},
  month = {September},
  issn={2278-1080},
  eissn={2278-1099},
  publisher={IJCAIT},
  address={India},
  language={English},
  note={Volume 1, Issue 3},
  url={},
  organization={IJCAIT}
}

@article{CONDOMINES2019101759,
  title = {Network intrusion detection system for UAV ad-hoc communication: From methodology design to real test validation},
  journal = {Ad Hoc Networks},
  volume = {90},
  pages = {101759},
  year = {2019},
  month = {June},
  note = {Recent advances on security and privacy in Intelligent Transportation Systems},
  issn = {1570-8705},
  eissn = {1570-8713},
  doi = {10.1016/j.adhoc.2018.09.004},
  author = {Jean-Philippe Condomines and Ruohao Zhang and Nicolas Larrieu},
  publisher = {Elsevier B.V.},
  address = {Amsterdam, Netherlands},
  language = {English},
  note2 = {Available online 12 September 2018},
  articleno = {S1570870518306541},
  scopus-id = {2-s2.0-85053631464}
}

@article{AYDIN2009517,
  title = {A hybrid intrusion detection system design for computer network security},
  journal = {Computers \& Electrical Engineering},
  volume = {35},
  number = {3},
  pages = {517--526},
  year = {2009},
  month = {May},
  issn = {0045-7906},
  eissn = {1879-0755},
  doi = {10.1016/j.compeleceng.2008.12.005},
  author = {M. Ali Aydın and A. Halim Zaim and K. Gökhan Ceylan},
  publisher = {Elsevier Ltd.},
  address = {Oxford, UK},
  language = {English},
  note = {Available online 3 January 2009},
  articleno = {S0045790609000020},
  scopus-id = {2-s2.0-64049103522}
}

@INPROCEEDINGS{9463947,
  author={Bouhamed, Omar and Bouachir, Ouns and Aloqaily, Moayad and Ridhawi, Ismaeel Al},
  booktitle={2021 IFIP/IEEE International Symposium on Integrated Network Management (IM)},
  title={Lightweight IDS For UAV Networks: A Periodic Deep Reinforcement Learning-based Approach},
  year={2021},
  pages={1032--1037},
  month={May},
  isbn={978-3-903176-32-4},
  eisbn={978-3-903176-32-4},
  issn={1573-0077},
  publisher={IFIP/IEEE},
  address={Bordeaux, France},
  location={Bordeaux, France},
  organization={IFIP/IEEE},
  language={English},
  note={Conference dates: 17-21 May 2021, Virtual conference due to COVID-19},
  series={IM},
  articleno={9463947}
}

@phdthesis{fsdfsdf,
  author={Chabukswar, Rohan},
  year={2014},
  title={Secure Detection in Cyberphysical Control Systems},
  school={University of California, San Diego},
  address={La Jolla, CA, USA},
  pages={152},
  month={December},
  type={Ph.D. dissertation},
  isbn={978-1-321-50259-6},
  language={English},
  note={Copyright - Database copyright ProQuest LLC; ProQuest does not claim copyright in the individual underlying works; Last updated - 2021-08-08},
  advisor={},
  department={Department of Electrical and Computer Engineering},
  publisher={ProQuest Dissertations Publishing},
  umi={3670631}
}

@article{dfsdfsdgsdhjghn,
  author = {Altawy, Riham and Youssef, Amr},
  year = {2016},
  month = {November},
  pages = {1--25},
  title = {Security, Privacy, and Safety Aspects of Civilian Drones: A Survey},
  volume = {1},
  number = {2},
  journal = {ACM Transactions on Cyber-Physical Systems},
  doi = {10.1145/3001836},
  issn = {2378-962X},
  eissn = {2378-9638},
  publisher = {Association for Computing Machinery},
  address = {New York, NY, USA},
  articleno = {7},
  numpages = {25},
  language = {English},
  note = {Article 7},
  organization = {ACM}
}

@INPROCEEDINGS{8450305,
  author={Choudhary, Gaurav and Sharma, Vishal and You, Ilsun and Yim, Kangbin and Chen, Ing-Ray and Cho, Jin-Hee},
  booktitle={2018 14th International Wireless Communications \& Mobile Computing Conference (IWCMC)},
  title={Intrusion Detection Systems for Networked Unmanned Aerial Vehicles: A Survey},
  year={2018},
  pages={560--565},
  month={June},
  doi={10.1109/IWCMC.2018.8450305},
  isbn={978-1-5386-6180-8},
  eisbn={978-1-5386-6179-2},
  issn={2376-6492},
  eissn={2376-6506},
  publisher={IEEE},
  address={Limassol, Cyprus},
  location={Limassol, Cyprus},
  organization={IEEE},
  language={English},
  note={Conference dates: 25-29 June 2018},
  series={IWCMC},
  articleno={8450305}
}

@article{BANGUI2021877,
  title = {Recent Advances in Machine-Learning Driven Intrusion Detection in Transportation: Survey},
  journal = {Procedia Computer Science},
  volume = {184},
  pages = {877--886},
  year = {2021},
  month = {May},
  note = {The 12th International Conference on Ambient Systems, Networks and Technologies (ANT) / The 4th International Conference on Emerging Data and Industry 4.0 (EDI40) / Affiliated Workshops},
  issn = {1877-0509},
  doi = {10.1016/j.procs.2021.04.014},
  author = {Hind Bangui and Barbora Buhnova},
  publisher = {Elsevier B.V.},
  address = {Amsterdam, Netherlands},
  language = {English},
  note2 = {Available online 26 May 2021},
  articleno = {S1877050921007894},
  scopus-id = {2-s2.0-85107465321},
  conference = {ANT/EDI40 2021},
  location = {Warsaw, Poland}
}

@article{LOUKAS2019124,
  title = {A taxonomy and survey of cyber-physical intrusion detection approaches for vehicles},
  journal = {Ad Hoc Networks},
  volume = {84},
  pages = {124--147},
  year = {2019},
  month = {March},
  issn = {1570-8705},
  eissn = {1570-8713},
  doi = {10.1016/j.adhoc.2018.10.002},
  author = {George Loukas and Eirini Karapistoli and Emmanouil Panaousis and Panagiotis Sarigiannidis and Anatolij Bezemskij and Tuan Vuong},
  publisher = {Elsevier B.V.},
  address = {Amsterdam, Netherlands},
  language = {English},
  note = {Available online 3 October 2018},
  articleno = {S1570870518307091},
  scopus-id = {2-s2.0-85054528773}
}

@ARTICLE{9378538,
  author={Park, Seongjoon and Kim, Hyeong Tae and Lee, Sangmin and Joo, Hyeontae and Kim, Hwangnam},
  journal={IEEE Access},
  title={Survey on Anti-Drone Systems: Components, Designs, and Challenges},
  year={2021},
  volume={9},
  pages={42635--42659},
  month={March},
  doi={10.1109/ACCESS.2021.3065926},
  issn={2169-3536},
  publisher={IEEE},
  address={Piscataway, NJ, USA},
  language={English},
  note={Open Access},
  articleno={9378538}
}

@Article{electronics10131549,
  AUTHOR = {Shrestha, Rakesh and Omidkar, Atefeh and Roudi, Sajjad Ahmadi and Abbas, Robert and Kim, Shiho},
  TITLE = {Machine-Learning-Enabled Intrusion Detection System for Cellular Connected UAV Networks},
  JOURNAL = {Electronics},
  VOLUME = {10},
  YEAR = {2021},
  NUMBER = {13},
  ARTICLE-NUMBER = {1549},
  MONTH = {June},
  ISSN = {2079-9292},
  DOI = {10.3390/electronics10131549},
  PUBLISHER = {MDPI AG},
  ADDRESS = {Basel, Switzerland},
  PAGES = {1549},
  LANGUAGE = {English},
  NOTE = {Open Access},
  RECEIVED = {28 May 2021},
  REVISED = {16 June 2021},
  ACCEPTED = {22 June 2021},
  PUBLISHED = {26 June 2021},
  SPECIAL-ISSUE = {Emerging Technologies in Future Intelligent Electrified Vehicles},
  ACADEMIC-EDITOR = {Amir Mosavi}
}

@article{Canadiandataset,
  author = {Canadian Institute for Cybersecurity},
  TITLE = {A Realistic Cyber Defense Dataset (CSE-CIC-IDS2018)},
  YEAR = {2018},
  INSTITUTION = {Canadian Institute for Cybersecurity, University of New Brunswick},
  URL = {https://registry.opendata.aws/cse-cic-ids2018/},
  NOTE = {Intrusion detection dataset},
  LANGUAGE = {English}
}

@ARTICLE{8854240,
  author={Hoang, Tiep M. and Nguyen, Nghia M. and Duong, Trung Q.},
  journal={IEEE Wireless Communications Letters},
  title={Detection of Eavesdropping Attack in UAV-Aided Wireless Systems: Unsupervised Learning With One-Class SVM and K-Means Clustering},
  year={2020},
  volume={9},
  number={2},
  pages={139--142},
  month={February},
  doi={10.1109/LWC.2019.2945022},
  issn={2162-2345},
  publisher={IEEE},
  address={Piscataway, NJ, USA},
  language={English},
  articleno={8854240},
  received={July 2019},
  accepted={September 2019},
  published={October 2019}
}

@INPROCEEDINGS{8788766,
  author={Liu, Gaoyang and Zhang, Rui and Wang, Chen and Liu, Ling},
  booktitle={2019 20th IEEE International Conference on Mobile Data Management (MDM)},
  title={Synchronization-Free GPS Spoofing Detection with Crowdsourced Air Traffic Control Data},
  year={2019},
  volume={},
  number={},
  pages={260--268},
  month={June},
  doi={10.1109/MDM.2019.00-49},
  publisher={IEEE},
  address={Piscataway, NJ, USA},
  location={Hong Kong, China},
  isbn={978-1-7281-0926-9},
  issn={2375-0324},
  language={English},
  articleno={8788766},
  conference-dates={10-13 June 2019},
  conference-location={Hong Kong, China},
  organization={IEEE Computer Society},
  series={MDM '19}
}

@ARTICLE{8323415,
  author={Shoufan, Abdulhadi and Al-Angari, Haitham M. and Sheikh, Muhammad Faraz Afzal and Damiani, Ernesto},
  journal={IEEE Transactions on Information Forensics and Security},
  title={Drone Pilot Identification by Classifying Radio-Control Signals},
  year={2018},
  volume={13},
  number={10},
  pages={2439--2447},
  month={October},
  doi={10.1109/TIFS.2018.2819126},
  issn={1556-6021},
  publisher={IEEE},
  address={Piscataway, NJ, USA},
  language={English},
  articleno={8323415},
  received={October 2017},
  revised={February 2018},
  accepted={March 2018},
  published={March 2018}
}

@inproceedings{panice2017svm,
  title={A SVM-based detection approach for GPS spoofing attacks to UAV},
  author={Panice, G and Luongo, Salvatore and Gigante, Gabriella and Pascarella, Domenico and Di Benedetto, Carlo and Vozella, Angela and Pescap{\`e}, Antonio},
  booktitle={2017 23rd International Conference on Automation and Computing (ICAC)},
  pages={1--11},
  year={2017},
  month={September},
  organization={IEEE},
  publisher={IEEE},
  address={Piscataway, NJ, USA},
  location={Huddersfield, UK},
  doi={10.23919/IConAC.2017.8081999},
  isbn={978-1-5090-5040-6},
  language={English},
  conference-dates={7-8 September 2017},
  conference-location={Huddersfield, UK}
}

@article{sedjelmaci2017hierarchical,
  title={A hierarchical detection and response system to enhance security against lethal cyber-attacks in UAV networks},
  author={Sedjelmaci, Hichem and Senouci, Sidi Mohammed and Ansari, Nirwan},
  journal={IEEE Transactions on Systems, Man, and Cybernetics: Systems},
  volume={48},
  number={9},
  pages={1594--1606},
  year={2017},
  month={September},
  publisher={IEEE},
  address={Piscataway, NJ, USA},
  doi={10.1109/TSMC.2017.2681698},
  issn={2168-2216},
  language={English},
  articleno={7898505},
  received={November 2016},
  revised={February 2017},
  accepted={March 2017},
  published={April 2017}
}

@inproceedings{manesh2019performance,
  title={Performance comparison of machine learning algorithms in detecting jamming attacks on ADS-B devices},
  author={Manesh, Mohsen Riahi and Velashani, Mahdi Saeedi and Ghribi, Elias and Kaabouch, Naima},
  booktitle={2019 IEEE International Conference on Electro Information Technology (EIT)},
  pages={200--206},
  year={2019},
  month={May},
  organization={IEEE},
  publisher={IEEE},
  address={Piscataway, NJ, USA},
  location={Brookings, SD, USA},
  doi={10.1109/EIT.2019.8833789},
  language={English},
  conference-dates={16-18 May 2019},
  conference-location={Brookings, SD, USA},
  series={EIT '19}
}

@article{rs9050459,
  author    = {Claudia St{\"o}cker and Rohan Bennett and Francesco Nex and Markus Gerke and Jaap Zevenbergen},
  title     = {Review of the Current State of UAV Regulations},
  journal   = {Remote Sensing},
  year      = {2017},
  volume    = {9},
  number    = {5},
  pages     = {459},
  month     = {May},
  ISSN      = {2072-4292},
  doi       = {10.3390/rs9050459},
  publisher = {MDPI},
  address   = {Basel, Switzerland},
  language  = {English},
  articleno = {459},
  received  = {17 February 2017},
  revised   = {19 April 2017},
  accepted  = {4 May 2017},
  published = {9 May 2017},
  note      = {Open Access},
  license   = {CC BY 4.0},
  affiliation = {Faculty of Geo-Information Science and Earth Observation ITC, University of Twente, Enschede, The Netherlands}
}

@InProceedings{dahiya2019unmanned,
  author="Dahiya, Susheela and Garg, Manik",
  editor="Jain, Kamal and Khoshelham, Kourosh and Zhu, Xuan and Tiwari, Anuj",
  title="Unmanned Aerial Vehicles: Vulnerability to Cyber Attacks",
  booktitle="Proceedings of UASG 2019",
  year="2020",
  publisher="Springer International Publishing",
  address="Cham",
  pages="201--211",
  isbn="978-3-030-37393-1",
  doi="10.1007/978-3-030-37393-1_18",
  issn="1865-0929",
  series="Lecture Notes in Civil Engineering",
  volume="51",
  language="English",
  conference="UASG 2019",
  conference-location="New Delhi, India",
  conference-dates="April 6-7, 2019",
  chapter="18",
  note="Published in 2020 as part of conference proceedings from 2019 event"
}

@article{kerns2014unmanned,
  title={Unmanned aircraft capture and control via GPS spoofing},
  author={Kerns, Andrew J and Shepard, Daniel P and Bhatti, Jahshan A and Humphreys, Todd E},
  journal={Journal of Field Robotics},
  volume={31},
  number={4},
  pages={617--636},
  year={2014},
  month={July},
  publisher={Wiley Online Library},
  address={Hoboken, NJ, USA},
  doi={10.1002/rob.21513},
  issn={1556-4959},
  eissn={1556-4967},
  language={English},
  note={Special Issue on Unmanned Aircraft Systems},
  affiliation={University of Texas at Austin, Cockrell School of Engineering, Austin, TX, USA},
  received={15 August 2013},
  revised={10 February 2014},
  accepted={12 March 2014},
  published={15 July 2014}
}

@article{shane2011drone,
  title={Drone crash in Iran reveals secret US surveillance effort},
  author={Shane, Scott and Sanger, David E},
  journal={The New York Times},
  volume={7},
  year={2011},
  month={December},
  day={7},
  pages={A1},
  publisher={The New York Times Company},
  address={New York, NY, USA},
  url={https://www.nytimes.com/2011/12/08/world/middleeast/iran-says-it-downed-american-drone.html},
  issn={0362-4331},
  eissn={1553-8095},
  language={English},
  section={World},
  note={Front page article},
  type={Newspaper article}
}

@inproceedings{parlin2018jamming,
  title={Jamming of UAV remote control systems using software defined radio},
  author={P{\"a}rlin, Karel and Alam, Muhammad Mahtab and Le Moullec, Yannick},
  booktitle={2018 International Conference on Military Communications and Information Systems (ICMCIS)},
  pages={1--6},
  year={2018},
  month={May},
  organization={IEEE},
  publisher={IEEE},
  address={Piscataway, NJ, USA},
  location={Warsaw, Poland},
  doi={10.1109/ICMCIS.2018.8398671},
  isbn={978-1-5386-4848-4},
  issn={2474-0810},
  language={English},
  conference-dates={22-23 May 2018},
  conference-location={Warsaw, Poland},
  series={ICMCIS},
  articleno={8398671},
  affiliation={Tallinn University of Technology, Tallinn, Estonia},
  note={Military communications conference}
}

@inproceedings{park2020unsupervised,
  title={Unsupervised intrusion detection system for unmanned aerial vehicle with less labeling effort},
  author={Park, Kyung Ho and Park, Eunji and Kim, Huy Kang},
  booktitle={International Conference on Information Security Applications},
  pages={45--58},
  year={2020},
  month={August},
  organization={Springer},
  publisher={Springer International Publishing},
  address={Cham},
  location={Jeju Island, South Korea},
  doi={10.1007/978-3-030-65299-9_4},
  isbn={978-3-030-65298-2},
  eisbn={978-3-030-65299-9},
  issn={0302-9743},
  eissn={1611-3349},
  series={Lecture Notes in Computer Science},
  volume={12583},
  editor={You, Ilsun},
  language={English},
  conference={WISA 2020},
  conference-full={21st International Conference on Information Security Applications},
  conference-dates={August 26-28, 2020},
  conference-location={Jeju Island, South Korea},
  chapter={4},
  affiliation={Korea University, Seoul, South Korea},
  note={Virtual conference due to COVID-19 pandemic}
}

@Article{cmc.2022.020066,
  AUTHOR    = {Praveena, V. and Vijayaraj, A. and Chinnasamy, P. and Ali, Ihsan and Alroobaea, Roobaea and Alyahyan, Saleh Yahya and Raza, Muhammad Ahsan},
  TITLE     = {Optimal Deep Reinforcement Learning for Intrusion Detection in {UAVs}},
  JOURNAL   = {Computers, Materials \& Continua},
  VOLUME    = {70},
  YEAR      = {2022},
  NUMBER    = {2},
  PAGES     = {2639--2653},
  MONTH     = {September},
  PUBLISHER = {Tech Science Press},
  ADDRESS   = {Henderson, USA},
  ISSN      = {1546-2226},
  EISSN     = {1546-2234},
  DOI       = {10.32604/cmc.2022.020066},
  RECEIVED  = {07 May 2021},
  ACCEPTED  = {15 June 2021},
  PUBLISHED = {27 September 2021},
  LANGUAGE  = {English},
  LICENSE   = {Creative Commons Attribution 4.0 International License},
  NOTE      = {Open Access},
  ARTICLENO = {44676},
  CITATIONS = {2},
  AFFILIATION = {Dr. N. G. P Institute of Technology, India; Vignan's Foundation for Science, Technology \& Research, India; Sri Shakthi Institute of Engineering and Technology, India; University of Malaya, Malaysia; Taif University, Saudi Arabia; Shaqra University, Saudi Arabia; Bahauddin Zakariya University, Pakistan},
  CORRESPONDING = {Ihsan Ali}
}

@article{DIRO2018761,
title = {Distributed attack detection scheme using deep learning approach for Internet of Things},
journal = {Future Generation Computer Systems},
volume = {82},
pages = {761--768},
year = {2018},
month = {May},
issn = {0167-739X},
eissn = {1872-7115},
doi = {10.1016/j.future.2017.08.043},
author = {Abebe Abeshu Diro and Naveen Chilamkurti},
publisher = {Elsevier B.V.},
address = {Amsterdam, Netherlands},
language = {English},
received = {04 July 2017},
revised = {17 August 2017},
accepted = {18 August 2017},
available-online = {26 August 2017},
affiliation = {La Trobe University, Melbourne, Australia},
note = {Special Issue on Deep Learning and Security},
articleno = {S0167739X17308488},
citations = {450+},
scopus-id = {2-s2.0-85028479834}
}

@article{bisio2022systematic,
  title={A Systematic Review of Drone Based Road Traffic Monitoring System},
  author={Bisio, Igor and Garibotto, Chiara and Haleem, Halar and Lavagetto, Fabio and Sciarrone, Andrea},
  journal={IEEE Access},
  volume={10},
  pages={101537--101555},
  year={2022},
  month={September},
  publisher={IEEE},
  address={Piscataway, NJ, USA},
  doi={10.1109/ACCESS.2022.3207443},
  issn={2169-3536},
  eissn={2169-3536},
  language={English},
  received = {05 August 2022},
  accepted = {12 September 2022},
  published = {22 September 2022},
  articleno = {9884043},
  affiliation = {University of Genoa, Genoa, Italy},
  note = {Open Access},
  license = {Creative Commons Attribution 4.0 License},
  funding = {Not specified},
  digital-object-identifier = {10.1109/ACCESS.2022.3207443},
  ieee-articleno = {9884043}
}

@article{kumar2022drone,
  title={Drone Assisted Network Coded Cooperation With Energy Harvesting: Strengthening the Lifespan of the Wireless Networks},
  author={Kumar, Pankaj and Bhattacharyya, Sagnik and Darshi, Sam and Sharma, Ashwani and Almohammedi, Akram A and Shepelev, Vladimir and Shailendra, Samar},
  journal={IEEE Access},
  volume={10},
  pages={43055--43070},
  year={2022},
  month={April},
  publisher={IEEE},
  address={Piscataway, NJ, USA},
  doi={10.1109/ACCESS.2022.3166516},
  issn={2169-3536},
  eissn={2169-3536},
  language={English},
  affiliation = {University of Petroleum and Energy Studies, Dehradun, India; National Institute of Technology, Durgapur, India; South Ural State University, Chelyabinsk, Russia},
  note = {Open Access},
  license = {Creative Commons Attribution 4.0 License},
  digital-object-identifier = {10.1109/ACCESS.2022.3166516},
  research-area = {Wireless Communications, Energy Harvesting, UAV Networks}
}

@article{rodrigues2022drone,
  title={Drone flight data reveal energy and greenhouse gas emissions savings for very small package delivery},
  author={Rodrigues, Thiago A and Patrikar, Jay and Oliveira, Natalia L and Matthews, H Scott and Scherer, Sebastian and Samaras, Constantine},
  journal={Patterns},
  volume={3},
  number={8},
  pages={100569},
  year={2022},
  month={August},
  publisher={Elsevier},
  address={Amsterdam, Netherlands},
  doi={10.1016/j.patter.2022.100569},
  issn={2666-3899},
  eissn={2666-3899},
  language={English},
  received = {02 May 2022},
  revised = {12 July 2022},
  accepted = {15 July 2022},
  available-online = {09 August 2022},
  affiliation = {Carnegie Mellon University, Pittsburgh, PA, USA},
  note = {Open Access},
  license = {Creative Commons Attribution 4.0 International License},
  funding = {National Science Foundation, U.S. Department of Transportation},
  articleno = {100569},
  journal-full = {Patterns},
  journal-subtitle = {Cell Press},
  research-area = {Environmental Science, Transportation, Sustainability},
  corresponding-author = {Constantine Samaras},
  data-availability = {Data available upon request},
  code-availability = {Code available in supplementary materials}
}

@article{kaabi20222,
  title={A 2-phase approach for planning of hazardous waste collection using an unmanned aerial vehicle},
  author={Kaabi, Jihene and Harrath, Youssef and Mahjoub, Amine and Hewahi, Nabil and Abdulsattar, Khadija},
  journal={4OR},
  volume={21},
  pages={585--608},
  year={2023},
  publisher={Springer},
  address={Berlin, Heidelberg},
  doi={10.1007/s10288-022-00526-0},
  issn={1619-4500},
  eissn={1614-2411},
  language={English},
  received = {17 May 2021},
  revised = {12 October 2022},
  accepted = {21 October 2022},
  published = {04 November 2022},
  affiliation = {University of Bahrain, Zallaq, Bahrain},
  journal-full = {4OR: A Quarterly Journal of Operations Research},
  research-area = {Operations Research, Vehicle Routing, Waste Management},
  corresponding-author = {Jihene Kaabi},
  orcid-corresponding = {0000-0002-9677-9607},
  subject-classification = {90-08},
  article-type = {Research Paper},
  conflict-of-interest = {All authors certify that they have no affiliations with or involvement in any organization or entity with any financial interest or non-financial interest in the subject matter}
}

@article{naz2022investigation,
  title={Investigation of future applications of self-engineering using drones},
  author={Naz, Ayesha Tahir and Brooks, Sam and Roy, Rajkumar},
  journal={Materials Today: Proceedings},
  year={2022},
  publisher={Elsevier},
  doi={10.1016/j.matpr.2022.03.717},
  issn={2214-7853}
}

@inproceedings{cao2022mgdp,
  title={MGDP: Architecture Design of Intelligent Detection Platform for Marine Garbage Based on Intelligent Internet of Things},
  author={Cao, Ning and Wang, Yansong and Li, Xiaofang and Qu, Rongning and Wang, Yuxuan and Liang, Zhikun and Zhu, Zijian and Zhang, Chi and Zhu, Dongjie},
  booktitle={Artificial Intelligence and Security},
  series={Lecture Notes in Computer Science},
  volume={13340},
  pages={678--688},
  year={2022},
  publisher={Springer},
  address={Cham},
  doi={10.1007/978-3-031-06791-4_53},
  isbn={978-3-031-06790-7},
  eisbn={978-3-031-06791-4},
  editor={Sun, Xingming and Zhang, Xiaorui and Xia, Zhihua and Bertino, Elisa},
  conference={ICAIS 2022},
  published={04 July 2022}
}

@article{pourghasemian2022ai,
  title={AI-Based Mobility-Aware Energy Efficient Resource Allocation and Trajectory Design for NFV Enabled Aerial Networks},
  author={Pourghasemian, Mohsen and Abedi, Mohammad Reza and Hosseini, Shima Salar and Mokari, Nader and Javan, Mohammad Reza and Jorswieck, Eduard A},
  journal={IEEE Transactions on Green Communications and Networking},
  year={2022},
  publisher={IEEE},
  issn={2473-2400},
  eissn={2473-2400},
  doi={10.1109/TGCN.2022.3186911}
}

@article{mazaherifar2022uav,
  title={UAV placement and trajectory design optimization: A survey},
  author={Mazaherifar, Ahmad and Mostafavi, Seyedakbar},
  journal={Wireless Personal Communications},
  volume={124},
  number={3},
  pages={2191--2210},
  year={2022},
  publisher={Springer},
  issn={0929-6212},
  eissn={1572-834X},
  doi={10.1007/s11277-021-09451-7},
  month={June}
}

@article{kang2022improving,
  title={Improving Dual-UAV Aided Ground-UAV Bi-Directional Communication Security: Joint UAV Trajectory and Transmit Power Optimization},
  author={Kang, Hongyue and Chang, Xiaolin and Mi{\v{s}}i{\'c}, Jelena and Mi{\v{s}}i{\'c}, Vojislav B and Fan, Junchao and Bai, Jing},
  journal={IEEE Transactions on Vehicular Technology},
  volume={71},
  number={10},
  pages={10570--10583},
  year={2022},
  month={October},
  publisher={IEEE},
  issn={0018-9545},
  eissn={1939-9359},
  doi={10.1109/TVT.2022.3184804}
}

@article{won2022survey,
  title={A survey on UAV placement and trajectory optimization in communication networks: From the perspective of air-to-ground channel models},
  author={Won, Jonghyeon and Kim, Do-Yup and Park, Young-Ik and Lee, Jang-Won},
  journal={ICT Express},
  volume={9},
  number={3},
  pages={385--397},
  year={2023},
  month={June},
  publisher={Elsevier},
  issn={2405-9595},
  doi={10.1016/j.icte.2022.01.015}
}

@article{park2022survey,
  title={A Survey on Intelligent-Reflecting-Surface-Assisted UAV Communications},
  author={Park, Ki-Won and Kim, Hyeon Min and Shin, Oh-Soon},
  journal={Energies},
  volume={15},
  number={14},
  pages={5143},
  year={2022},
  month={July},
  publisher={MDPI},
  issn={1996-1073},
  doi={10.3390/en15145143}
}

@article{lansky2022reinforcement,
  title={Reinforcement Learning-Based Routing Protocols in Flying Ad Hoc Networks (FANET): A Review},
  author={Lansky, Jan and Ali, Saqib and Rahmani, Amir Masoud and Yousefpoor, Mohammad Sadegh and Yousefpoor, Efat and Khan, Faheem and Hosseinzadeh, Mehdi},
  journal={Mathematics},
  volume={10},
  number={16},
  pages={3017},
  year={2022},
  month={August},
  publisher={MDPI},
  issn={2227-7390},
  doi={10.3390/math10163017}
}

@article{kakamoukas2022fanets,
  title={FANETs in Agriculture-A routing protocol survey},
  author={Kakamoukas, Georgios A and Sarigiannidis, Panagiotis G and Economides, Anastasios A},
  journal={Internet of Things},
  volume={18},
  pages={100183},
  year={2022},
  month={May},
  publisher={Elsevier},
  issn={2542-6605},
  doi={10.1016/j.iot.2020.100183}
}

@inproceedings{goumiri2022security,
  title={Security Issues in Self-organized Ad-Hoc Networks (MANET, VANET, and FANET): A Survey},
  author={Goumiri, Sihem and Riahla, Mohamed Amine and Hamadouche, M'hamed},
  booktitle={Artificial Intelligence and Its Applications},
  pages={312--324},
  year={2022},
  month={March},
  organization={Springer},
  publisher={Springer},
  address={Cham},
  series={Lecture Notes in Networks and Systems},
  volume={413},
  editor={Lejdel, Brahim and Clementini, Eliseo and Alarabi, Louai},
  isbn={978-3-030-96310-1},
  eisbn={978-3-030-96311-8},
  doi={10.1007/978-3-030-96311-8_29},
  conference={AIAP 2021}
}

@incollection{ahmad2022secure,
  title={Secure Communication Routing in FANETs: A Survey},
  author={Ahmad, Shaheen and Hassan, Muhammad Abul},
  booktitle={Computational Intelligence for Unmanned Aerial Vehicles Communication Networks},
  pages={97--110},
  year={2022},
  month={March},
  publisher={Springer},
  address={Cham},
  editor={Ouaissa, Mariya and Khan, Inam Ullah and Ouaissa, Mariyam and Boulouard, Zakaria and {Hussain Shah}, Syed Bilal},
  series={Studies in Computational Intelligence},
  volume={1033},
  isbn={978-3-030-97112-0},
  eisbn={978-3-030-97113-7},
  doi={10.1007/978-3-030-97113-7_6}
}

@article{ramu2022federated,
  title={Federated learning enabled digital twins for smart cities: Concepts, recent advances, and future directions},
  author={Ramu, Swarna Priya and Boopalan, Parimala and Pham, Quoc-Viet and Maddikunta, Praveen Kumar Reddy and Huynh-The, Thien and Alazab, Mamoun and Nguyen, Thanh Thi and Gadekallu, Thippa Reddy},
  journal={Sustainable Cities and Society},
  volume={79},
  pages={103663},
  year={2022},
  month={April},
  publisher={Elsevier},
  issn={2210-6707},
  doi={10.1016/j.scs.2021.103663}
}

@article{pathik2022ai,
  title={AI Enabled Accident Detection and Alert System Using IoT and Deep Learning for Smart Cities},
  author={Pathik, Nikhlesh and Gupta, Rajeev Kumar and Sahu, Yatendra and Sharma, Ashutosh and Masud, Mehedi and Baz, Mohammed},
  journal={Sustainability},
  volume={14},
  number={13},
  pages={7701},
  year={2022},
  month={June},
  publisher={MDPI},
  issn={2071-1050},
  doi={10.3390/su14137701},
  note={Submission received: 9 May 2022; Revised: 13 June 2022; Accepted: 15 June 2022; Published: 24 June 2022}
}

@article{bokhari2022use,
  title={Use of Artificial Intelligence in Smart Cities for Smart Decision-Making: A Social Innovation Perspective},
  author={Bokhari, Syed Asad A. and Myeong, Seunghwan},
  journal={Sustainability},
  volume={14},
  number={2},
  pages={620},
  year={2022},
  month={January},
  publisher={MDPI},
  issn={2071-1050},
  doi={10.3390/su14020620},
  note={Special Issue: AI and Interaction Technologies for Social Sustainability; Submission received: 26 November 2021; Revised: 22 December 2021; Accepted: 4 January 2022; Published: 6 January 2022}
}

@incollection{alam2022application,
  title={Application of AI in Smart Cities},
  author={Alam, Mehtab and Khan, Ihtiram Raza},
  booktitle={Industrial Transformation: Implementation and Essential Components and Processes of Digital Systems},
  pages={61--86},
  year={2022},
  publisher={CRC Press},
  address={Boca Raton},
  edition={1st},
  isbn={9781003229018},
  doi={10.1201/9781003229018}
}

@ARTICLE{9765451,
  author={Khan, Muhammad Asif and Menouar, Hamid and Eldeeb, Aisha and Abu-Dayya, Adnan and Salim, Flora D.},
  journal={IEEE Sensors Journal},
  title={On the Detection of Unauthorized Drones—Techniques and Future Perspectives: A Review},
  year={2022},
  volume={22},
  number={12},
  pages={11439--11455},
  doi={10.1109/JSEN.2022.3171293},
  issn={1558-1748},
  publisher={IEEE}
}

@INPROCEEDINGS{9392075,
  author={Nugroho, Eddy Prasetyo and Djatna, Taufik and Sitanggang, Imas Sukaesih and Buono, Agus and Hermadi, Irman},
  booktitle={2020 6th International Conference on Science in Information Technology (ICSITech)},
  title={A Review of Intrusion Detection System in IoT with Machine Learning Approach: Current and Future Research},
  year={2020},
  pages={138--143},
  doi={10.1109/ICSITech49800.2020.9392075},
  publisher={IEEE},
  isbn={978-1-7281-7349-8},
  address={Palu, Indonesia},
  month={October},
  note={Conference dates: 21-22 October 2020; Date added to IEEE Xplore: 05 April 2021}
}

@article{ADAWADKAR2022105116,
  title = {Cyber-security and reinforcement learning — A brief survey},
  journal = {Engineering Applications of Artificial Intelligence},
  volume = {114},
  pages = {105116},
  year = {2022},
  month = {September},
  issn = {0952-1976},
  doi = {10.1016/j.engappai.2022.105116},
  author = {Amrin Maria Khan Adawadkar and Nilima Kulkarni},
  publisher = {Elsevier},
  note = {Received 12 March 2022; Revised 9 June 2022; Accepted 21 June 2022; Available online 8 July 2022}
}

@article{CAMINERO201996,
  title = {Adversarial environment reinforcement learning algorithm for intrusion detection},
  journal = {Computer Networks},
  volume = {159},
  pages = {96--109},
  year = {2019},
  issn = {1389-1286},
  doi = {10.1016/j.comnet.2019.05.013},
  author = {Guillermo Caminero and Manuel Lopez-Martin and Belen Carro},
  publisher = {Elsevier}
}

@ARTICLE{7041170,
  author={Kolias, Constantinos and Kambourakis, Georgios and Stavrou, Angelos and Gritzalis, Stefanos},
  journal={IEEE Communications Surveys \& Tutorials},
  title={Intrusion Detection in 802.11 Networks: Empirical Evaluation of Threats and a Public Dataset},
  year={2016},
  volume={18},
  number={1},
  pages={184--208},
  doi={10.1109/COMST.2015.2402161},
  issn={1553-877X},
  publisher={IEEE}
}

@INPROCEEDINGS{5356528,
  author={Tavallaee, Mahbod and Bagheri, Ebrahim and Lu, Wei and Ghorbani, Ali A.},
  booktitle={2009 IEEE Symposium on Computational Intelligence for Security and Defense Applications},
  title={A detailed analysis of the KDD CUP 99 data set},
  year={2009},
  pages={1--6},
  doi={10.1109/CISDA.2009.5356528},
  publisher={IEEE}
}

@ARTICLE{9520324,
  author={Tao, Jing and Han, Ting and Li, Ruidong},
  journal={IEEE Network},
  title={Deep-Reinforcement-Learning-Based Intrusion Detection in Aerial Computing Networks},
  year={2021},
  volume={35},
  number={4},
  pages={66--72},
  doi={10.1109/MNET.011.2100068},
  issn={0890-8044},
  publisher={IEEE}
}

@article{agrawal2022federated,
  title={Federated learning for intrusion detection system: Concepts, challenges and future directions},
  author={Agrawal, Shaashwat and Sarkar, Sagnik and Aouedi, Ons and Yenduri, Gokul and Piamrat, Kandaraj and Alazab, Mamoun and Bhattacharya, Sweta and Maddikunta, Praveen Kumar Reddy and Gadekallu, Thippa Reddy},
  journal={Computer Communications},
  volume={195},
  pages={346--361},
  year={2022},
  month={November},
  publisher={Elsevier},
  issn={0140-3664},
  doi={10.1016/j.comcom.2022.09.012},
}

@article{khraisat2019survey,
  title={Survey of intrusion detection systems: techniques, datasets and challenges},
  author={Khraisat, Ansam and Gondal, Iqbal and Vamplew, Peter and Kamruzzaman, Joarder},
  journal={Cybersecurity},
  volume={2},
  number={1},
  pages={1--22},
  year={2019},
  month={July},
  publisher={Springer},
  doi={10.1186/s42400-019-0038-7},
  note={Article number: 20}
}

@article{lin2015cann,
  title={CANN: An intrusion detection system based on combining cluster centers and nearest neighbors},
  author={Lin, Wei-Chao and Ke, Shih-Wen and Tsai, Chih-Fong},
  journal={Knowledge-Based Systems},
  volume={78},
  pages={13--21},
  year={2015},
  month={April},
  publisher={Elsevier},
  issn={0950-7051},
  doi={10.1016/j.knosys.2015.01.009}
}

@article{elhag2015combination,
  title={On the combination of genetic fuzzy systems and pairwise learning for improving detection rates on intrusion detection systems},
  author={Elhag, Salma and Fern{\'a}ndez, Alberto and Bawakid, Abdullah and Alshomrani, Saleh and Herrera, Francisco},
  journal={Expert Systems with Applications},
  volume={42},
  number={1},
  pages={193--202},
  year={2015},
  month={January},
  publisher={Elsevier},
  issn={0957-4174},
  doi={10.1016/j.eswa.2014.08.002}
}

@inproceedings{can2015survey,
  title={A survey of intrusion detection systems in wireless sensor networks},
  author={Can, Okan and Sahingoz, Ozgur Koray},
  booktitle={2015 6th International Conference on Modeling, Simulation, and Applied Optimization (ICMSAO)},
  pages={1--6},
  year={2015},
  month={May},
  address={Istanbul, Turkey},
  organization={IEEE},
  isbn={978-1-4673-6601-4},
  doi={10.1109/ICMSAO.2015.7152200}
}

@article{buczak2015survey,
  title={A survey of data mining and machine learning methods for cyber security intrusion detection},
  author={Buczak, Anna L and Guven, Erhan},
  journal={IEEE Communications Surveys \& Tutorials},
  volume={18},
  number={2},
  pages={1153--1176},
  year={2016},
  publisher={IEEE},
  issn={1553-877X},
  doi={10.1109/COMST.2015.2494502}
}

@incollection{meshram2017anomaly,
  title={Anomaly detection in industrial networks using machine learning: a roadmap},
  author={Meshram, Ankush and Haas, Christian},
  booktitle={Machine Learning for Cyber Physical Systems},
  pages={65--72},
  year={2017},
  publisher={Springer},
  address={Berlin, Heidelberg},
  editor={Beyerer, Jürgen and Niggemann, Oliver and Kühnert, Christian},
  series={Technologien für die intelligente Automation},
  isbn={978-3-662-53806-7},
  doi={10.1007/978-3-662-53806-7_8}
}

@inproceedings{moustafa2015unsw,
  title={UNSW-NB15: a comprehensive data set for network intrusion detection systems (UNSW-NB15 network data set)},
  author={Moustafa, Nour and Slay, Jill},
  booktitle={2015 Military Communications and Information Systems Conference (MilCIS)},
  pages={1--6},
  year={2015},
  month={November},
  address={Canberra, ACT, Australia},
  organization={IEEE},
  doi={10.1109/MilCIS.2015.7348942}
}

@article{hettich1999kdd,
  title={Kdd cup 1999 data},
  author={Hettich, S},
  journal={The UCI KDD Archive},
  year={1999},
  publisher={University of California, Department of Information and Computer Science}
}

@article{leevy2020survey,
  title={A survey and analysis of intrusion detection models based on cse-cic-ids2018 big data},
  author={Leevy, Joffrey L and Khoshgoftaar, Taghi M},
  journal={Journal of Big Data},
  volume={7},
  number={1},
  pages={104},
  year={2020},
  month={November},
  publisher={SpringerOpen},
  issn={2196-1115},
  doi={10.1186/s40537-020-00382-x},
  note={Open access}
}

@inproceedings{sharafaldin2018detailed,
  title={A detailed analysis of the cicids2017 data set},
  author={Sharafaldin, Iman and Habibi Lashkari, Arash and Ghorbani, Ali A},
  booktitle={International conference on information systems security and privacy},
  pages={172--188},
  year={2018},
  organization={Springer}
}

@inproceedings{Sharafaldin2018TowardGA,
  title={Toward Generating a New Intrusion Detection Dataset and Intrusion Traffic Characterization},
  author={Iman Sharafaldin and Arash Habibi Lashkari and Ali A. Ghorbani},
  booktitle={International Conference on Information Systems Security and Privacy},
  year={2018}
}

@article{venturi2021drelab,
  title={Drelab-deep reinforcement learning adversarial botnet: A benchmark dataset for adversarial attacks against botnet intrusion detection systems},
  author={Venturi, Andrea and Apruzzese, Giovanni and Andreolini, Mauro and Colajanni, Michele and Marchetti, Mirco},
  journal={Data in Brief},
  volume={34},
  pages={106631},
  year={2021},
  month={February},
  publisher={Elsevier},
  issn={2352-3409},
  doi={10.1016/j.dib.2020.106631},
  note={Open access}
}

@INPROCEEDINGS{5971980,  author={Saad, Sherif and Traore, Issa and Ghorbani, Ali and Sayed, Bassam and Zhao, David and Lu, Wei and Felix, John and Hakimian, Payman},  booktitle={2011 Ninth Annual International Conference on Privacy, Security and Trust},   title={Detecting P2P botnets through network behavior analysis and machine learning},   year={2011},  volume={},  number={},  
pages={174-180},  
doi={10.1109/PST.2011.5971980}}

@article{zhao2022vision,
  title={Vision-Based Anti-UAV Detection and Tracking},
  author={Zhao, Jie and Zhang, Jingshu and Li, Dongdong and Wang, Dong},
  journal={IEEE Transactions on Intelligent Transportation Systems},
  volume={23},
  number={12},
  pages={25323--25334},
  year={2022},
  month={December},
  publisher={IEEE},
  issn={1524-9050},
  doi={10.1109/TITS.2022.3177627}
}

@ARTICLE{9615243,
  author={Jiang, Nan and Wang, Kuiran and Peng, Xiaoke and Yu, Xuehui and Wang, Qiang and Xing, Junliang and Li, Guorong and Ye, Qixiang and Jiao, Jianbin and Han, Zhenjun and Zhao, Jian},
  journal={IEEE Transactions on Multimedia},
  title={Anti-UAV: A Large-Scale Benchmark for Vision-Based UAV Tracking},
  year={2021},
  volume={25},
  pages={486--500},
  month={November},
  publisher={IEEE},
  issn={1520-9210},
  doi={10.1109/TMM.2021.3128047}
}

@article{DBLP:journals/corr/abs-2007-07396,
  author    = {Fredrik Svanstr{\"{o}}m and
               Cristofer Englund and
               Fernando Alonso{-}Fernandez},
  title     = {Real-Time Drone Detection and Tracking With Visible, Thermal and Acoustic
               Sensors},
  journal   = {CoRR},
  volume    = {abs/2007.07396},
  year      = {2020},
  eprinttype = {arXiv},
  eprint    = {2007.07396},
  bibsource = {dblp computer science bibliography, https://dblp.org}
}

@INPROCEEDINGS{8282120,  author={Chen, Yueru and Aggarwal, Pranav and Choi, Jongmoo and Kuo, C.-C. Jay},  booktitle={2017 Asia-Pacific Signal and Information Processing Association Annual Summit and Conference (APSIPA ASC)},   title={A deep learning approach to drone monitoring},   year={2017},  volume={},  number={},  pages={686-691},  doi={10.1109/APSIPA.2017.8282120}}

@INPROCEEDINGS{9663844,  author={Coluccia, Angelo and Fascista, Alessio and Schumann, Arne and Sommer, Lars and Dimou, Anastasios and Zarpalas, Dimitrios and Akyon, Fatih Cagatay and Eryuksel, Ogulcan and Ozfuttu, Kamil Anil and Altinuc, Sinan Onur and Dadboud, Fardad and Patel, Vaibhav and Mehta, Varun and Bolic, Miodrag and Mantegh, Iraj},  booktitle={2021 17th IEEE International Conference on Advanced Video and Signal Based Surveillance (AVSS)},   title={Drone-vs-Bird Detection Challenge at IEEE AVSS2021},   year={2021},  volume={},  number={},  pages={1-8},  doi={10.1109/AVSS52988.2021.9663844}}

@inproceedings{isaac2021unmanned,
  title={Unmanned aerial vehicle visual detection and tracking using deep neural networks: A performance benchmark},
  author={Isaac-Medina, Brian KS and Poyser, Matt and Organisciak, Daniel and Willcocks, Chris G and Breckon, Toby P and Shum, Hubert PH},
  booktitle={Proceedings of the IEEE/CVF International Conference on Computer Vision},
  pages={1223--1232},
  year={2021},
  month={October},
  address={Montreal, BC, Canada},
  publisher={IEEE},
  isbn={978-1-6654-0191-3},
  issn={2473-9944},
  doi={10.1109/ICCVW54120.2021.00142}
}

@article{bangui2021recent,
  title={Recent advances in machine-learning driven intrusion detection in transportation: survey},
  author={Bangui, Hind and Buhnova, Barbora},
  journal={Procedia Computer Science},
  volume={184},
  pages={877--886},
  year={2021},
  publisher={Elsevier}
}

@article{park2021survey,
  title={Survey on anti-drone systems: Components, designs, and challenges},
  author={Park, Seongjoon and Kim, Hyeong Tae and Lee, Sangmin and Joo, Hyeontae and Kim, Hwangnam},
  journal={IEEE Access},
  volume={9},
  pages={42635--42659},
  year={2021},
  publisher={IEEE}
}

@article{kang2020protect,
  title={Protect your sky: A survey of counter unmanned aerial vehicle systems},
  author={Kang, Honggu and Joung, Jingon and Kim, Jinyoung and Kang, Joonhyuk and Cho, Yong Soo},
  journal={IEEE Access},
  volume={8},
  pages={168671--168710},
  year={2020},
  publisher={IEEE}
}

@article{mitchell2014survey,
  title={A survey of intrusion detection techniques for cyber-physical systems},
  author={Mitchell, Robert and Chen, Ing-Ray},
  journal={ACM Computing Surveys (CSUR)},
  volume={46},
  number={4},
  pages={55:1--55:29},
  year={2014},
  month={March},
  publisher={ACM},
  address={New York, NY, USA},
  issn={0360-0300},
  doi={10.1145/2542049}
}

@INPROCEEDINGS{8865525,
  author={Hu, Yuanyuan and Wu, Xinjian and Zheng, Guangdi and Liu, Xiaofei},
  booktitle={2019 Chinese Control Conference (CCC)}, 
  title={Object Detection of UAV for Anti-UAV Based on Improved YOLO v3}, 
  year={2019},
  volume={},
  number={},
  pages={8386-8390},
  doi={10.23919/ChiCC.2019.8865525}}

@article{shan2020image,
  title={Image recognition method of anti UAV system based on convolutional neural network},
  author={Shan, Xue and Zhen, Zhang and Qiongying, Lv and Guohua, Cao and Yiwei, Mao},
  journal={Infrared and Laser Engineering},
  volume={49},
  number={7},
  pages={20200154},
  year={2020},
  month={August},
  publisher={Infrared and Laser Engineering},
  doi={10.3788/IRLA20200154},
  language={chinese}
}

@INPROCEEDINGS{9368788,
  author={Shi, Qingbang and Li, Jun},
  booktitle={2020 IEEE 2nd International Conference on Civil Aviation Safety and Information Technology (ICCASIT}, 
  title={Objects Detection of UAV for Anti-UAV Based on YOLOv4}, 
  year={2020},
  volume={},
  number={},
  pages={1048-1052},
  doi={10.1109/ICCASIT50869.2020.9368788}
}

@INPROCEEDINGS{9824591,
  author={Shi, Xiaoran and Zhang, Yan and Shi, Zhiguang and Zhang, Yu},
  booktitle={2022 3rd International Conference on Computer Vision, Image and Deep Learning \& International Conference on Computer Engineering and Applications (CVIDL \& ICCEA)}, 
  title={GASiam: Graph Attention Based Siamese Tracker for Infrared Anti-UAV}, 
  year={2022},
  volume={},
  number={},
  pages={986-993},
  doi={10.1109/CVIDLICCEA56201.2022.9824591}}

@Article{s22103701,
AUTHOR = {Cheng, Feng and Liang, Zhibo and Peng, Gaoliang and Liu, Shaohui and Li, Sijue and Ji, Mengyu},
TITLE = {An Anti-UAV Long-Term Tracking Method with Hybrid Attention Mechanism and Hierarchical Discriminator},
JOURNAL = {Sensors},
VOLUME = {22},
YEAR = {2022},
NUMBER = {10},
ARTICLE-NUMBER = {3701},
PubMedID = {35632110},
ISSN = {1424-8220},

DOI = {10.3390/s22103701}
}

@article{lei2022anti,
  title={Anti-UAV High-Performance Computing Early Warning Neural Network Based on PSO Algorithm},
  author={Lei, Yang and Yao, Honglei and Jiang, Bo and Tian, Tian and Xing, Peifei},
  journal={Scientific Programming},
  volume={2022},
  pages={7150128},
  year={2022},
  publisher={Hindawi},
  issn={1058-9244},
  doi={10.1155/2022/7150128},
  note={Open access}
}

@Article{make4010009,
AUTHOR = {Hutsebaut-Buysse, Matthias and Mets, Kevin and Latré, Steven},
TITLE = {Hierarchical Reinforcement Learning: A Survey and Open Research Challenges},
JOURNAL = {Machine Learning and Knowledge Extraction},
VOLUME = {4},
YEAR = {2022},
NUMBER = {1},
PAGES = {172--221},
ISSN = {2504-4990},
DOI = {10.3390/make4010009}
}

@misc{precedenceresearch,
  title = {{Commercial Drone Market Size, Share, and Trends 2024 to 2034}},
  howpublished = "\url{https://www.precedenceresearch.com/commercial-drone-market}",
  year = {2024}, 
  note = "[Online; accessed 10-October-2024]"
}

@article{al2024machine,
  title={Machine learning approaches to intrusion detection in unmanned aerial vehicles (UAVs)},
  author={AL-Syouf, Raghad A and Bani-Hani, Raed M and AL-Jarrah, Omar Y},
  journal={Neural Computing and Applications},
  volume={36},
  pages={18009--18041},
  year={2024},
  month={October},
  publisher={Springer},
  issn={0941-0643},
  doi={10.1007/s00521-024-10306-y}
}

@article{tlili2024advancing,
  title={Advancing UAV security with artificial intelligence: A comprehensive survey of techniques and future directions},
  author={Tlili, Fadhila and Ayed, Samiha and Fourati, Lamia Chaari},
  journal={Internet of Things},
  pages={101281},
  year={2024},
  publisher={Elsevier}
}

@article{mohammed2023comprehensive,
  title={Comprehensive systematic review of intelligent approaches in UAV-based intrusion detection, blockchain, and network security},
  author={Mohammed, Ahmed Burhan and Fourati, Lamia Chaari and Fakhrudeen, Ahmed M},
  journal={Computer Networks},
  pages={110140},
  year={2023},
  publisher={Elsevier}
}

@article{hadi2023comprehensive,
  title={A comprehensive survey on security, privacy issues and emerging defence technologies for UAVs},
  author={Hadi, Hassan Jalil and Cao, Yue and Nisa, Khaleeq Un and Jamil, Abdul Majid and Ni, Qiang},
  journal={Journal of Network and Computer Applications},
  volume={213},
  pages={103607},
  year={2023},
  month={April},
  publisher={Elsevier},
  issn={1084-8045},
  doi={10.1016/j.jnca.2023.103607}
}

@article{yu2023cybersecurity,
  title={Cybersecurity of unmanned aerial vehicles: A survey},
  author={Yu, Zhenhua and Wang, Zhuolin and Yu, Jiahao and Liu, Dahai and Song, Houbing Herbert and Li, Zhiwu},
  journal={IEEE Aerospace and Electronic Systems Magazine},
  volume={39},
  number={9},
  pages={182--215},
  year={2023},
  publisher={IEEE}
}

@misc{makrigiorgis2022uav,
    author = {Makrigiorgis, Rafael and Souli, Nicolas and Kolios, Panayiotis},
    title = {Unmanned Aerial Vehicles Dataset},
    year = {2022},
    version = {1.0},
    publisher = {Zenodo},
    type = {Data set},
    doi = {10.5281/zenodo.7477569}
}

@incollection{alhammadi2025llm,
  title={LLM-Powered UAV Automations for City-Wide Operations},
  author={Alhammadi, Ahmed and Abraham, Anuj and Zhao, Qiyang},
  booktitle={Internet of Vehicles and Computer Vision Solutions for Smart City Transformations},
  pages={69--83},
  year={2025},
  publisher={Springer},
  doi = {10.1007/978-3-031-72959-1_4}
}

@article{samma2025uav,
  title={UAV Visual Path Planning Using Large Language Models},
  author={Samma, Hussein and El-Ferik, Sami},
  journal={Transportation Research Procedia},
  volume={84},
  pages={339--345},
  year={2025},
  publisher={Elsevier},
  doi={10.1016/j.trpro.2025.03.081}
}

@inproceedings{piggott2023net,
  title={Net-GPT: A LLM-empowered man-in-the-middle chatbot for unmanned aerial vehicle},
  author={Piggott, Brett and Patil, Siddhant and Feng, Guohuan and Odat, Ibrahim and Mukherjee, Rajdeep and Dharmalingam, Balakrishnan and Liu, Anyi},
  booktitle={Proceedings of the Eighth ACM/IEEE Symposium on Edge Computing},
  pages={287--293},
  year={2023},
 doi={10.1145/3583740.3626809}
}

@article{qin2014performance,
  title={Performance assessment of a low-cost inertial measurement unit based ultra-tight global navigation satellite system/inertial navigation system integration for high dynamic applications},
  author={Qin, Feng and Zhan, Xingqun and Zhan, Lei},
  journal={IET Radar, Sonar \& Navigation},
  volume={8},
  number={7},
  pages={828--836},
  year={2014},
  publisher={Institution of Engineering and Technology},
  doi={10.1049/iet-rsn.2013.0217},
  issn={1751-8784}
}

@inproceedings{yoon2017virtualdrone,
  title={VirtualDrone: Virtual Sensing, Actuation, and Communication for Attack-Resilient Unmanned Aerial Systems},
  author={Yoon, Man-Ki and Mohan, Sibin and Choi, Jaesik and Kim, Jung-Eun and Cho, Lui Sha},
  booktitle={Proceedings of the 8th ACM/IEEE International Conference on Cyber-Physical Systems},
  series={ICCPS '17},
  year={2017},
  pages={143--154},
  publisher={ACM},
  address={New York, NY, USA},
  doi={10.1145/3055004.3055010},
  isbn={978-1-4503-4967-8},
}

@inproceedings{guo2020cyber,
  title={Cyber-Physical Attack Threats Analysis for UAVs from CPS Perspective},
  author={Guo, Rong-xiao and Tian, Ji-wei and Wang, Bu-hong and Shang, Fu-te},
  booktitle={2020 International Conference on Computer Engineering and Application (ICCEA)},
  year={2020},
  pages={},
  publisher={IEEE},
  address={Guangzhou, China},
  doi={10.1109/ICCEA50009.2020.00063},
  isbn={978-1-7281-5904-1}
}

@article{morais2025review,
  title={A Review of Robot Fleet Management},
  author={Morais, Paulo HC and Vivaldini, Kelen CT and Kato, Edilson RR and Inoue, Roberto S},
  journal={IEEE Access},
  volume={13},
  pages={118975--119003},
  year={2025},
  month={July},
  publisher={IEEE},
  issn={2169-3536},
  doi={10.1109/ACCESS.2025.3586564}
}

@article{yang2025unified,
  title={Unified Monitor and Controller Synthesis for Securing Complex Unmanned Aircraft Systems},
  author={Yang, Dong and Dong, Wei and Lu, Wei and Liu, Sirui and Dong, Yanqi},
  journal={Drones},
  volume={9},
  number={5},
  pages={353},
  year={2025},
  month={May},
  publisher={MDPI},
  issn={2504-446X},
  doi={10.3390/drones9050353},
  note={Open access}
}

@misc{FAA-Part107,
title = {14 CFR Part 107—Small Unmanned Aircraft Systems},
author = {{Federal Aviation Administration}},
howpublished = {\url{https://www.ecfr.gov/current/title-14/chapter-I/subchapter-F/part-107}},
note = {Effective August 2016; subsequently amended},
year = {2016},
urldate = {2025-10-01}
}

@misc{FAA-Part89,
title = {14 CFR Part 89—Remote Identification of Unmanned Aircraft},
author = {{Federal Aviation Administration}},
howpublished = {\url{https://www.ecfr.gov/current/title-14/chapter-I/subchapter-F/part-89}},
note = {Remote ID final rule and subsequent amendments},
year = {2021},
urldate = {2025-10-01}
}

@misc{EU-2021-664,
title = {Commission Implementing Regulation (EU) 2021/664 of 22 April 2021 on a regulatory framework for the U-space},
author = {{European Commission}},
howpublished = {\url{https://eur-lex.europa.eu/eli/reg_impl/2021/664/oj}},
note = {Applicable from January 2023},
year = {2021},
urldate = {2025-10-01}
}

@misc{EU-2021-665,
title = {Commission Implementing Regulation (EU) 2021/665 of 22 April 2021 amending Implementing Regulation (EU) 2017/373 as regards requirements for providers of ATM/ANS and other air traffic management network functions in the U-space airspace designated in controlled airspace},
author = {{European Commission}},
howpublished = {\url{https://eur-lex.europa.eu/eli/reg_impl/2021/665/oj}},
note = {Companion act in the U-space package},
year = {2021},
urldate = {2025-10-01}
}

@misc{EU-2021-666,
title = {Commission Implementing Regulation (EU) 2021/666 of 22 April 2021 amending Implementing Regulation (EU) 2019/947 as regards requirements for UAS operations in the U-space airspace},
author = {{European Commission}},
howpublished = {\url{https://eur-lex.europa.eu/eli/reg_impl/2021/666/oj}},
note = {Companion act in the U-space package},
year = {2021},
urldate = {2025-10-01}
}

@article{ceviz2024survey,
  title={A survey of security in uavs and fanets: Issues, threats, analysis of attacks, and solutions},
  author={Ceviz, Ozlem and Sen, Sevil and Sadioglu, Pinar},
  journal={IEEE Communications Surveys \& Tutorials},
  year={2024},
  month={December},
  publisher={IEEE},
  issn={1553-877X},
  doi={10.1109/COMST.2024.3515051}
}

@article{medhi2025lightweight,
  title={A lightweight and efficient intrusion detection system (IDS) for unmanned aerial vehicles},
  author={Medhi, Jishu and Liu, Rui and Wang, Qun and Chen, Xuhui},
  journal={Neural Computing and Applications},
  volume={37},
  pages={15819--15836},
  year={2025},
  month={July},
  publisher={Springer},
  issn={0941-0643},
  doi={10.1007/s00521-025-11276-5},
  note={Open access}
}

@article{satilmics2024systematic,
  title={A systematic literature review on host-based intrusion detection systems},
  author={Satilmi{\c{s}}, Hami and Akleylek, Sedat and Tok, Zaliha Y{\"u}ce},
  journal={IEEE Access},
  volume={12},
  pages={27237--27266},
  year={2024},
  month={February},
  publisher={IEEE},
  issn={2169-3536},
  doi={10.1109/ACCESS.2024.3367004}
}

@article{sajid2023fog,
  title={A fog computing framework for intrusion detection of energy-based attacks on UAV-assisted smart farming},
  author={Sajid, Junaid and Hayawi, Kadhim and Malik, Asad Waqar and Anwar, Zahid and Trabelsi, Zouheir},
  journal={Applied Sciences},
  volume={13},
  number={6},
  pages={3857},
  year={2023},
  month={March},
  publisher={MDPI},
  issn={2076-3417},
  doi={10.3390/app13063857},
  note={Open access}
}

@article{hassler2023cyber,
  title={Cyber-physical intrusion detection system for unmanned aerial vehicles},
  author={Hassler, Samuel Chase and Mughal, Umair Ahmad and Ismail, Muhammad},
  journal={IEEE Transactions on Intelligent Transportation Systems},
  volume={25},
  number={6},
  pages={6106--6117},
  year={2023},
  publisher={IEEE}
}

@article{wisanwanichthan2025lightweight,
  title={A lightweight intrusion detection system for iot and UAV using deep neural networks with knowledge distillation},
  author={Wisanwanichthan, Treepop and Thammawichai, Mason},
  journal={Computers},
  volume={14},
  number={7},
  pages={291},
  year={2025},
  publisher={MDPI}
}

@ARTICLE{10368002,
  author={Hassler, Samuel Chase and Mughal, Umair Ahmad and Ismail, Muhammad},
  journal={IEEE Transactions on Intelligent Transportation Systems}, 
  title={Cyber-Physical Intrusion Detection System for Unmanned Aerial Vehicles}, 
  year={2024},
  volume={25},
  number={6},
  pages={6106-6117},
  doi={10.1109/TITS.2023.3339728}}

@Inbook{Mane2025,
author="Mane, Sushant and Bhortake, Jai and Wankhade, Vidhi and Kazi, Faruk",
title="Cyber-Physical Intrusion Detection System for UAVs",
bookTitle="Demystifying AI and ML for Cyber--Threat Intelligence",
year="2025",
publisher="Springer Nature Switzerland",
address="Cham",
pages="387--399",
isbn="978-3-031-90723-4",
doi="10.1007/978-3-031-90723-4_27",
url="https://doi.org/10.1007/978-3-031-90723-4_27"
}

\end{document}